\documentclass{aa}

\usepackage{graphicx}
\usepackage{txfonts}
\usepackage{booktabs}
\usepackage{multirow}
\usepackage{makecell}
\usepackage{nicefrac}
\usepackage[colorlinks=true,allcolors=blue]{hyperref}
\usepackage{siunitx}

\newcommand{\Halpha}{{\text{H}\ensuremath{\alpha}}}
\newcommand{\vmax}{\ensuremath{V_{\mathrm{max}}}}

\begin{document}

\title{Little Red Dots at $z\sim2$ in EIGER reveal a gentle decline with respect to their peak number density at $z\sim5$}
\titlerunning{$z=2$ LRDs in EIGER}

\authorrunning{S. Kapoor et al.}

\newcommand{\orcid}[1]{}

\author{Shrriya~Kapoor\inst{\ref{inst:ista}}
\and Jorryt~Matthee\orcid{0000-0003-2871-127X}\inst{\ref{inst:ista}}\thanks{Corresponding author: jorryt.matthee@ista.ac.at}
\and Alberto~Torralba\orcid{0000-0001-5586-6950}\inst{\ref{inst:ista}}
\and Ivan~G.~Kramarenko\inst{\ref{inst:ista}}
\and Rongmon~Bordoloi\orcid{0000-0002-3120-7173} \inst{\ref{inst:northcarolina}}
\and Jenny~E.~Greene\orcid{0000-0002-5612-3427}\inst{\ref{inst:princeton}}
\and Edoardo~Iani\orcid{0000-0001-8386-3546}\inst{\ref{inst:ista}}
\and Daichi~Kashino\orcid{0000-0001-9044-1747}\inst{\ref{inst:naoj}}
\and Zhaoran~Liu\orcid{0009-0002-8965-1303}\inst{\ref{inst:MIT_kavli}}
\and Ruari~Mackenzie\orcid{0000-0003-0417-385X}\inst{\ref{inst:EPFL}}
\and Sara~Mascia\orcid{0000-0002-9572-7813}\inst{\ref{inst:ista}}
\and Rohan~P.~Naidu\orcid{0000-0003-3997-5705}\inst{\ref{inst:MIT_kavli}, \ref{inst:hawaii}}
\and Rob~Simcoe\orcid{0000-0003-3769-9559}\inst{\ref{inst:MIT_kavli}}
}

\institute{
    Institute of Science and Technology Austria (ISTA), Am Campus 1, 3400 Klosterneuburg, Austria.\label{inst:ista}
    \and Department of Physics and Astronomy, North Carolina State University, Raleigh, 27695, North Carolina, USA. \label{inst:northcarolina}
    \and Department of Astrophysical Sciences, Princeton University, Princeton, NJ 08544, USA.\label{inst:princeton}
    \and National Astronomical Observatory of Japan, 2-21-1 Osawa, Mitaka, Tokyo, 181-8588, Japan \label{inst:naoj}
    \and MIT Kavli Institute for Astrophysics and Space Research, Massachusetts Institute of Technology, Cambridge, MA 02139, USA.\label{inst:MIT_kavli}
    \and Institute of Physics, Laboratory of Astrophysics, EPFL, Observatoire de Sauverny, 1290 Versoix, Switzerland \label{inst:EPFL}
    \and Institute for Astronomy, University of Hawai'i, 2680 Woodlawn Drive, Honolulu, HI 96822, USA \label{inst:hawaii}
    }

\abstract{We report the discovery of a sample of little red dots (LRDs) at $z \approx 2$ identified from deep JWST/NIRCam imaging and wide-field slitless spectroscopy over $140$ arcmin$^2$ from the EIGER survey. With an improved blind broad-line identification algorithm, we select 19 sources at spectroscopic redshifts $z = 1.55-3.18$ identified via rest-frame near-infrared lines (Paschen-$\beta$, \ion{He}{i}+Pa$\gamma$ and \ion{O}{i}). Based on a range of spectro-photometric and morphological criteria, we classify five of these sources as LRDs and the other 14 as classical active galactic nuclei (AGNs). This classification is corroborated by some X-ray detections among the AGNs. Classical AGNs dominate the number counts above optical luminosities M$_{5100}<-22.5$, whereas the LRD fraction among broad-line sources reaches 100 \% at M$_{5100}\approx-20$. The LRDs span the range in Balmer break strengths seen in the higher redshift populations. Blue-shifted \ion{He}{i} absorption is detected in the two reddest sources. The \ion{He}{i}/Pa$\gamma$ ratio cleanly separates LRDs from classical AGNs and seems to anti-correlate with Balmer break strength, which may be tracing \ion{He}{i} self-absorption at higher gas column densities. Our LRD sample has a similar optical luminosity range as their high-redshift counterparts, corresponding to black hole masses of $\sim10^{6}$ M$_{\odot}$ at the Eddington luminosity. We measure LRD number densities of $\approx 7\times10^{-6}$\,cMpc$^{-3}$ at $z = 1.9-2.5$, which indicates that LRDs are $\sim 30 \times$ rarer than the AGN population at comparable luminosities. Our results confirm the previously reported decline in the LRD number density with respect to $z \approx 5$ based on photometric surveys, although we find the decline to be more gentle than earlier emphasized.}

\maketitle

\section{Introduction}\label{sec:introduction}
Among the many surprises unveiled by JWST is the serendipitously discovered population of `Little Red Dots' \citep[LRDs; e.g.][]{Labbe23,Matthee24,kokorev2024a,Akins24,PerezGonzalez26}, whose physical nature remains elusive \citep[see][for a recent review]{Ubler26}. LRDs are primarily found at redshifts $z\sim5-7$ with number densities $\sim10^{-4}$ to $10^{-5}$ cMpc$^{-3}$. Their red appearance in NIRCam imaging is driven by a combination of strong H$\alpha$ line-emission and very red UV to optical colors, often associated with a strong Balmer break \citep{Setton24b}. These compact, red sources typically show broad H$\alpha$ lines \citep{Hviding25}, which is indicative that they are powered by an active galactic nucleus (AGN; see \citealt{InayoshiHo25} for a review), although alternative powering engines as supermassive stars are also being explored \citep[e.g.][]{Nandal2026, Zwick2026, Chisholm26}. Given that LRDs are far more abundant than quasars at $z\sim5$ and have much lower luminosities, it is likely that we are witnessing a newly identified, poorly understood phase in the formation of supermassive black holes.

The unusual spectral energy distribution (SED) has been a key fuel for the debate on LRDs. Besides the `V-shape' in the rest-frame UV to optical spectrum (with a relatively blue UV slope and strong reddening from $\sim0.3-0.5$ $\mu$m), the LRDs lack the usual X-ray and hot dust emission expected for an AGN \citep[e.g.][]{Yue24,Maiolino2025LbolLx,Xiao25}, nor do they show significant optical variability on $\lesssim10$ yr timescales \citep{Burke25,Kokubo25,Liu26}. The H$\alpha$ lines are very strong relative to the continuum \citep{Yanagisawa26}, and the Balmer lines show steep Balmer decrements \citep{Nikopoulos26} and (in many cases) absorption features \citep{Deugenio25,Matthee26}. These are very rare in classical AGN populations \citep{Shangguan26}, but commonly seen in stellar phenomena \citep[e.g.][]{Martins26}. Models invoking obscuration by dense, partially excited hydrogen gas with a high covering fraction (rather than obscuration by dust) have been successful in explaining various of these features simultaneously \citep[e.g.][]{Inayoshi24,Naidu25,deGraaff25,Torralba25b,Sneppen26}. However, the nature of the LRDs and their role in the formation of galaxies and supermassive black holes remains unclear, with debate around the powering engine and the origin and geometry of the dense gas \citep{Santarelli26,Brazzini26,Madau26,RomanGarza26,Nandal26}.

Given that seemingly distinct physical scenarios can yield similar spectra, additional constraints on the nature of LRDs are highly valuable. Clustering measurements have shown that LRDs typically reside in relatively low mass galaxies and halos \citep{Lin24,Arita25,Matthee25clustering,Pan26}. The steep decline in the bright end of the LRD luminosity function (LF) at $z\sim5$ favors a relatively low mass powering engine with a narrow distribution of Eddington-level luminosities \citep{YMa25b}. Thus, similar to how key insights about galaxy formation have been derived from the evolution of galaxy stellar mass functions \citep[e.g.][]{Peng10,Lilly13}, detailed measurements of the redshift evolution of LRD LFs down to lower redshifts promise to offer complementary constraints on their nature.

A challenge is that photometric selections of compact LRDs \citep[e.g.][]{kokorev2024a,Akins24,Carranza25} are difficult to carry out uniformly across redshift as strong emission-lines and (Balmer) breaks impact broad-band colors depending on the redshift in a complex way and high-resolution data are required to remove interlopers \citep[e.g.][]{Bisigello25,YMa25,Pan26}. These studies find somewhat lower number of LRDs at $z\sim2$ than at $z\sim5-7$, but they are currently still uncertain without extensive spectroscopic follow-up. LRDs have been identified in data from very wide spectroscopic surveys such as the Sloan Digital Sky Survey (SDSS) and the Dark Energy Spectroscopic Instrument (DESI) survey in the $z\approx0.5$ Universe (e.g. \citealt{Lin25_Lowz,Lin26,Park26}, see also \citealt{Izotov08} for early studies). Their number density estimates are $\sim10^{-9}$ cMpc$^{-3}$, orders of magnitude lower than at high redshift, but inherently uncertain due to the complex spectroscopic selection function of these surveys.

Surveys using wide-field slitless spectroscopy (WFSS) with JWST/NIRCam have been very efficient in identifying LRDs through their broad H$\alpha$ line emission at $z\sim5$ \citep{Matthee24,Lin24,Zhang25,CoveloPaz24,Lin25}. In principle, emission-line surveys are more straightforward to carry out uniformly over a larger redshift baseline and WFSS on NIRCam covers H$\alpha$ over $z\approx2.8-6.5$. However, both the sensitivity and the field of view of the NIRCam grism decline significantly towards the bluest and reddest wavelengths of the coverage. WFSS on NIRISS covers H$\alpha$ at cosmic noon, but its smaller field of view and lower resolution challenge the identification of broad line emitters. Alternatively, \cite{Loiacono25} and \cite{NaiduALT24} identified LRDs at $z\approx2.4$ through broad \ion{He}{i} and Paschen-$\gamma$ (Pa$\gamma$) emission. Although these lines are significantly fainter than H$\alpha$ (H$\rm\alpha/Pa\gamma\approx9$ in the $z=2.26$ `Rosetta Stone' LRD identified by \citealt{Juodzbalis24b}), they are covered at $z\sim2$ in JWST's most sensitive WFSS wavelengths.

In this paper, we use deep NIRCam WFSS data from the EIGER survey \citep{Kashino23} to specifically survey for broad \ion{He}{i}, Pa$\gamma$ and Pa$\beta$ emission at $z\sim1.5-3$. The search method builds upon the work of \cite{Matthee24} who partly used EIGER data to identify broad H$\alpha$, but here we use new, optimized detection algorithms, more data and we can benefit from the increased understanding of the broadband 1--5\,$\mu$m SEDs of LRDs. We first detail the data used in Sect. $\ref{sec:data}$. We explain the newly developed broad-line identification algorithm and our means of separating LRDs from other broad-line sources in Sect. $\ref{sec:method}$. The properties of the identified LRDs at $z\approx2$ are described in Sect. $\ref{sec:sample}$. In Sect. $\ref{sec:LF}$ we present the LF and compare this to other measurements in the literature. In Sect. $\ref{sec:nevo}$ we discuss the number density evolution of LRDs. We summarize our results in Sect. $\ref{sec:summary}$. Throughout this work we use a $\Lambda$CDM cosmology as described by \citet{Planck18}, with $\Omega_\Lambda=0.69$, $\Omega_\text{M}=0.31$, and $H_0=67.7$ km\,s$^{-1}$\,Mpc$^{-1}$. Magnitudes are in the AB system.

\section{Data} \label{sec:data}

Our analysis is primarily based on the EIGER survey \citep{Kashino23,Matthee23}, which consists of imaging and deep wide-field slitless spectroscopy obtained with NIRCam \citep{Rieke23} on the JWST. EIGER (PID 1243, PI: Lilly) observed a $\sim4$ by 4 arcmin$^2$ mosaic centered on six luminous quasars at $z\sim6-7$ with the primary goals of studying the relation between galaxies and the intergalactic medium \citep{Kashino26}, characterizing the environments of quasars \citep{Eilers24}, as well as investigating the relation between metal absorption lines and foreground galaxies \citep{Bordoloi24}, including galaxies at cosmic noon. Among the various NIRCam grism surveys that JWST performed \citep[e.g.][]{Oesch23,Sun25,KakiichiCOS3D,Wang26}, EIGER stands out for its combination of depth ($\approx3-10$ hours of exposure time per pixel) and area (143 arcmin$^2$) over six independent fields.

\begin{figure}
    \centering
    \includegraphics[width=\linewidth]{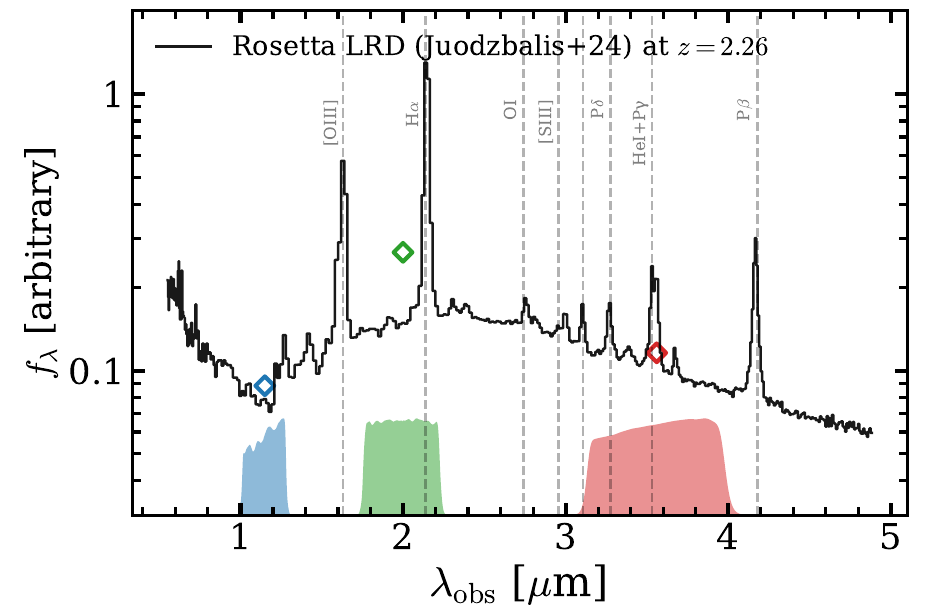}
    \caption{Illustration of the spectrum of the reference `Rosetta Stone' LRD at $z=2.26$ \citep{Juodzbalis24b}. We mark the strongest emission-lines in the spectrum, particularly highlighting Pa$\beta$ and \ion{He}{i}+Pa$\gamma$ that at $z\sim2$ fall in the F356W filter for which we have grism data. We also show the transmission curves of the F115W, F200W and F356W filters whose data is available in the EIGER survey in blue, green and red, respectively. Colored open symbols illustrate the apparent magnitudes in these passbands.  }
    \label{fig:zcoverage}
\end{figure}

The spectral resolution of $R\sim1600$ combined with the sensitivity of the deep NIRCam grism data in the F356W filter has proven to be highly efficient in identifying relatively faint, broad H$\alpha$ emission lines at $z\sim4-5$ that are serendipitously observed in these fields \citep{Matthee24,Lin24}. Here we extend this broad line search to lower redshifts by searching the grism data for broad \ion{He}{i} and Paschen lines in the rest-frame near-infrared, that are redshifted into the $3.15-3.95 \mu$m coverage at redshifts $z\approx2$ (see Fig. $\ref{fig:zcoverage}$). In addition to grism data, there is uniform deep NIRCam imaging available in the F115W and F200W filters taken simultaneously with the WFSS data, as well as direct imaging in the F356W filter. These data are used to characterize the morphologies and the spectral energy distributions of the identified sources. For a detailed description of the observations, data reduction and photometric catalog production we refer to \cite{Kashino26}. In this paper we use 2D grism spectra extracted at the positions of all $\approx120,000$ NIRCam-detected sources identified across the six fields to search for broad emission-lines.

\begin{figure*}
    \centering
    \includegraphics[width=\linewidth]{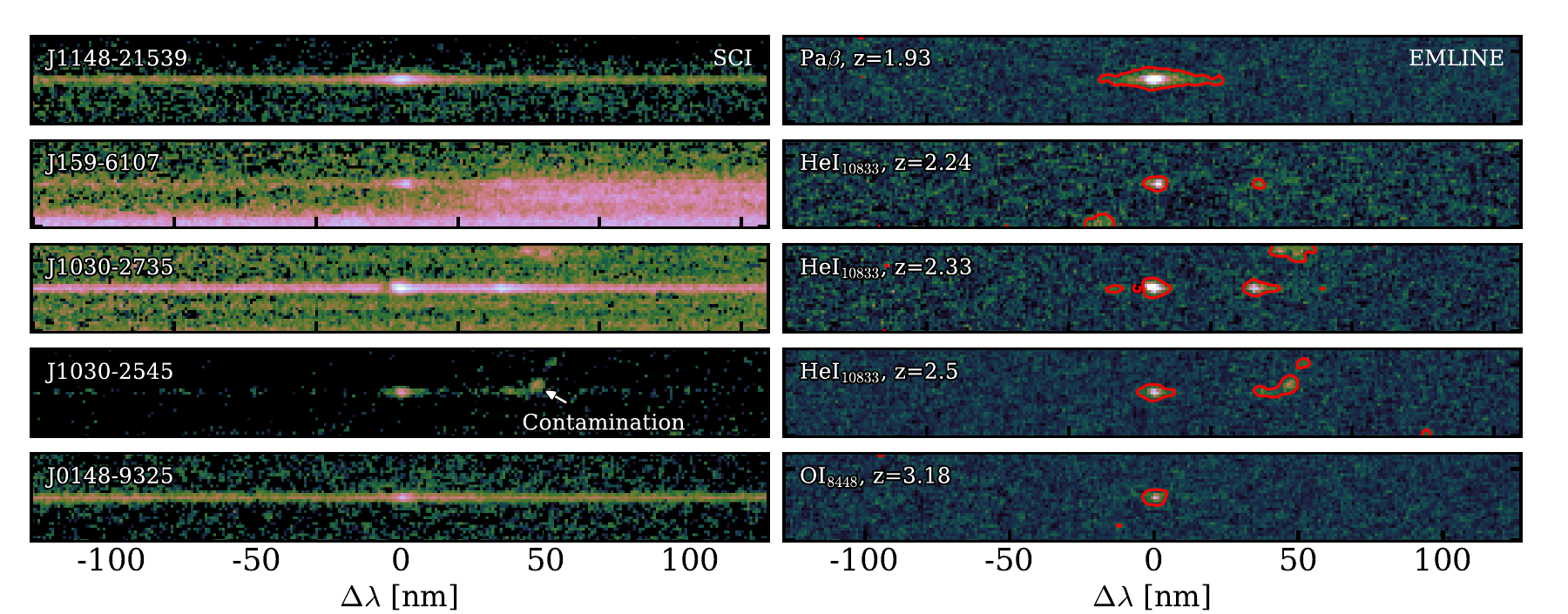}\\
    \caption{The 2D NIRCam WFSS spectra of the 5 LRDs identified in this work based on the blind identification of broad emission-lines in these data. The spectra are centered on the strongest emission-line in each source. The left column shows original reduced grism spectrum (`SCI'), which may be contaminated (as for example seen in J159$\_$6107). The `EMLINE' spectra in the right column show the continuum-removed spectra that were used for initial emission-line identification. Red contours are shown at the $4\sigma$ significance level to highlight the (faint) broad-line features. Similar figures for the broad-line sources in our sample that are not LRDs are shown in Fig. $\ref{fig:searchmethod_nonLRD}$.}
    \label{fig:searchmethod}
\end{figure*}

\section{Search method} \label{sec:method}

\subsection{Broad-line identification in grism data}
Building upon the methodology developed in \cite{Matthee24}, we identify candidate broad-line emitters on continuum-subtracted emission-line spectra (`EMLINE') extracted at the position of NIRCam-detected sources using the code Allegro (Kramarenko \& Matthee in prep.). Candidate broad emission-lines are identified in these EMLINE spectra using Source Extractor \citep{Bertin96}. These candidate lines should be within a distance $\Delta y=1.5$ ($\approx0.1''$) from the center of the trace to reduce contamination by emission-lines from other sources. We impose a signal-to-noise (S/N) threshold of $>15$ measured within a Kron radius (as opposed to a circular aperture-based S/N threshold that is less optimal for broad lines) and we impose an elongation criterion ($A/B> 1.4$, where $A$ and $B$ are the major and minor axes derived from the lines intensity distribution) with an angle $|\theta|<10^{\circ}$ to only select for lines that are extended along the spectral direction ($\theta=0$). Some of the identified broad-lines are shown in Fig. $\ref{fig:searchmethod}$, where we show the EMLINE spectra used for line-identification, as well as the original `SCI' spectra (i.e. not continuum-subtracted) that we used for fitting the line-profiles and fluxes. Some SCI spectra are heavily contaminated by other sources (e.g. J159$\_$6107), but this does not prevent the line-identification.

\begin{table*}
\caption{General properties of the broad-line emitters.}
\label{tab:spectral}
\centering
\setlength{\tabcolsep}{4pt}
\renewcommand{\arraystretch}{1.2}
\begin{tabular}{llcccccccccc}
\hline
ID & $z_\mathrm{spec}$ & RA & Dec & Line &
$M_{5100}$  & $L_\mathrm{\ion{He}{i}+Pa\gamma}$ & \ion{He}{i}/Pa$\gamma$ & QSO & X-rays & Radio & Outflow \\
 & & J2000 & J2000 & &
 & $10^{41}$ erg\,s$^{-1}$ & & & & & \\
\hline
LRDs & & & &  & & & & & & &\\
J1148$\_$21539 & 1.931 & 11:48:05.23 & +52:50:33.8 &  Pa$\beta$ & $-20.4$  & $6.0\pm0.5$ & $1.0\pm0.2$ & & & & \\
J159$\_$6107 & 2.243 & 10:36:54.30 & $-$2:30:28.3 &  \ion{He}{i} & $-19.4$ & $4.0\pm0.5$ & $1.6\pm0.5$ & & & & \\
J1030$\_$2735 & 2.328 & 10:30:18.01 & +5:22:44.9 &  \ion{He}{i} & $-21.5$  & $16.9\pm1.2$ & $1.2\pm0.2$ & & & & \\
J1030$\_$2545 & 2.498 & 10:30:20.54 & +5:24:31.8 &  \ion{He}{i} & $-19.7$ & $4.9\pm0.5$ & $1.5\pm0.4$  & & & & \\
J0148$\_$9325 & 3.177 & 01:48:39.33 & +5:58:17.0 &  \ion{O}{i}$_{8446}$ & $-22.3$ &  & & & & &  \\ \hline

Non-LRDs & & & &  & & & & & & & \\
J0100$\_$18107 & 1.552 & 01:00:07.59 & +28:03:57.5 &  Pa$\beta$ & $-21.6$   &  & & & & \checkmark & \checkmark \\
J1120$\_$11834 & 2.046 & 11:19:57.65 & +6:42:24.5 &   \ion{He}{i} & $-21.7$   & $15.1\pm0.7$ & $3.5\pm0.6$  & \checkmark & &  & \checkmark\\
J0148$\_$7135 & 2.130 & 01:48:33.40 & +6:00:49.3 &   \ion{He}{i} & $-22.9$   & $53.2\pm1.4$ & $3.6\pm0.4$ & & \checkmark & & \checkmark \\
J0148$\_$15902 & 2.178 & 01:48:39.56 & +6:00:37.7 &   \ion{He}{i} & $-21.5$   & $11.4\pm0.7$ & $3.1\pm0.6$  & & & & \checkmark\\
J1148$\_$21459 & 2.192 & 11:48:07.85 & +52:52:50.1 &   \ion{He}{i} & $-21.6$   & $24.2\pm0.5$ & $26.6\pm5.0$ & & \checkmark & & \checkmark \\
J1148$\_$22359 & 2.201 & 11:48:08.82 & +52:52:49.0 &   \ion{He}{i} & $-22.4$   & $26.0\pm0.4$ & $10.6\pm1.0$ & & & & \checkmark \\
J0100$\_$1751 & 2.219 & 01:00:24.11 & +28:01:04.0 &   \ion{He}{i} & $-21.5$   & $25.3\pm0.9$ & $5.0\pm0.9$  & & & & \checkmark\\
J159$\_$6127 & 2.247 & 10:36:53.66 & $-$2:30:49.4 &   \ion{He}{i} & $-22.6$   & $19.9\pm1.8$ & $2.7\pm0.6$ & & & &\checkmark\\
J1148$\_$3444 & 2.334 & 11:48:24.95 & +52:50:47.3 &   \ion{He}{i} & $-24.4$   & $86.6\pm2.1$ & $3.4\pm0.3$ & \checkmark & \checkmark &  & \checkmark\\
J0148$\_$12798 & 2.357 & 01:48:43.75 & +5:59:39.9 &   \ion{He}{i} & $-22.3$   & $28.2\pm0.4$ & $4.7\pm0.3$  & \checkmark & &  & \checkmark\\
J1030$\_$12378 & 2.378 & 10:30:34.46 & +5:26:34.1 &   \ion{He}{i} & $-22.1$   & $76.0\pm2.5$ & $2.4\pm0.2$ & & \checkmark & &\\
J0148$\_$18339 & 2.400 & 01:48:37.67 & +6:00:36.1 &   \ion{He}{i} & $-21.3$   & $4.0\pm0.4$ & $3.4\pm1.0$ & & & & \checkmark \\
J1030$\_$9732 & 2.409 & 10:30:25.66 & +5:25:20.4 &   \ion{He}{i} & $-20.2$ &  $20.8\pm0.4$ & $8.0\pm1.0$ & \checkmark & & & \\
J0148$\_$10704 & 2.435 & 01:48:39.93 & +5:58:17.9 &   \ion{He}{i} & $-25.3$   & $39.1\pm0.7$ & & \checkmark & & & \\ \hline
\end{tabular}
\tablefoot{Line corresponds to the brightest emission-line in the NIRCam WFSS data used to identify the source. The absolute rest-frame optical magnitude M$_{5100}$ is based on template fits to the photometry (Sect. $\ref{sec:LRDident}$). The \ion{He}{i} and Pa$\gamma$ luminosities and fluxes are based on their total fluxes (i.e. narrow+broad). We add flags whether sources are best-fit with QSO templates, X-ray and/or radio detected or show outflows in the \ion{He}{i} or Pa$\beta$ profiles. No deep X-ray data is available in the J159 and J0148 fields.}
\end{table*}

The only photometric selection that we impose is that we remove all objects for which the emission-line strength measured in the F356W grism data is too high for an object's F356W magnitude, as this can only be due to contamination (i.e. the emission-line originates from another object). Then, we visually inspect the 834 candidate broad-lines using the \texttt{Specvizitor}\footnote{\url{https://github.com/ivkram/specvizitor}} tool. The goal of this inspection is to remove sources for which the identified emission-line is not real, or not associated with the galaxy in whose spectrum it was identified. Most of the candidate broad-lines are identified as contamination, such as residuals from the continuum-subtraction at the edge of the spectra from contaminants, residuals from stars or elliptical galaxies with absorption features that mimic broad emission lines. We note that more stringent criteria (smaller $|\theta|$ or $\Delta y$ thresholds) would significantly reduce the number of contaminants, but we would also risk losing some real sources. For example, one can notice that the real Paschen-$\gamma$ line in J1030$\_$2545 (Fig. $\ref{fig:searchmethod}$) is impacted by emission-line contamination, which shifts its centroid in the spatial direction and increases its $\theta$. The sample of candidate broad-lines that survived this visual inspection step was 120. We create optimally-extracted 1D spectra, with the extraction aperture based on their extent along the spatial direction in the EMLINE spectra.

Other sources of contaminants to our broad-line sample are galaxies with real emission-lines originating from \ion{H}{ii} regions that happen to be spatially extended along the dispersion direction. Specifically, for an extended source, we expect that the observed line-width of a spectrally-unresolved line is the convolution of the grism line-spread function (LSF; $R=1600$) and the spatial extent along the dispersion direction (where a NIRCam long-wavelength pixel of 0.063$''$ corresponds to $\approx9.8$ {\AA} in wavelength). We call this the extended-source LSF and use this to verify real broad lines. Typically we find that the extended LSF implies an effective resolution $R\sim400-600$. To distinguish velocity broadening from spatial broadening, we fit the identified candidate broad-lines with the following models: a single, narrow Gaussian convolved with the extended-source LSF, a combination of a narrow and a broad Gaussian with the point-source LSF, as well as a combination of a narrow Gaussian with extended-source LSF and a broad Gaussian with point-source LSF. We use the Bayesian Information Criterion (BIC) to determine whether line-profiles that require broad emission (FWHM$>500$ km s$^{-1}$) beyond the spatial extent of the objects along the dispersion direction are warranted (with a BIC difference of $>10$ compared to models without broad emission). For sources with closely separated \ion{He}{i} + Pa$\gamma$ emission we model independent broad and narrow components for both lines. We fit the H$\alpha$ + [\ion{N}{ii}] profile exclusively with narrow [\ion{N}{ii}] emission. We note that this fitting setup is used for broad-line identification and below (Sect. $\ref{sec:linefitting}$) we describe the specific detailed fitting used for line characterization. In fits with broad component, the width of the narrow component can freely vary with a minimum being the spectral resolution. For our final sample of LRDs, fits that include broad components are strongly preferred with $\Delta$ BIC $>10$ (14 for J159\_6107, $\approx50-150$ for other sources). While our fitting approach is relatively successful in removing pseudo-broad line emitters, we noticed upon visual inspection that in the cases of complex source morphologies (closely separated clumps or disturbed systems visible in the NIRCam images) our extended-source LSFs were not simply quantified by a Gaussian kernel, impacting the constraining power of the BIC analysis. Moreover, the standard continuum-filtering procedure employed in NIRCam grism analysis could add correlated noise to our spectra, which is challenging to quantify in a BIC analysis. We therefore supplemented this analysis with careful visual inspection of their source morphologies across all bands (including the high-resolution imaging data in F200W) and of the 2D grism spectra. After this stage, our total broad-line sample consists of 53 sources. About 60 \% of the sources are removed based on the line-profile fitting and the rest through visual inspection. We note that a simple compactness criterion (e.g. F356W FWHM $<0.3''$) would have sufficed to remove the majority of sources removed through these steps, confirming such selections are fairly complete to broad-line sources. We also remove the six targeted quasars at $z\approx6-7$ for which we identify broad lines as H$\beta$.

\subsection{Redshift identification}
In the F356W grism data used, we cover the following emission-lines: Paschen-$\alpha$ at $z=0.68-1.10$, Paschen-$\beta$ (Pa$\beta$) at $z=1.45-2.08$, \ion{He}{i}+Pa$\gamma$ at $z=1.91-2.64$, [\ion{S}{iii}] at $z=2.30-3.14$, \ion{O}{i}$_{8446}$ at $z=2.73-3.68$ and H$\alpha$ at $z=3.80-5.02$. The redshift identification is split in two steps. In the first step, we use an automated search for secondary (i.e. fainter) lines in the grism coverage to yield a unique redshift solution. In this automated search, secondary lines are detected with the same algorithm optimally designed for broad-lines. All sources with \ion{He}{i} in the sample are also detected in the Pa$\gamma$ line. There are also redshift windows where we additionally cover the relatively strong [\ion{S}{iii}] line. In some cases, we also detect faint [\ion{Fe}{ii}] lines at $z\sim1-3$.

The second step is to identify redshifts for objects with single emission-lines identified from the 2D spectra. In most cases, this specifically means that we need to choose whether a line is H$\alpha$ (as [\ion{N}{ii}] is usually not detected; \citealt{Matthee24}), Pa$\beta$ or Pa$\alpha$. By exploring the spectra of these sources in detail, we detect various faint [\ion{Fe}{ii}] lines (J0100$\_$18107) in an optimally extracted 1D spectrum. For another Paschen-$\beta$ emitter (J1148$\_$21539), we cover \ion{He}{i} + Pa$\gamma$ confirming the redshift. For the remaining single-line detections, we rely on the fact that the equivalent width (EW) of the H$\alpha$ line can reach much higher values than Paschen-series lines (see e.g. Fig. $\ref{fig:zcoverage}$) and the observed EW scales with (1+$z$), such that we classify virtually all single line-emitters as H$\alpha$ lines. Our limited photometric data further supports these redshift identifications.

Out of the 53 broad-line sources, 26 are H$\alpha$ emitters at $z=3.86-5.02$ (including all sources identified in the EIGER data in \citealt{Matthee24}), 16 are identified based on their \ion{He}{i} emission ($z=2.04-2.50$), 8 on [\ion{S}{iii}] emission ($z=2.38-3.03$) and 2 on Pa$\beta$ ($z=1.55-1.93$). One source that deserves special attention is J0148$\_$9325 at $z=3.177$, for which the primary line in our spectrum is \ion{O}{i}$_{8446}$. Here, we determined the redshift thanks to detections of Paschen-11, Paschen-10, Paschen-9 and [\ion{Fe}{ii}]$_{9179}$ that we identified through their similar appearance to the spectrum of the $z=3.5$ LRD identified by \cite{Kokorev25}. No broad Pa$\alpha$ emission was identified. In the rest of the paper we focus on the sources identified through Pa$\beta$, \ion{He}{i} and \ion{O}{i} emission. As the forbidden lines in LRDs are (very) narrow \citep[e.g.][]{Juodzbalis24b}, we ignore sources with broad forbidden [\ion{S}{iii}] lines as those are all classical AGNs. Further, there is no 1:1 comparison sample of LRDs with similar redshift coverage, as is the case for the \ion{He}{i}-selected sample. The key properties of the sources studied in more detail are listed in Table $\ref{tab:spectral}$.

\subsection{LRD classification} \label{sec:LRDident}
Having identified our sample of broad line-emitters at $z\sim2$, here we detail how we separate LRDs from non-LRDs (i.e. classical AGNs) among these sources. Usually, LRDs are defined based on compactness, broad H$\alpha$ lines and a `V-shape', an inflection in their continuum colors from the UV to the optical, although usually only two of these criteria automatically lead to satisfying the third \citep[e.g.][]{Hviding25}. Alternative selections focusing on extreme subsets \citep{Weibel26} or more tolerant colors \citep{Rinaldi26} have also been explored.

At $z\sim2$, \cite{YMa25} previously explored a color selection mimicking a V-shape and a compactness criterion based on ground-based data. As illustrated in Fig. $\ref{fig:zcoverage}$, however, the strong H$\alpha$ emission in LRDs boosts the F200W band photometry at $z=2.2$, whereas F115W captures continuum flux around the Balmer break and the F356W band contains relatively weak lines. For somewhat higher redshifts, on the other hand, H$\alpha$ no longer boosts the F200W photometry and the F115W fully covers blueward of the Balmer break, indicating that the NIRCam colors are strongly redshift dependent within the $z\approx1.5-3$ bin.

The challenge for our classification is that we lack broad-band PRISM spectra from NIRSpec (i.e. we have no H$\alpha$ spectroscopy or cannot spectroscopically measure the continuum colors) and our photometric measurements are limited to only three NIRCam filters (HST/ACS photometry in the F606W and F775W band is available for 4 and 6 out of the 19 sources, respectively, see \citealt{Kashino26} and Table \ref{tab:photometry}). We therefore use a combination of template fitting, \ion{He}{i} and Pa$\gamma$ emission-line properties, and we discuss the morphologies and X-ray and radio data.

We first explore the following templates:
\begin{itemize}
    \item A set of ten dust-reddened quasar templates. These are based on the empirical composite UV to NIR spectrum from \cite{vandenberk2001} and \cite{Glikman06}. We create a set of 10 templates with dust attenuation in the range E$(B-V)$ from 0 to 1, assuming the \cite{Calzetti00} reddening law.
    \item A set of seven template LRDs based on stacks of NIRSpec PRISM spectra of high-redshift LRDs presented in \cite{Matthee26} that cover the rest-frame UV to optical. These consist of four stacks in bins of the redness of the UV to optical color that capture the diversity in LRD spectra at $z\sim3-7$. We interpolate between these four stacks to better sample the variations in Balmer break strengths.
\end{itemize}

Each template is redshifted to the spectroscopic redshift of the source and we calculate the $\chi^2$ based on synthetic photometry in the available filters that calculates the flux density in the filter based on the transmission-weighted mean flux density of the template in question. The normalization is a free parameter. We add a 5 $\%$ error to the photometry to account for uncertainties in zero points, aperture corrections and possibly underestimated correlated noise. We include all available photometry.

Out of the 19 sources at $z=1.55-3.18$, LRD templates are preferred for 13 sources ($|\Delta {\rm BIC}| > 10$, typically $\gtrsim50$), whereas the templates are inconclusive for J159$\_6127$. In most cases at $z\sim2$, the LRD templates do not cover the rest-frame near-infrared, resulting in the F356W photometry being ignored. This severely limits the distinguishing power from these template fits. The key distinguishing features between the best-fit dust-reddened quasar templates and the LRD templates are the V-shape in the LRD spectra (e.g. \citealt{Labbe23,Setton24b}). Additionally, the high EWs of the H$\alpha$ or [\ion{O}{iii}] lines in LRDs in many cases boost the photometry in the F200W filter \citep[see also][]{Torralba26b}, which is challenging to explain with quasar templates that have lower EWs.

Upon inspection of the 13 sources that are best-matched with LRD templates, we note that 8 of them show strong blue-skewed wings in their emission-lines that we associate with powerful outflows. In many cases, the outflow velocities extend up to $>1000$ km s$^{-1}$, suggesting they are AGN-driven. Such strong features of outflows are not seen in LRDs and we also notice that these outflows are typically found in the sources that have lower line EWs. Moreover, among the four broad-line sources that are detected in the Chandra Source Catalog 2.1 \citep{ChandraCatEvans2024}, two are best-fitted by an LRD template, whereas LRDs are known to be extremely X-ray faint \citep{Maiolino2025LbolLx}. The one radio-detected source among our sample (from the LOFAR Two-metre Sky Survey; \citealt{Shimwell26}) is also best-fit with an LRD template. These results suggest that our template fits alone are not sufficiently capable of separating LRDs from classical AGNs, probably due to limitations in the templates as well as the low number of available filters. By comparing the synthetic photometry in detail (see Figs. $\ref{fig:sources_nonLRDs}$ and $\ref{fig:sources_nonLRDs2}$), we note that the F356W photometry matches the quasar templates particularly well, whereas in many cases the scaled template of the Rosetta Stone LRD (which does cover the rest-frame near-infrared) lies below the F356W flux. We therefore speculate that the quasar templates could have been preferred in more sources if our LRD templates extended into the rest-frame NIR.

\begin{figure}
    \centering
    \includegraphics[width=0.99\linewidth]{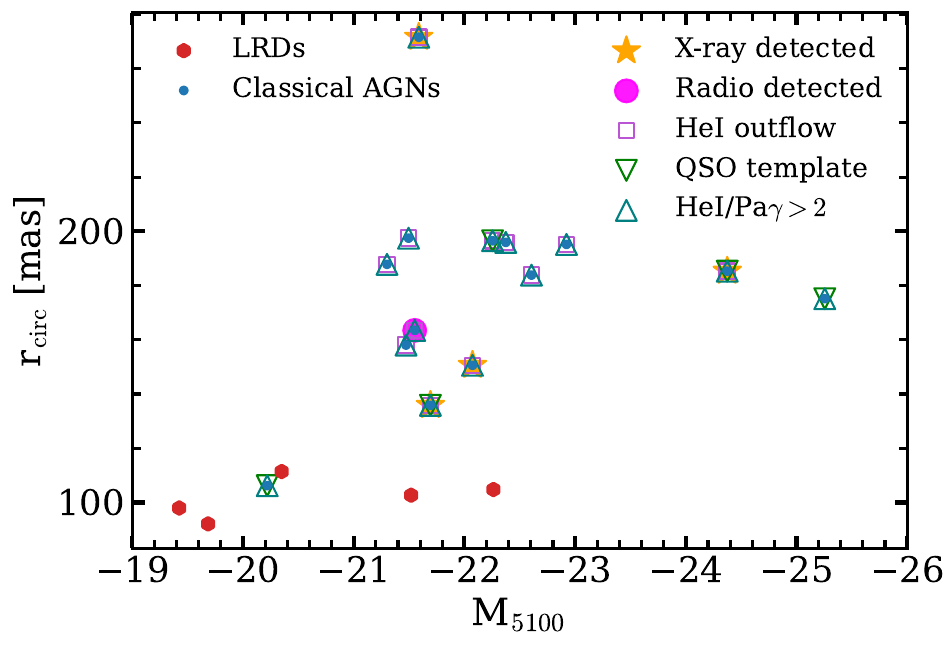}
    \caption{The optical luminosity and size for our broad-line sample. The size is measured on the JWST/NIRCam F356W image. The sources classified as LRDs are shown in red hexagons, whereas classical AGNs are shown with a blue dot. One can clearly identify the transitionary J1030\_9732 among the parameter space populated by LRDs. We highlight the sources that are X-ray detected (yellow star), radio-detected (magenta circle), show strong outflows in \ion{He}{i} (purple squares), are best-fit by a QSO template (green downward pointing triangle) and with a high \ion{He}{i}/Pa$\gamma$ ratio (blue upward pointing triangle). Note that not all sources have X-ray coverage and that the X-ray sensitivity varies from field-to-field.}
    \label{fig:M5100_Rcirc}
\end{figure}

\begin{figure*}
    \centering
    \begin{tabular}{cc}
        \includegraphics[width=0.5\linewidth]{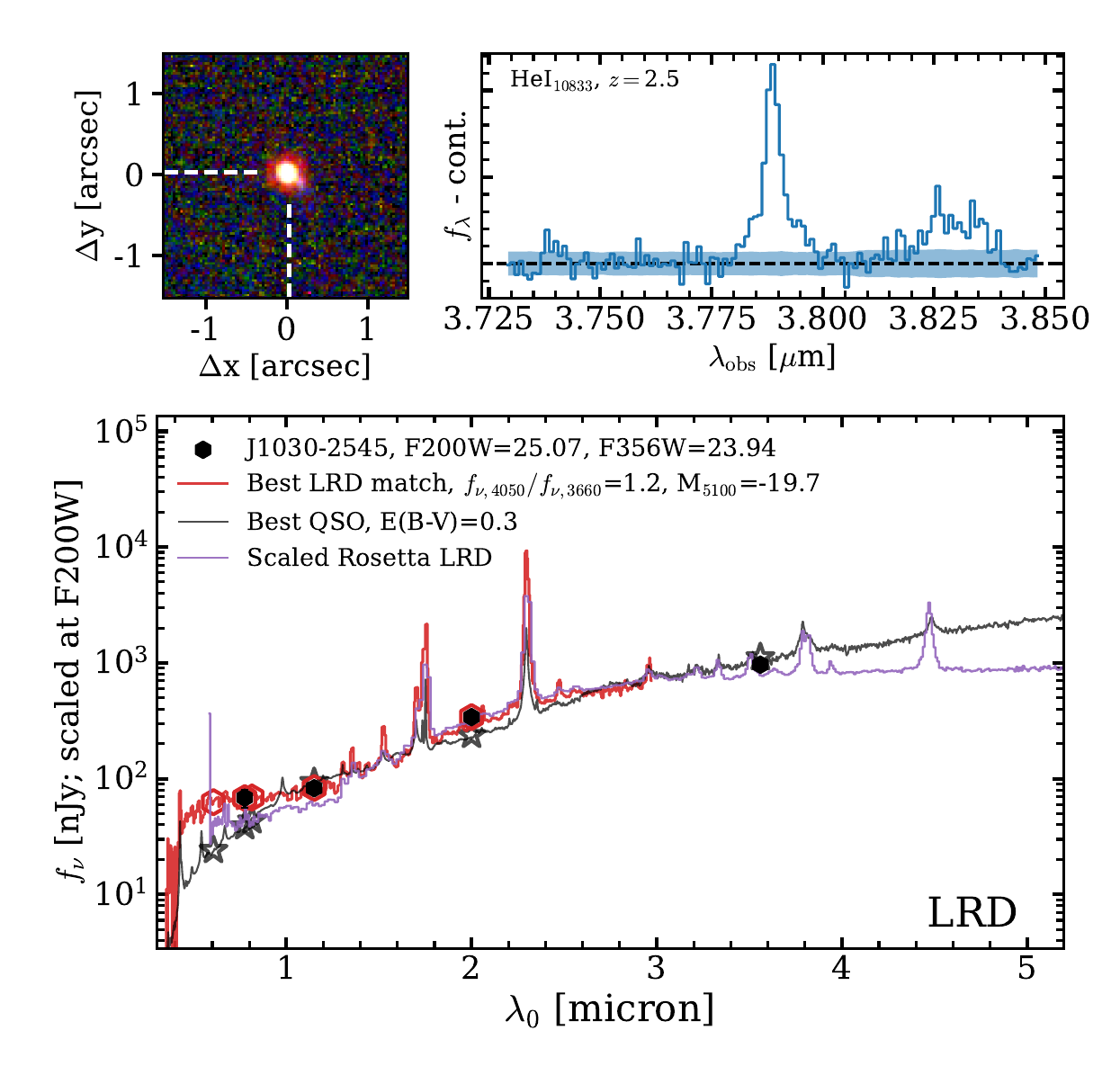} &
 \hspace{-0.8cm}   \includegraphics[width=0.5\linewidth]{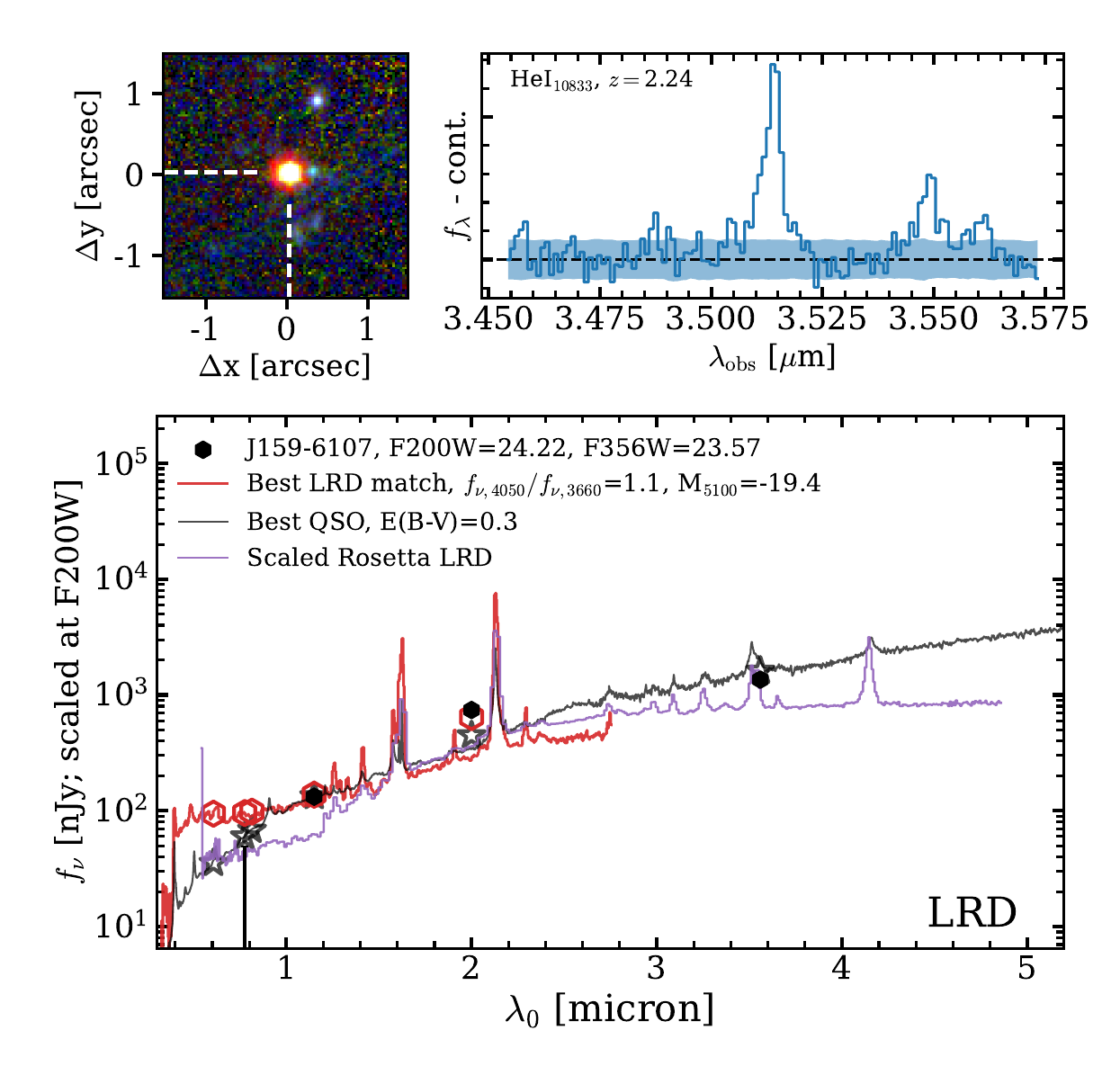} \\
    \includegraphics[width=0.5 \linewidth]{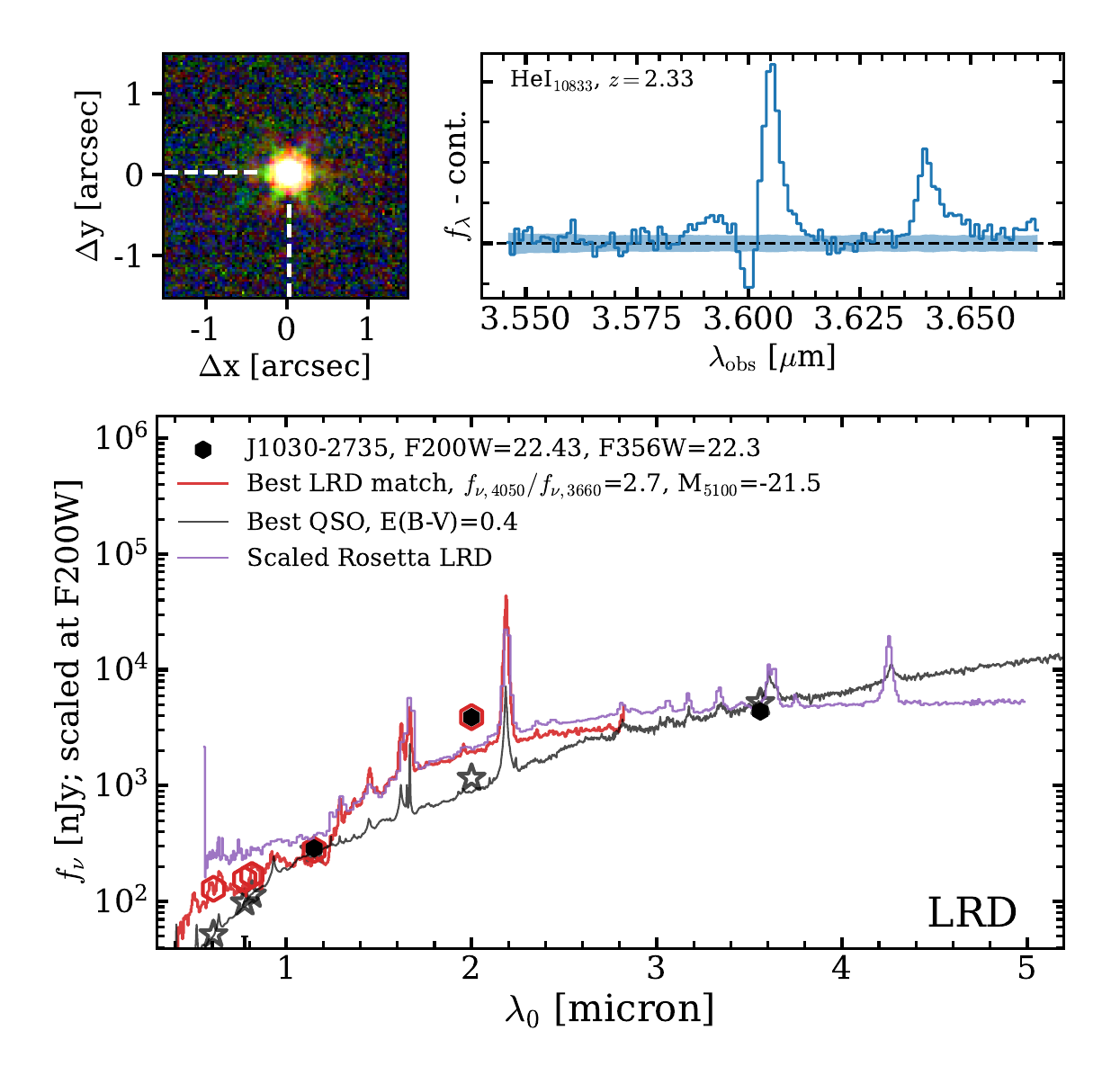} &
\hspace{-0.8cm}    \includegraphics[width=0.5 \linewidth]{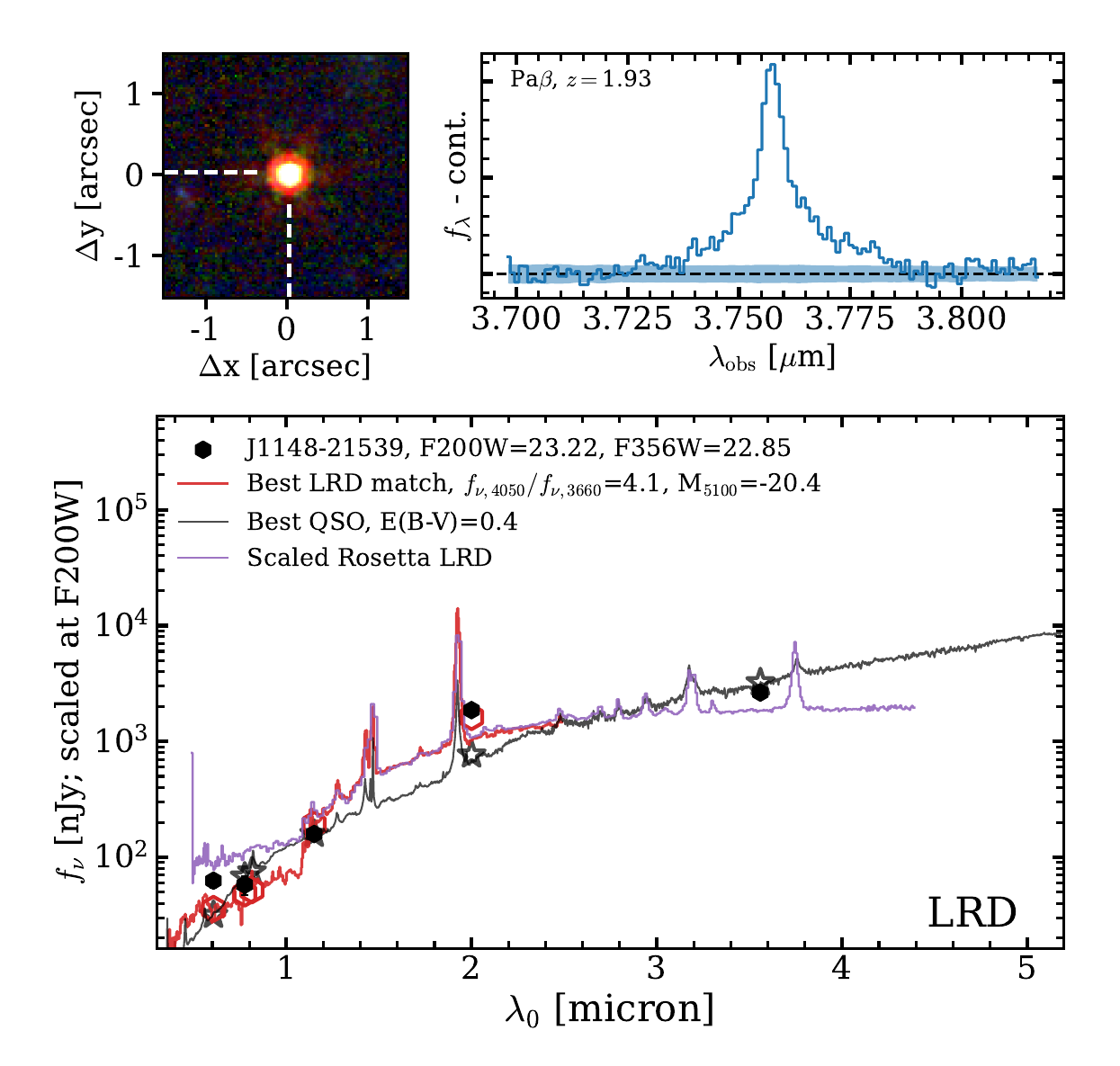} \\
    \end{tabular}
    \caption{Overview of the key observables of the four LRDs at $z=1.9-2.5$ for which we cover \ion{He}{i}+Pa$\gamma$. For each LRD, we show a false-color image based on NIRCam F356W/F200W/F115W data, a zoom-in on the primary-identified emission-line in the continuum-subtracted NIRCam grism spectrum and the spectral energy distribution. Black points show the observed photometry, whereas red, open hexagons show synthetic photometry in the best-fit LRD template and the black, open stars show the synthetic photometry in the best-fit QSO template. For comparison, we also show the prism spectrum of the $z=2.26$ `Rosetta Stone' LRD \citep{Juodzbalis24b} scaled to the redshift and F200W magnitude of each source. }
    \label{fig:sources1}
\end{figure*}

\begin{figure*}
    \centering
    \begin{tabular}{cc}
 \includegraphics[width=0.5 \linewidth]{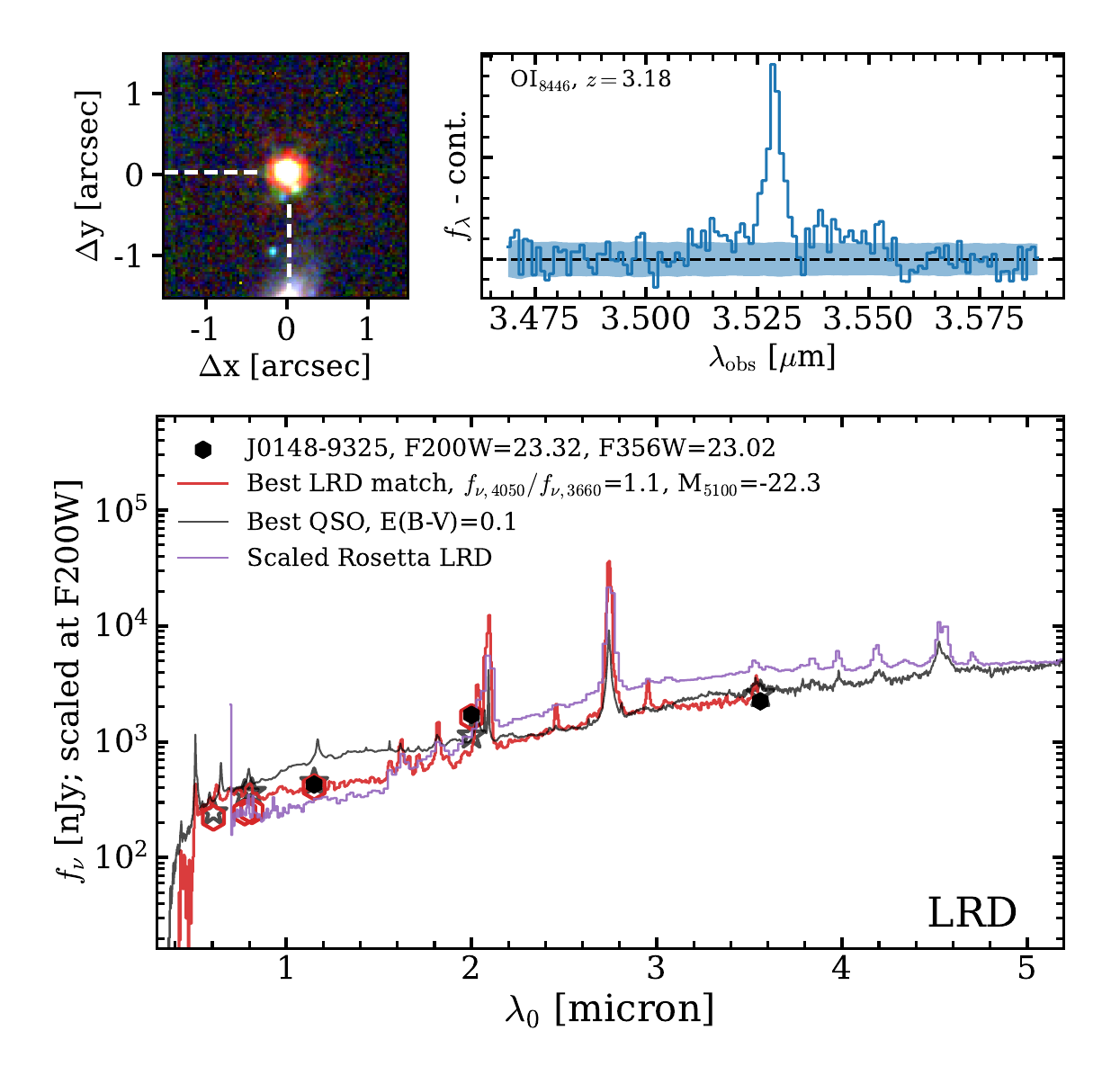} &
 \hspace{-0.8cm}       \includegraphics[width=0.5\linewidth]
    {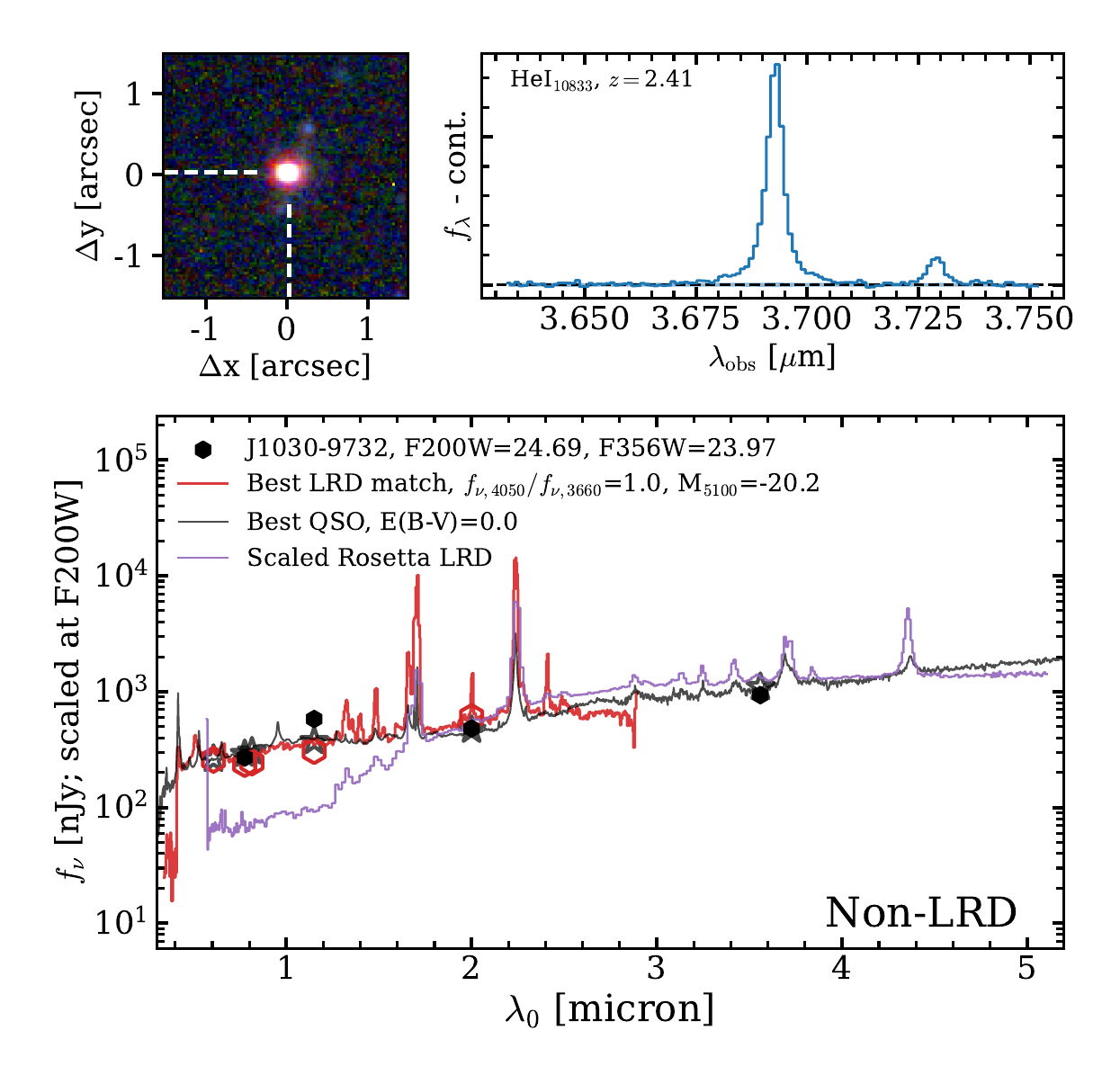} \\
    \end{tabular}
    \caption{Overview of two noteworthy objects identified in our search (as Fig. $\ref{fig:sources1}$). J0148$\_$9325 (left) is classified as a luminous LRD at $z=3.18$. The strongest detected line for this object is \ion{O}{i}$_{8446}$ and various high-order Paschen and [\ion{Fe}{ii}] lines confirm the redshift. We do not cover Pa$\beta$, \ion{He}{i} + Pa$\gamma$, [\ion{S}{iii}] or H$\alpha$ at this redshift. J1030$\_$9732 (right) stands out as the faintest broad-line emitter that we did not conclusively classify as an LRD, because its spectrum is better fitted with a blue quasar template, but its morphological appearance, particularly with faint clumpy neighbors, resembles LRDs. Photometric symbols are as described in Fig. \ref{fig:sources1}.  }
    \label{fig:sources2}
\end{figure*}

To improve our classification, we also investigate the \ion{He}{i}/Pa$\gamma$ line ratios that are excitation diagnostics \citep[e.g.][]{Brinchmann23} as well as their rest-frame near-infrared sizes. In Fig. $\ref{fig:M5100_Rcirc}$ we show the sizes and optical magnitudes of the BL emitters. For the size measurement, we use the circularized radius r$_{\rm circ}=\sqrt{a b}$ that is calculated from the semi-major and semi-minor axes ($a$ and $b$, respectively) measured on the F356W image, without PSF deconvolution. Stars with comparable brightness as the broad-line sources have r$_{\rm circ}\approx100$ mas in these images. We highlight X-ray detections, a radio detection, sources with strong asymmetric emission in \ion{He}{i}, sources that are best-matched with a QSO template as well as sources with \ion{He}{i}/Pa$\gamma$ ratio above $>2$ that is usually considered the demarcation of AGN excitation (see Sect. $\ref{sec:HeIPaGratio}$ where we discuss this in detail). The BL emitters separate in two groups: compact, faint sources that are best-fit by LRD templates, have low \ion{He}{i}/Pa$\gamma$ ratios and no X-ray or fast outflow detection. The more luminous and extended sources all show high \ion{He}{i}/Pa$\gamma$ ratios and/or outflow signatures. Some are best-fit with quasar templates and some show X-ray detections \citep[e.g.][]{Marchesi21}. We note that X-ray data is far from uniform in sensitivity across the fields and is totally missing in the J159 and J0148 fields. Based on these measurements, we classify 9 sources that best-fit with LRD templates as classical AGNs based on their larger sizes and \ion{He}{i} line profile. The only faint, compact source that is classified as a classical AGN is J1030$\_$9732 because of its blue colors and high \ion{He}{i}/Pa$\gamma$ ratio. This source could be a transitory object, similar to blue broad-line sources as GS3073 at high-redshift \citep{Ubler23,Brazzini26}, see also \cite{Loiacono26} who similarly discuss this source (ID 3646 in their paper).

The five LRDs that we identified are shown in Figs. $\ref{fig:sources1}$ and $\ref{fig:sources2}$ and we investigate their properties in more detail below. We show the SEDs of the other non-LRDs in Figs. $\ref{fig:sources_nonLRDs}$ and $\ref{fig:sources_nonLRDs2}$. Our sample of LRDs includes the LRD previously identified by \cite{Loiacono25} (J1030$\_$2735), but others have so far not been published. In Fig. $\ref{fig:sources2}$ we also show the transitionary source J1030$\_$9732 because of its uncertain classification. Interestingly, we notice that two sources (J159$\_$6107 and J159$\_$6127; one source our faintest LRD and the other a luminous classical AGN) are separated by only 23$''$ on sky (corresponding to a projected separation of $\approx190$ proper kpc) and 370 km s$^{-1}$ in the line of sight direction. This distance corresponds roughly to the virial radius of halos hosting AGNs at $z\approx2.5$ \citep{Aird21}, suggesting that the LRD is hosted in a satellite halo that is about to merge with the more luminous AGN. Generally, we also note that between $z=1.9-2.5$, two of the four LRDs are in the J1030 field, and three EIGER fields do not show any LRD at all. This highlights the strong impact that cosmic variance has on the number counts.

\section{Sample properties} \label{sec:sample}
\subsection{Colors and morphologies} \label{sec:colors}
Here we focus on characterizing the properties of the LRDs that we identified. We use our best-fit templates to estimate the Balmer break strength of our sources. The uncertainty that we associate with these Balmer break strengths ($f_{\nu, 4050}/f_{\nu, 3650}$) captures the range of Balmer break strengths within the samples used to construct the templates. Our sample spans a wide range in Balmer break strengths, from $f_{\nu, 4050}/f_{\nu, 3650} \approx1 - 4$.

Beyond the dominant point source that is powering the broad emission-lines, we notice that most LRDs have nearby associated clumps, that can most clearly be distinguished in the F115W imaging data (rest-frame $\approx0.35 \mu$m) where the compact source is usually less dominant. Specifically, J0148$\_$9325, J1030$\_$2545 and J159$\_$6107 have faint, distinct clumps within 0.5$''$, that are probably associated regions of star formation \citep[e.g.][]{Rinaldi25,Baggen26}. No separate clumps are seen around J1030$\_$2735, but its F115W morphology appears somewhat resolved. J1148$\_$21539 similarly has a resolved morphology in the F115W data and further has a companion at a separation of 1.2$''$ ($\approx10$ kpc). We note that none of these clumps impacts the line-profiles as they are not along the dispersion axis.

\begin{figure*}
\centering
    \begin{tabular}{cc}
    \includegraphics[height=5.3cm]{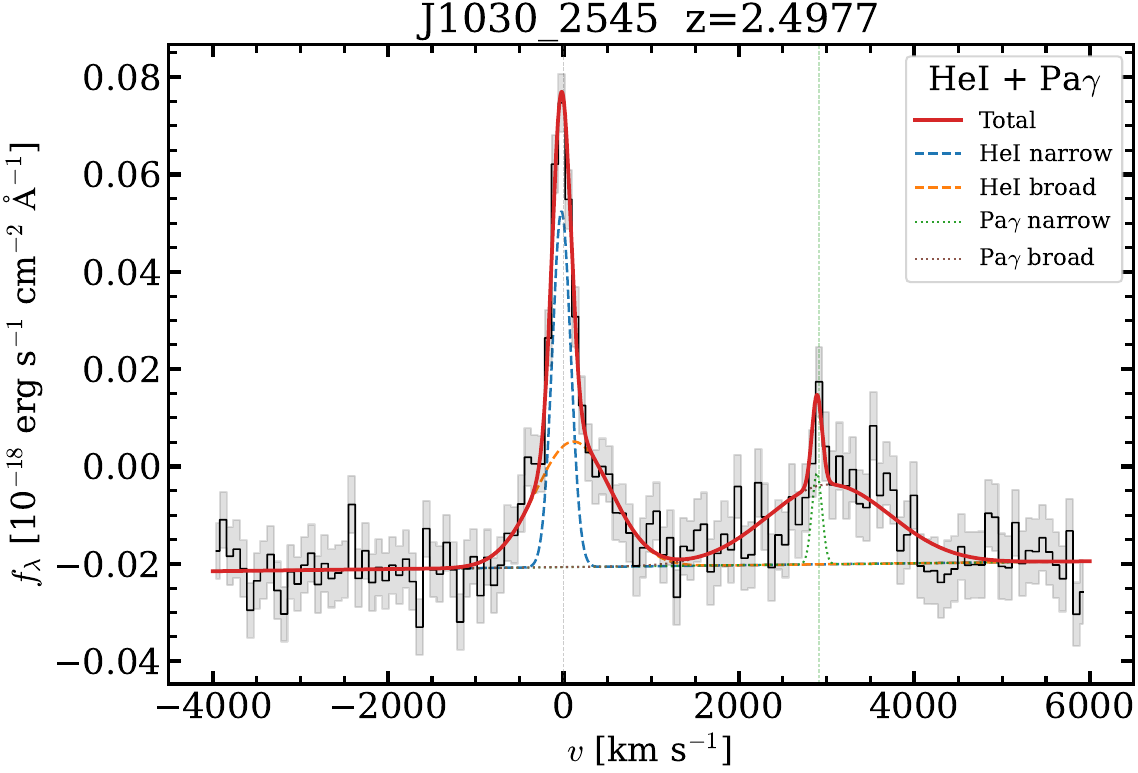} &
    \includegraphics[height=5.3cm]{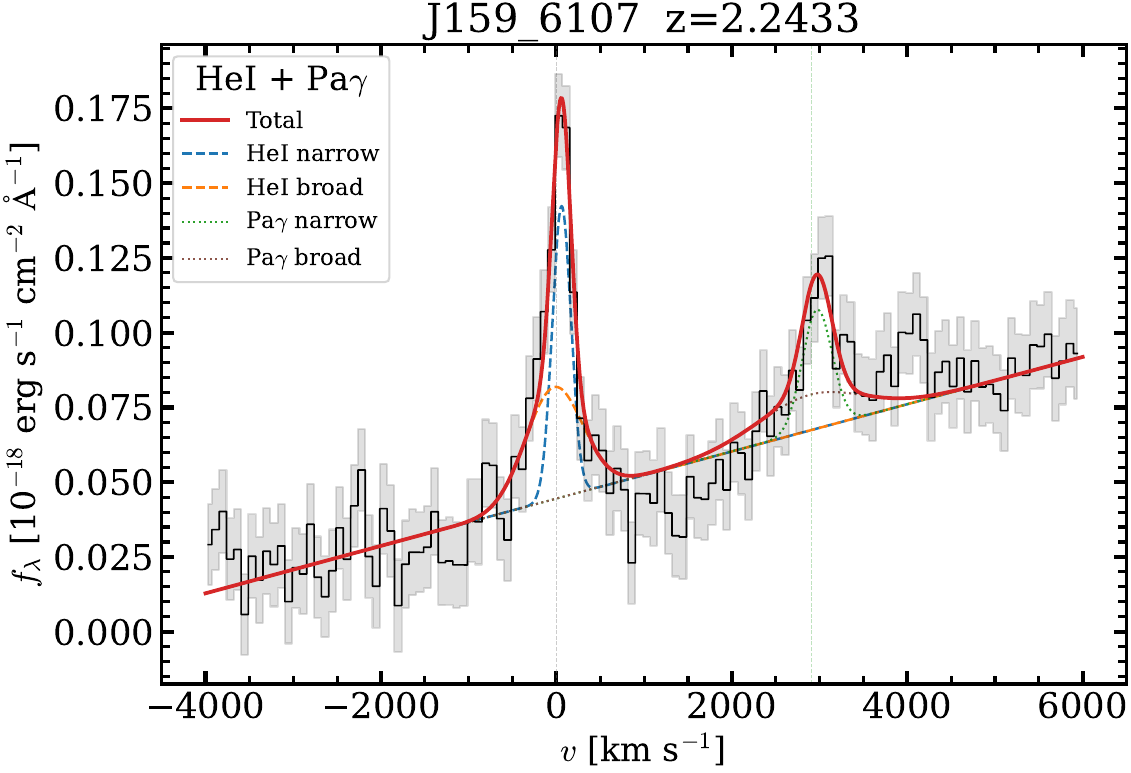} \\    \includegraphics[height=5.3cm]{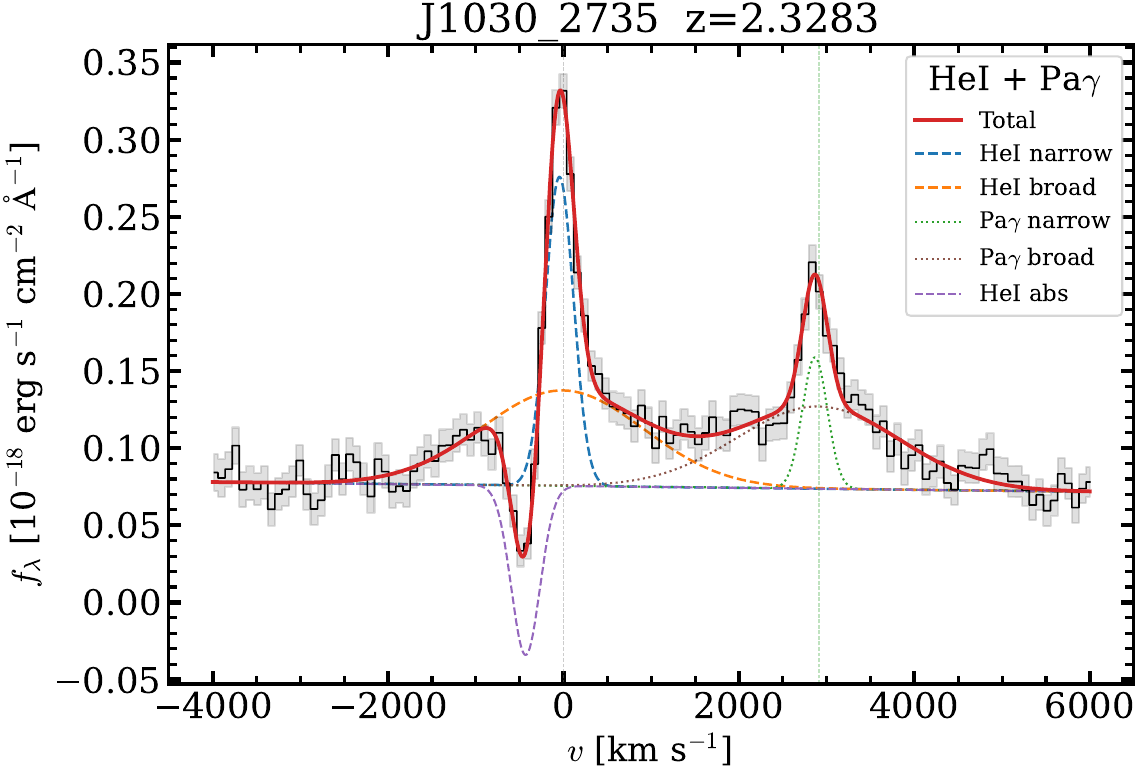} &
    \includegraphics[height=5.3cm]{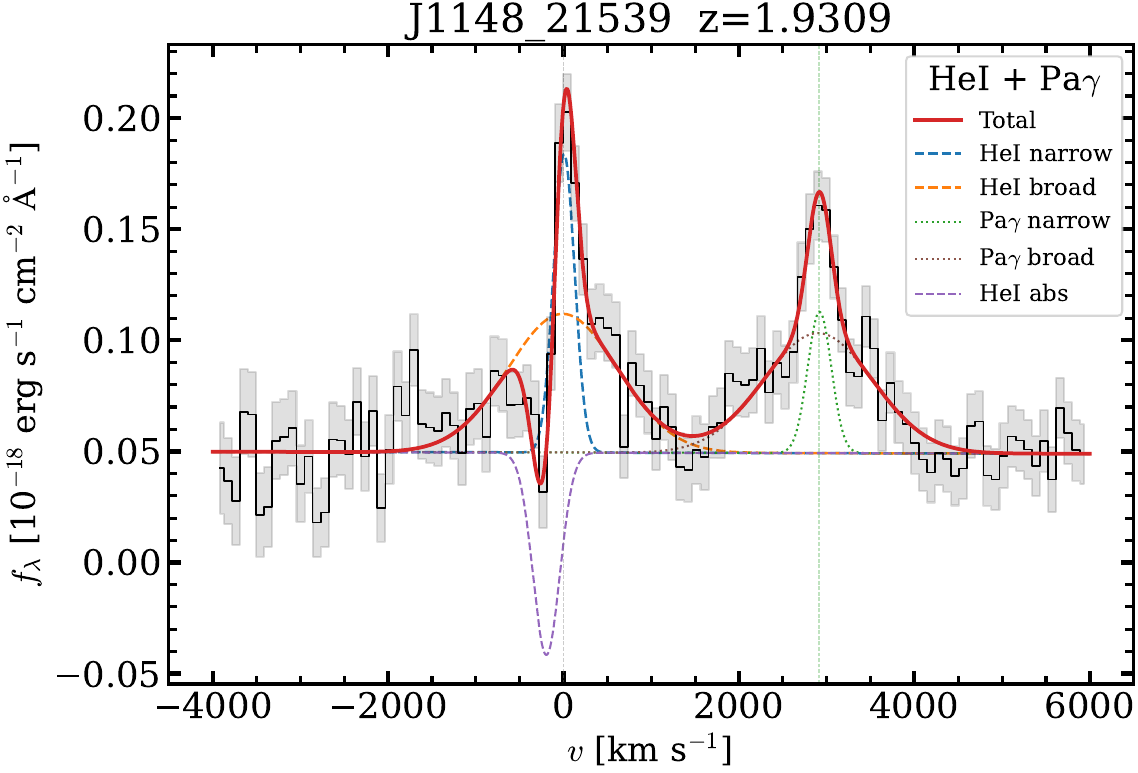} \\
    \end{tabular}
    \caption{The \ion{He}{i}+Pa$\gamma$ profiles of the four LRDs in the sample with coverage of these lines. Red lines show the total fitted profiles, that are composed of narrow and broad Pa$\gamma$ (dotted green and purple, respectively), narrow and broad \ion{He}{i} emission (blue and orange dashed lines). In the case of J1030$\_$2735 and J1148$\_$21539, we also find significant detections of blue-shifted \ion{He}{i} absorption. We note that the line-profiles were fitted on the SCI spectra (see Fig.~\ref{fig:searchmethod}). The continuum emission in J159$\_$6107 is clearly dominated by contamination.} \label{fig:lineprofiles}
\end{figure*}

\subsection{Line-fitting} \label{sec:linefitting}

We obtain fluxes of the relevant emission lines (\ion{He}{i} $\lambda 10830$, Pa$\gamma$, Pa$\beta$) for our sources by fitting a multi-Gaussian model to the extracted 1D `SCI' spectra. While `SCI' spectra may have contamination by continuum emission from other sources (see e.g. J159$\_$6107 in Fig. $\ref{fig:searchmethod}$ that shows clear contamination), we prefer to model this continuum rather than the `EMLINE' spectra, where the broad wings of the line-profiles may be impacted by the running median continuum-removal procedure, in particular in the case of complex combinations of lines such as \ion{He}{i} + Pa$\gamma$. Although high-quality spectra of LRDs suggest that their broad components have exponential wings \citep[e.g.][]{Torralba25b,Kokorev25,Brazzini26}, our spectra are not sensitive enough to distinguish such shapes from simple Gaussians. We therefore use a model consisting of a broad (conservatively $\rm FWHM > 600$ km s$^{-1}$) and a narrow Gaussian component (whose width is allowed to vary as long as it is broader than the LSF). We allow small velocity offsets between both components ($\pm 300$~\unit{km.s^{-1}}), and in the case of the sources with \ion{He}{i}+Pa$\gamma$ we impose the broad and narrow components of both lines to share the same offset, respectively. Additionally, we model the possible blue-shifted \ion{He}{i} $\lambda 10830$ absorption as an additional Gaussian with negative flux. For every object with \ion{He}{i}+Pa$\gamma$ coverage, we fit a version of our model with and without the \ion{He}{i} absorption component, and find that the model including the absorption is preferred for only two objects ($\Delta\rm BIC > 10$; J1030$\_$2735 and J1148$\_$21539, both classified as LRDs). The continuum is modeled as a power law ($f_{\lambda,\rm cont}\propto\lambda^{\alpha}$). Some examples of the multi-Gaussian line fits for LRDs are shown in Fig.~\ref{fig:lineprofiles}. The Pa$\beta$ profile of J1148\_21539 is shown in Fig. $\ref{fig:lineprofiles_PaB}$ and the line-profiles of the non-LRDs are shown in Fig. $\ref{fig:profiles_nonLRDs}$.

For our sample of LRDs, we find typical Gaussian FWHM for the broad components of 1400--2100~$\rm km\,s^{-1}$ with typical uncertainties of $\approx300$ km s$^{-1}$. The line-profiles of the \ion{He}{i}, Pa$\gamma$ and Pa$\beta$ lines are all somewhat different: we find a narrow-to-broad ratio of the fitted lines in the range 0.1--1.0 for Pa$\gamma$, 0.4--0.8 for \ion{He}{i} and 0.3--0.6 for Pa$\beta$. The narrow components of the \ion{He}{i} and Pa$\gamma$ lines have similar widths, with FWHM $\approx300\pm70$ km s$^{-1}$. The broad components also have similar FWHM within the uncertainties. These results are broadly consistent with the flux-ratios measured by \citet{Loiacono25} for J1030\_2735, and by \citet{Juodzbalis24b} with narrow-to-broad ratios of 0.38 and 0.09 for \ion{He}{i} and Pa$\gamma$, respectively.
As discussed above, \ion{He}{i} absorption is significantly detected in two of the LRDs: J1030$\_$2735 and J1148$\_$21539, with offset velocities $-180\pm60\rm\ km\,s^{-1}$ and $-440\pm20\rm\ km\,s^{-1}$, respectively. These velocity offsets are high in comparison to the typical central velocities of \Halpha{} absorbers \citep[$\sim-200$ to $+50$~$\rm km\,s^{-1}$;][]{Matthee26,Davis26}, but this is consistent with other studies that find that the shape of \ion{He}{i} absorbers is very different from the Balmer absorbers \citep[e.g.][]{Juodzbalis24b,Lin25_Lowz}. We note that low-redshift studies show a similar variety in \ion{He}{i} profiles \citep{Lin25_Lowz}, some including very narrow and faint absorption that is extremely challenging to detect with the resolution and sensitivity of our data. No absorption is detected in the spectrum of the transitionary J1030\_9732.

\subsection{Bolometric luminosity and Balmer breaks} \label{sec:Lbolvalues}
While we only have three-band photometry available for the majority of our sources and in many cases (some of) the bands are impacted by strong lines such as H$\alpha$ or [\ion{O}{iii}], we find that the photometry can distinguish well among the range of LRD templates that we used. We use these templates to estimate the absolute magnitude in the rest-frame optical M$_{5100}$, as well as the Balmer break (see Table $\ref{tab:spectral}$).

The absolute magnitudes range from $M_{5100} = -19.4$ to M$_{5100}=-22.3$, which is a similar luminosity range as LRD candidates identified at $z\sim2$ from wide-area ground-based surveys \citep{YMa25} and which also overlaps with samples at $z\sim5$ based on JWST data \citep[e.g.][]{kokorev2024a}. In Fig. $\ref{fig:M5100_Rcirc}$ we show that the rest-frame optical luminosities of the most luminous LRDs overlap with those of the faintest classical AGNs that we identified, but the LRDs dominate at luminosities fainter than M$_{5100} > -21$. The bright-end of the LF appears to drop rather steeply, as we did not identify LRDs brighter than M$_{5100} \lesssim -22$, whereas we identified about a handful of more luminous, classical AGNs in the same covered volume.

The bolometric conversion in LRDs has been a key topic of investigation \citep[e.g.][]{Loiacono25,Greene26}, especially given their relative faintness in the X-rays and infrared \citep[e.g.][]{Akins24,Delvecchio25}. Here, for simplicity, we assume the bolometric conversion $L_{\rm bol}/\lambda L_{5100} = 5.4$ based on empirical data on LRDs \citep{Greene26}, which is roughly half of the conversion in classical AGNs \citep{Kaspi2000}. This corresponds to log$_{10}$(L$_{\rm bol}$/erg s$^{-1}$) $= -0.4 $M$_{5100} +36$. Our sample therefore spans bolometric luminosities from $(5.88 - 80.34)\times10^{43}$ erg s$^{-1}$. For reference, these luminosities correspond to the Eddington luminosities for objects with masses of $(4.7-64)\times10^{5}$ M$_{\odot}$.

The sources show a diversity in SED shapes, with J1030$\_$2735 (see also \citealt{Loiacono25}) and J1148$\_$21539 showing strong Balmer breaks. These breaks coincide with absorption in the \ion{He}{i} line (Fig. $\ref{fig:lineprofiles}$), reminiscent of the correlation between the Balmer break strength and the presence of absorption in the Balmer lines seen in the high-redshift LRDs \citep{Matthee26}. J1030$\_$2545, J0148$\_$9325 and J159$\_$6107 have relatively weak Balmer breaks. While their overall SED shape is relatively similar to the quasar template, we note that the preference for the LRD templates is primarily driven by the excess in the F200W filter due to the high EW of H$\alpha$ and [\ion{O}{iii}].

\subsection{Properties from the \ion{He}{i} and Pa$\gamma$ emission-line ratios} \label{sec:HeIPaGratio}
The rest-frame near infrared emission lines in LRDs are sensitive to the dust attenuation and the gas and ionization conditions. For example, \cite{Chang26} argue that Paschen-$\alpha$ emission should be boosted relative to the other transitions due to radiative decay in the process of H$\beta$ scattering. Moreover, in addition to Balmer line-ratios, the Paschen line ratios also provide complementary constraints on the relative importance of dust attenuation and optical depth effects. The \ion{He}{i}$_{1.083}$ line is a metastable transition and therefore very sensitive to the helium column density \citep[e.g.][]{BWang24a}.

\begin{figure}
    \centering
    \includegraphics[width=1.03\linewidth]{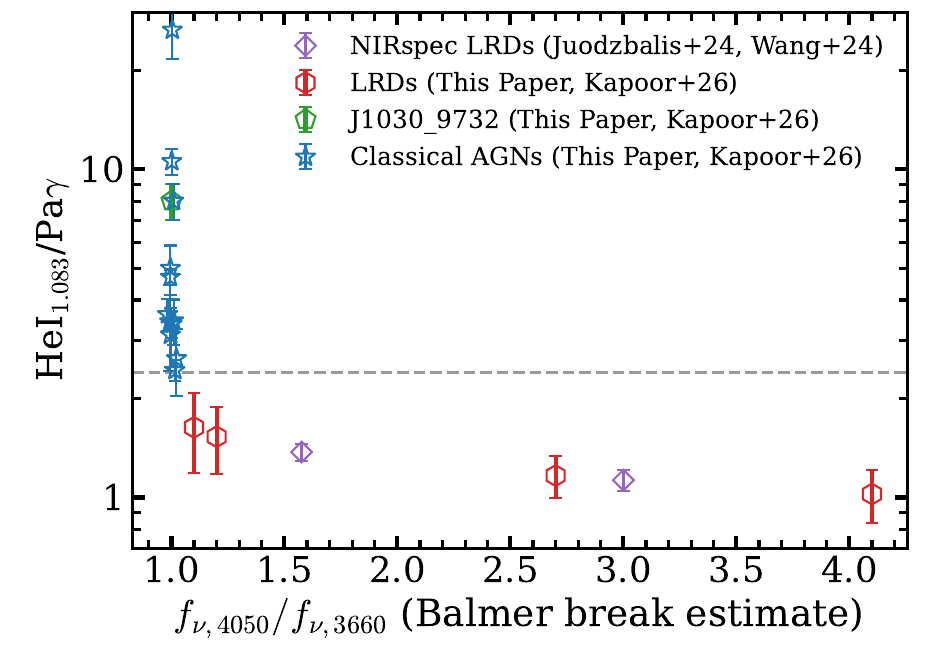}
    \caption{The integrated \ion{He}{i}$_{1.083}$ to Pa$\gamma$ line ratio versus the estimated strength of the Balmer break. LRDs are marked with red hexagons and non-LRDs are shuffled randomly around at a Balmer break strength of 1 for reference. The transitionary source J1030\_9732 whose classification is uncertain is marked with a green pentagon. The dashed horizontal line corresponds to the \ion{He}{i}/Pa$\gamma$ ratio identified to delineate AGN-powered sources (\ion{He}{i}/Pa$\gamma >2.3$) from sources powered by star formation by \cite{Brinchmann23}. The LRDs have relatively strong Pa$\gamma$, and the \ion{He}{i}/Pa$\gamma$ ratio appears to anti-correlate with the Balmer break strength. }
    \label{fig:HeIPaG_BBreak}
\end{figure}

As Fig. $\ref{fig:HeIPaG_BBreak}$ shows, the LRDs (including the $z=2.26, 3.1$ LRDs identified by \citealt{Juodzbalis24b} and \citealt{BWang24a}, respectively) separate from the classical AGNs by their low \ion{He}{i}/Pa$\gamma$ ratio. All normal AGNs have \ion{He}{i}/Pa$\gamma>2.3$, which happens to correspond to the maximum line-ratio identified in \ion{H}{ii} regions ionized by star formation \citep{Brinchmann23} since the \ion{He}{i}/Pa$\gamma$ is a tracer of the ionization parameter due to the differences in the excitation energies. The LRDs are below this line-ratio, highlighting that whatever is dominating the ionizing radiation that power the \ion{H}{ii} regions in the compact centers of the LRDs have a relatively soft ionizing spectrum (see also \citealt{BWang26}). Moreover, the source J1030\_9732 with uncertain classification since its spectrum is bluer than the LBDs, has a high \ion{He}{i}/Pa$\gamma$ ratio, reflecting its transitionary nature. For LRDs, we find that the \ion{He}{i}/Pa$\gamma$ line-ratio tentatively anti-correlates with the estimated strength of the Balmer break (Pearson P-value $\approx0.06$ propagating flux uncertainties). If this trend is confirmed with larger samples, this could imply that the LRDs with weaker breaks have a somewhat higher ionization parameter and/or harder spectra, or that it is caused by an optical depth effect where objects with stronger Balmer breaks have higher gas column densities. This would particularly manifest in stronger \ion{He}{i} self-absorption, lowering the \ion{He}{i}/Pa$\gamma$ ratio. Since star-forming galaxies tend to show relatively low \ion{He}{i}/Pa$\gamma$ ratios and an increasingly strong host galaxy contribution is expected to weaken the Balmer break strength \citep{Barro25,WSun26}, we note that this correlation likely traces the properties of the central engine rather than the relative contribution of the host galaxy emission.

Due to the relatively narrow wavelength coverage of the grism data, we usually only detect a single Paschen line for each LRD. However, thanks to the fortunate redshift of J1148\_21539, we cover \ion{He}{i} + Pa$\gamma$ as well as Pa$\beta$, for which we measure total Pa$\gamma$/Pa$\beta$ flux ratio of $0.42\pm0.05$.
Within the sensitivity of our data (S/N per pixel at the line half maxima $\approx10$ and $5$ for Pa$\beta$ and Pa$\gamma$, respectively), the line-profiles of these lines are consistent with each other, with a broad-to-total flux ratio of $\approx80$ \%. The observed line-ratio is significantly below the case B expectation for a $10^4$ K gas with electron density of 100 cm$^{-3}$ and could indicate a strong attenuation of $A_V\approx4$. However, optical depth effects that could have stronger impact on the lower order Pa$\beta$ line could lead to a lower intrinsic ratio (`case C'), which would mimic the effect of attenuation. This is not implausible given the detection of blue-shifted \ion{He}{i} absorption (see Fig. $\ref{fig:lineprofiles}$) and the strong estimated Balmer break (the strongest in our sample), but follow-up spectroscopy of higher order Paschen and of Balmer lines could differentiate these scenarios conclusively.

\begin{figure}
    \centering
    \includegraphics[width=0.99\linewidth]{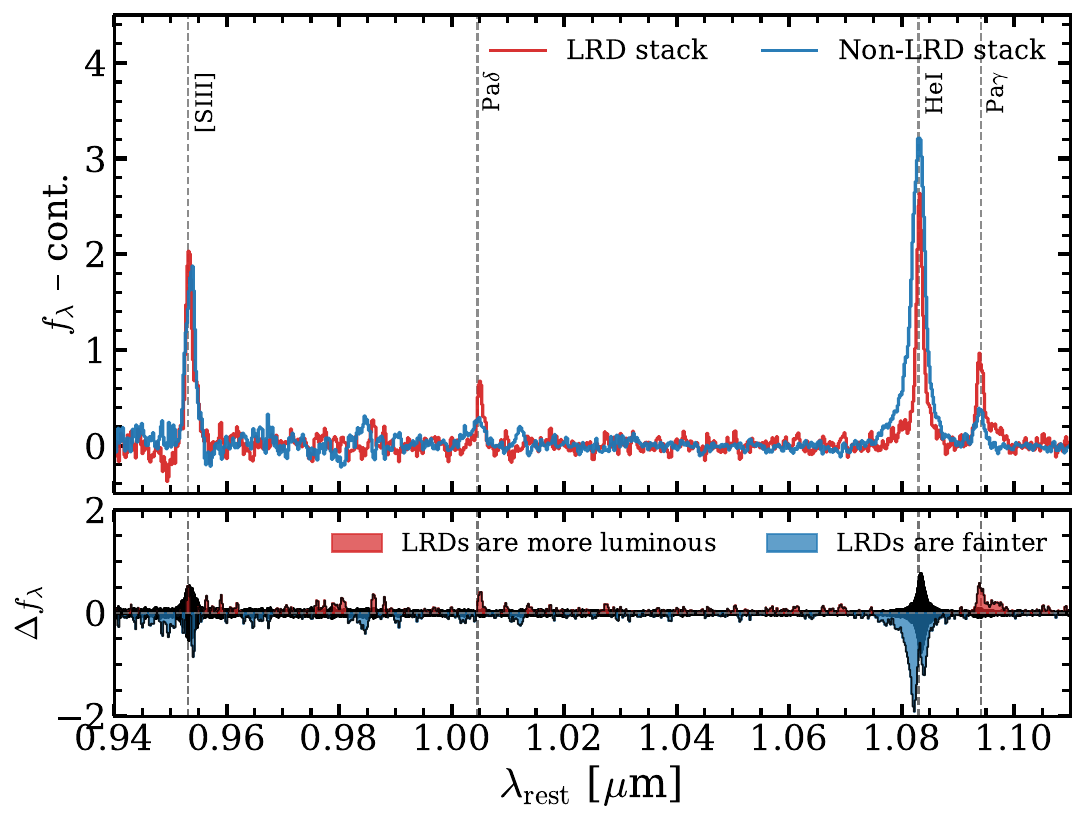}
    \caption{The median-stacked spectra of LRDs (red) and non-LRDs (blue) in this sample, based on continuum-subtracted data from our NIRCam grism spectroscopy in the F356W filter. We only highlight the rest-frame wavelength range covered by the majority of our sample. The stacks highlight the relatively strong and blue-asymmetric \ion{He}{i} emission in the non-LRDs that contrasts with the relatively weak \ion{He}{i} emission in LRDs, especially compared to the Paschen emission.}
    \label{fig:stacks}
\end{figure}

The stacked spectra of the LRDs and the non-LRDs further illustrate the relative weakness of the \ion{He}{i} line in LRDs. As shown in Fig. $\ref{fig:stacks}$, we also detect the fainter Pa$\delta$ line in our stack. These stacks are made in the rest-frame, correcting for differences in luminosity distance but without further normalization. For LRDs, the Pa$\delta$/Pa$\gamma$ ratio is $\approx0.6$, which is in line with case B recombination expectations indicative of little dust attenuation in line with ALMA constraints on high-redshift LRDs \citep[e.g.][]{Casey25,KChen25,Baggen26b}. Besides the stronger \ion{He}{i}, the non-LRDs also show asymmetric, blue-shifted \ion{He}{i} emission which indicates that these sources are characterized by strong outflows. The [\ion{S}{iii}] lines appear similar among LRDs and non-LRDs, which suggests this emission could trace star formation in the host galaxies, similar to as has been argued for [\ion{O}{iii}] in the rest-frame optical \citep{Torralba25b,WSun26}. Note that a faint ``elephant ear'' negative residual can be seen around the [\ion{S}{iii}] line, which is due to the continuum filtering technique of the grism data \citep[e.g.][]{Kotiwale26}.

\section{Luminosity function} \label{sec:LF}
\subsection{Methodology} \label{sec:methodology}
Our simple blind broad emission-line selection in wide-field slitless spectroscopic data enables us to measure the number densities and derive the luminosity function of LRDs at $z\sim2$ complementarily to earlier studies based on photometry \citep[e.g.][]{YMa25}. At $z\sim4-7$, NIRCam grism surveys have derived LRD LFs based on broad H$\alpha$-selected sources with a similar methodology \citep[e.g.][]{Matthee24,Lin24,Zhuang26}. Since we do not cover the H$\alpha$ line in our data, we here focus on the rest-frame optical luminosity function as this enables comparisons to photometrically-selected samples. Based on typical line-ratios between H$\alpha$ and \ion{He}{i}+Pa$\gamma$ \citep{Juodzbalis24b,Lin25_Lowz} in other LRDs, we estimate that the H$\alpha$ luminosities overlap the samples selected at high-redshift, similar to their optical luminosities.

We derive the optical LF using the $1/\vmax{}$ estimator \citep{Schmidt1968}. This method introduces the maximum volume available for each source $i$ in the survey ($\vmax{}_{,\,i}$), which we calculate as follows:

\begin{equation}
    \vmax{}_{,\,i} = \int_{\Omega} \int_{z_{\mathrm{min}}}^{z_{\mathrm{max}}} \frac{dV_c}{d\Omega dz} f_c \left( L_{\ion{He}{i},\,i}, z, \alpha, \delta \right) \, d\Omega \, dz, \label{eq:vmax}
\end{equation}
where $\frac{dV_c}{d\Omega dz}$ is the differential comoving volume, $f_c \left( L_{\ion{He}{i},\,i}, z, \alpha, \delta \right)$ is the completeness function, $d\Omega$ is the solid angle element, and $z_{\mathrm{min}}=1.907$ ($z_{\mathrm{max}}=2.600$) is the lower (upper) limit of the redshift range of \ion{He}{i} emitters in our NIRCam F356W grism data. For bright sources with $f_c \sim 100$\%, we find $\vmax{} \approx 49,000$--$51,000$~cMpc$^3$ depending on the EIGER field, with a total volume of $\vmax{} \approx 305,000$~cMpc$^3$ over 143 arcmin$^2$. The luminosity function is then given by

\begin{equation}
    \phi_{1/V_{\max}} (\langle M_{5100} \rangle) = \frac{1}{\Delta M_{5100}} \sum_i \frac{1}{V_{\max,\,i}},
\end{equation}

where $\langle M_{5100} \rangle$ is the average $M_{5100}$ of a bin, $\Delta M_{5100}$ is the width of the bin, and the sum runs over all sources $i$ in that bin. The LF uncertainties are estimated assuming the Poisson distribution as follows \citep[e.g.,][]{Johnston2011}:

\begin{equation}
    \Delta \phi_{1/V_{\max}} (\langle M_{5100} \rangle) = \sqrt{\frac{1}{\Delta M_{5100}^2} \sum_i \frac{1}{V_{\mathrm{max},\,i}^2}}.
\end{equation}

We estimate cosmic variance errors following the \cite{Trenti08} cosmic variance calculator that takes the survey geometry into account. We assume LRDs are hosted in halos with masses $10^{11}$ M$_{\odot}$ \citep{Arita25}. Because we use six independent fields, the cosmic variance errors are negligible (0.048 dex) compared to the Poisson uncertainties that totally dominate the total error budget. Cosmic variance would only be important if LRDs would be hosted in extremely massive halos with mass $10^{14}$ M$_{\odot}$, which is highly unlikely.

Additionally, we derive an upper limit of the 84\% confidence interval of the LF at the bright luminosity end where no sources are detected following \citet{Gehrels86} and assuming $ f_c = 100 \%$.

\begin{table}
\centering
\caption{Luminosity function of LRDs and non-LRDs at $z=1.9-2.5$.}
\label{tab:ndsen}
\begin{tabular}{lccc}
\hline
M$_{5100}$ & N  & $\langle f_{c} \rangle_{V}$ & $\Phi$ (cMpc$^{-3}$ mag$^{-1}$) \\ \hline
LRDs & & & \\
$-$22.5 & 0 & - & $<6.04\times10^{-6}$\\
$-$20.9 [$-$22,$-$20] & 2 & 0.93 & $(3.53\pm2.50)\times 10^{-6}$ \\
$-$19.6 [$-$20,$-$19] & 2 & 0.84 & $(7.81\pm5.54)\times 10^{-6}$ \\ \hline
LRDs incl. J1030\_9732 & & & \\
$-$22.5 & 0 & - & $<6.04\times10^{-6}$\\
$-$20.7 [$-$22,$-$20] & 3 & 0.94 & $(5.25\pm3.04)\times 10^{-6}$ \\
$-$19.6 [$-$20,$-$19] & 2 & 0.84 & $(7.81\pm5.54)\times 10^{-6}$ \\ \hline
Non-LRDs & & & \\
$-$26.4 & 0 & - & $<6.04\times10^{-6}$\\
$-$24.8 [$-$25.5,$-$23.5] & 2 & 0.95 & $(3.47\pm2.46)\times10^{-6}$ \\
$-$22.8 [$-$23.5,$-$22.5] & 2 & 0.94 & $(6.95\pm4.93)\times10^{-6}$ \\
$-$22.0 [$-$22.5,$-$21.5] & 5 & 0.94 & $(17.39\pm7.82)\times10^{-6}$ \\
$-$21.1 [$-$21.5,$-$19.5] & 4 & 0.90 & $(7.28\pm3.66)\times10^{-6}$ \\ \hline
\end{tabular}
\tablefoot{The luminosity function is given as a function of optical luminosity M$_{5100}$ for LRDs (with and without the transitionary source J1030\_9732) and non-LRDs. The method to calculate the completeness, $f_{c}$, is detailed in Sect. $\ref{sec:methodology}$. We list the median luminosity and the ranges of each bin. Errors include Poisson noise and cosmic variance.}
\end{table}

The completeness function $f_c$ is a critical part of our measurement of $\vmax{}_{,\,i}$. The probability of detecting an object depends on its \ion{He}{i} line luminosity and the noise level at a given position and redshift in the grism data. We calculate the completeness building upon the methodology developed for narrow [\ion{O}{iii}] emitters in \cite{Matthee23}. We inject a fake \ion{He}{i} line in 2D EMLINE grism spectra extracted at random positions in the field of view. The injected line profile is modeled as a sum of narrow and broad Gaussian components and is representative of the detected LRDs. We perform recovery experiments with our detection algorithm as a function of the line flux and find that $f_c$ is self-similar and only depends on the ratio between the line flux and the propagated pixel error in the grism data (see also \citealp{Herenz19}). The resulting universal completeness curve is shown in Fig.~$\ref{fig:completeness}$ (red). Using this curve, we find that the volume-averaged completeness for our sources is high, $\langle f_c \rangle_V \approx 84$--$95$ \%. In addition to LRDs, we also derive the completeness for our broad-line sample that are not LRDs. As these generally have broader line-profiles, their completeness is typically lower at fixed luminosity (Fig.~$\ref{fig:completeness}$, blue), but this is offset by their higher luminosity. Our completeness simulations do not account for our final quality checks that are designed to remove pseudo broad-lines due to extended and/or complex source morphologies. While these excluded a large number of sources, we note that no sources were removed with similar combination of NIRCam colors, line equivalent widths and compact (unresolved) morphologies as our final sample, suggesting that no real source was excluded by these additional steps. In fact, a simple compactness criterion would have also sufficed to remove the majority of such pseudo broad-line sources.

\subsection{Optical LF}
\begin{figure}
    \centering
    \includegraphics[width=0.99\linewidth]{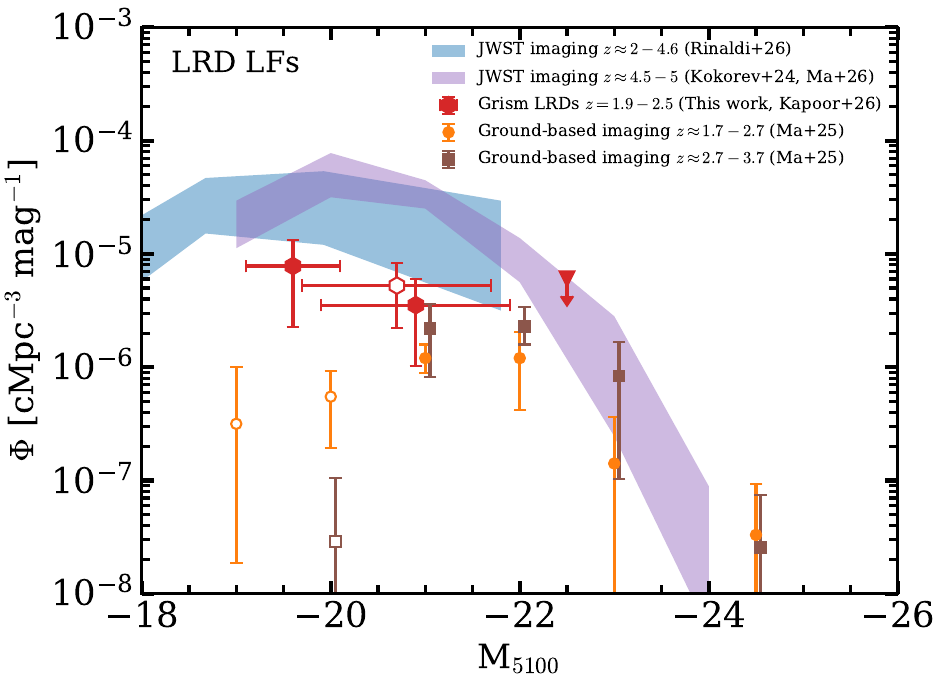}
    \caption{The optical luminosity function of LRDs at $z\approx2-5$. We compare our measurements (red hexagons and upper limit) to the measurements from \cite{YMa25} at $z\approx1.7-2.7$ and $z\approx2.7-3.7$ (orange and brown points, respectively, with open symbols highlighting incomplete luminosities) based on wide area ground-based data, as well as JWST-based measurements at $z\approx2-4.6$ ($\langle z\rangle =3.7$) from \cite{Rinaldi26} and at $z\approx4.5-5$ derived by \cite{YMa25b} based on \cite{kokorev2024a}. The open red hexagon shows our number density including the transitionary source J1030\_9732 whose classification is ambiguous.}
    \label{fig:M5100_LF}
\end{figure}

Although we caution that the number of LRDs identified here is small, we measure number densities of $\approx4-10 \times10^{-6}$ cMpc$^{-3}$ mag$^{-1}$, with higher abundances found for fainter LRDs (see Table $\ref{tab:ndsen}$). As discussed in Sect. $\ref{sec:LRDident}$, the classification of the LRD nature of J1030\_9732 is uncertain. Therefore we present LFs both including and excluding this source as an LRD. The changes to the number densities are fairly small and our general conclusions are not affected whether we include or exclude this source. In Fig.~$\ref{fig:M5100_LF}$ we show that our measurements confirm that the number density of LRDs is lower at $z\approx2$ than at $z\approx5$ \citep{kokorev2024a,YMa25b} at luminosities M$_{5100}\approx-21$. Considering the uncertainties, our number densities are in slight tension (i.e. 4 times higher, but with large uncertainties) with those reported in an overlapping redshift range in \cite{YMa25} at similar luminosities, and somewhat lower than those at $z\approx2-4.6$ measured by \cite{Rinaldi26}, which is probably due to their large redshift range. We note that our number densities at M$_{5100}\approx-22$ are consistent with the estimate from \cite{Loiacono25} at $z\sim2.5$ based on the EIGER J1030 data and the RUBIES and JADES surveys. Given that number densities are either estimated from large-area ground-based surveys or a combination of multiple independent JWST fields, we expect that cosmic variance cannot explain the differences. At faint magnitudes M$_{5100}\approx-19$, our measured number density is much higher than those reported in \citet{YMa25}, but this is likely explained by low completeness of the ground-based photometry. Our number densities are in broad agreement with those reported by \cite{Bisigello25} based on a photometric search in wide-area Euclid data. However, detailed comparison is challenging because their number densities are reported as a function of UV luminosity and the selection cannot easily be reproduced due to the use of photometric redshifts and model rest-frame UV and optical continuum slopes.

We explore the cause of the difference at the brighter end to the result from \cite{YMa25} in detail by mimicking their photometric selection criteria with synthetic photometry of our best-fit LRD templates (described in Sect. $\ref{sec:LRDident}$). While we find that only half of our sources would be selected by their $J-Ks$ color cut, one of these sources is at faint magnitudes where the ground-based photometric search from \cite{YMa25} is incomplete and the other is at $z=1.93$, where the completeness modeling in \cite{YMa25} indeed estimates a very low completeness based on high-redshift LRD. This is because at $z=1.93$ the $J$ band still captures flux redward of the Balmer break and the $K_s$ band does not contain the H$\alpha$ line that is very strong in LRDs. \cite{YMa25} also require a detection in the rest-frame UV (i.e. $r$ band). Based on our template fits, we estimate $r$ band magnitudes $\sim26-27$ (for J1148$\_$21539 verified with HST photometry), which would still be detected in ground-based data. Therefore, the differences between the \cite{YMa25} results and ours are not easily explained as being due to photometric selections, and may simply reflect our relatively small number statistics. We note that any significant number of contaminants in their sample would make this discrepancy larger.

Recently, \cite{Loiacono26} identified LRDs in the EIGER J1030 field based on photometric selection criteria as well as non-detections in the X-rays. Their independently identified sample overlaps with our sample in this field and additionally includes J1030$\_$9732 that we mark as transitionary source and do not include in the LF. Primarily due to cosmic variance (as the J1030 field happens to have most $z\approx2$ LRDs among all the EIGER fields), their number densities are about a factor two higher than ours.

\subsection{Bolometric LF}
\begin{figure}
    \centering
    \includegraphics[width=0.99\linewidth]{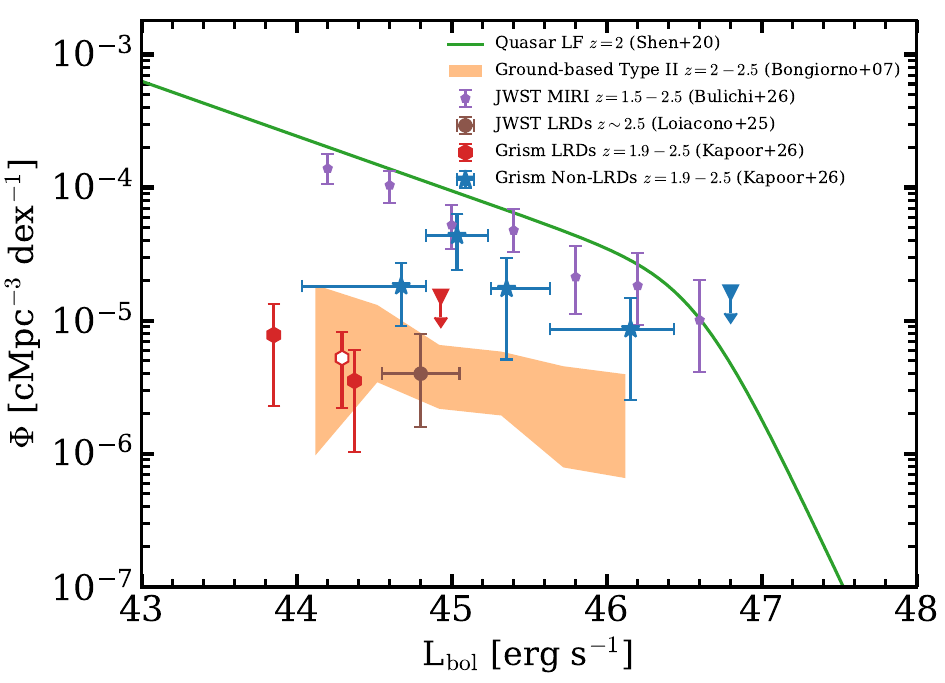}
    \caption{The bolometric LF of our samples of broad \ion{He}{i}-selected LRDs (red hexagons), non-LRDs (blue stars) as compared to the (green line) quasar LF \citep{Shen20} and (purple pentagons) JWST/MIRI-selected AGNs \citep{Bulichi26} at $z\approx2$. The open red hexagon shows our number density including the transitionary source J1030\_9732 whose classification is ambiguous. For LRDs (non-LRDs) we applied the bolometric conversion L$_{\rm bol}$/$\lambda$L$_{5100} = 5.4 (9.0)$ based on \cite{Greene26}. The brown point shows the LRD number density at $z\sim2.5$ estimated by \cite{Loiacono25}, shifted to the same bolometric conversion as employed here. The orange regions shows the broad-line quasar LF from the VVDS survey at $z\sim2$ \citep{Bongiorno07} to highlight that broad-line AGN do not constitute the whole quasar sample at $z\sim2$.}
    \label{fig:Lbol_LF}
\end{figure}

We compare the number densities of our samples of LRDs and non-LRDs to other AGN luminosity functions at $z\approx2$ by converting the optical luminosities to bolometric luminosities. For our non-LRDs -- several of which show X-ray detections and rising SEDs towards the rest-frame near-infrared -- we apply the standard bolometric conversion of L$_{\rm bol}$/$\lambda$L$_{5100} = 9.0$ \citep{Kaspi2000}. For LRDs, bolometric conversions are likely lower given their X-ray and IR faintness \citep{Ananna24,Setton2025,Casey25} and we adopt L$_{\rm bol}$/$\lambda$L$_{5100} = 5.4$ \citep{Greene26}, although we note this is based on only two sources and therefore carries significant uncertainty. In Fig. $\ref{fig:Lbol_LF}$ we compare our bolometric LFs to the canonical quasar LF \citep{Shen20} at $z=2$ and that measured at $z=1.5-2.5$ by \cite{Bulichi26} based on JWST/MIRI observations that are sensitive to both obscured and un-obscured AGNs. Our sample of LRDs hardly overlaps with the non-LRDs in terms of the bolometric luminosity (this result would only slightly change when similar bolometric conversions would be adopted, see Fig. $\ref{fig:M5100_Rcirc}$). At fixed luminosity, the LRDs are about 30 times rarer than quasars and other AGNs at $z\approx2$, which is the opposite of the situation at $z\approx5$ \citep{Greene26}, reflecting the strong differences in the evolution of the two populations. In fact, the LRDs have similar number densities as extremely luminous quasars with bolometric luminosities 300 times higher. Our sample of non-LRDs is also somewhat rarer than full AGN samples, but only by a factor three. We speculate that our broad \ion{He}{i}-selected sample is not complete for all types of AGNs and misses particularly strongly obscured AGNs that do not show broad lines \citep[e.g.][]{ShenKelly12,Buchner15}. Our number densities are higher than the number densities of (UV-selected) broad-line AGNs in the VVDS survey by \cite{Bongiorno07}, which is in agreement with zCOSMOS measurements in \citet{Schulze15}, simply showcasing such incomplete AGN selections are not unique to our methodology.

\begin{figure}
    \centering
    \includegraphics[width=1.01\linewidth]{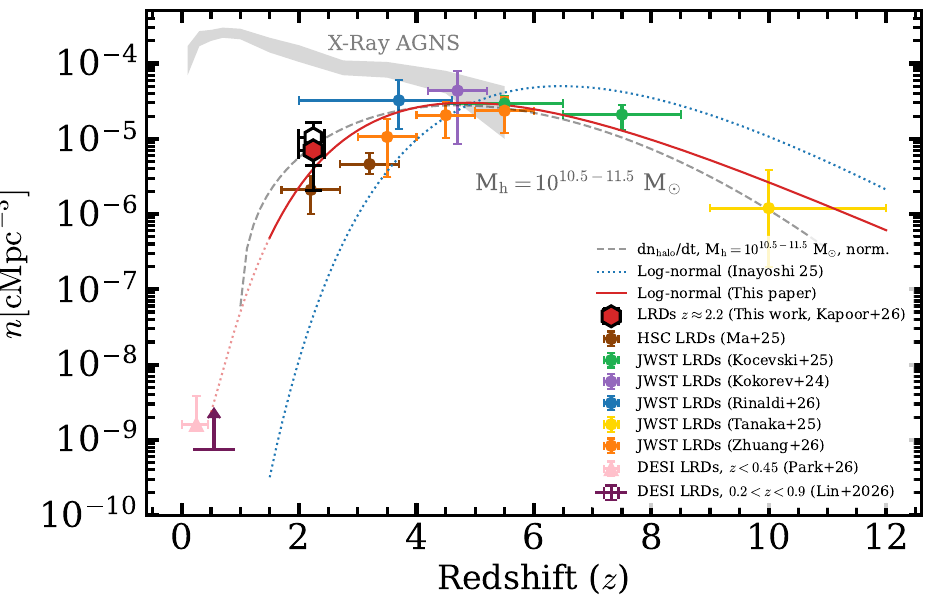}
    \caption{The number density evolution of LRDs, integrated down to optical magnitudes of M$_{5100}\approx-20$. We combine our new measurement at $z\approx2$ (red hexagon; open hexagon including J1030\_9732) with literature measurements based on JWST data \citep{kokorev2024a,Kocevski24,Rinaldi26,Zhuang26} at $z>3$ and wide-field ground-based imaging at $z\approx2-3$ \citep{YMa25}. Measurements from the DESI survey at $z\approx0.5$ \citep{Lin26,Park26} are to be considered lower limits due to uncertainties in the selection function. For comparison, we show the evolution of the X-ray AGN number density evolution \citep{Ueda14} in a gray shaded region. The dashed gray line shows the time derivative of the halo mass function at masses $10^{10.5 - 11.5}$ M$_{\odot}$ (normalized to the peak in the LRD number density by multiplying dn$_{\rm halo}$/dt by a 10$^5$ yr timescale). The blue dotted line shows the log-normal fit presented in \citealt{Inayoshi25}, whereas the red line shows our updated estimate. }
    \label{fig:numberdensity}
\end{figure}

\section{The gentle number density evolution of LRDs} \label{sec:nevo}
Our new rest-frame near-infrared emission-line selection of LRDs at $z\approx2$ independently confirms the reported decline in the number density of LRDs with respect to $z\approx5$ \citep{YMa25} that was based on a different selection method. However, compared to these earlier results, we find that the number density evolution is much more gentle than earlier emphasized \citep[e.g.][]{Inayoshi25}.

In Fig. $\ref{fig:numberdensity}$ we show a compilation of measurements of the LRD LF across $z\approx0-10$. The literature estimates of the LRD number densities are based on SDSS and DESI at $z\approx0-1$ \citep{Lin25_Lowz,Lin26,Park26}, where they are strictly reported as lower limits given uncertainties in the selection functions of those surveys, on JWST and ground-based data at $z\approx2$ (this work and \citealt{YMa25}) and JWST exclusively at $z\approx4-8$ \citep{kokorev2024a,Kocevski24,Rinaldi26,Zhuang26}, with first estimates enabled by MIRI data at $z\approx10$ \citep{Tanaka25}. We compare LRD number densities integrated down to the typical magnitudes of M$_{5100}\approx -20$ where possible and otherwise adopt similar conversion between UV and optical luminosity as adopted by \citet{YMa25b} such that we compare LRDs with similar luminosity. As a function of redshift, the LRD number density is characterized by a broad peak around $z\approx4-6$, with a $\approx5$-fold decline by $z\approx2$ and multiple orders of magnitude below $z<1$. The number density at $z\approx10$ is about ten times lower than the number density at $z\approx2$, although low number statistics challenge this comparison.

We quantify the redshift evolution of the LRD number density at M$_{5100}\approx-21$ using the parametrization as a log-normal as in \cite{Inayoshi25}:

\begin{equation}
\phi_{\rm LRD, M_{5100} \approx-21}(z) = \phi_0 f(z) \exp\left[  -\frac{\left\{\ln (1+z)-\mu_z \right\}^2}{2\sigma_z^2}\right],
\end{equation}
where we estimate $\phi_0 = 3\times10^{-6}$ cMpc$^{-3}$, $\mu_z=\ln(1+z_0)$ with a characteristic redshift $z_0=5$, $\sigma_z=0.28$ and where $f(z)$ is a cosmology-dependent term that accounts for the comoving volume dV/d$z$ and the redshift-time relation d$z$/dt that is approximated by the equation
\begin{equation}
f(z)=\frac{(1+z)^{3/2}}{[s(1+z)^{1/2}-1]^2},
\end{equation}
with s=0.903 at $z>1.5$. In Fig. $\ref{fig:numberdensity}$ we show this curve as a red line above $z>1.5$. Compared to \cite{Inayoshi25}, shown as a blue dotted line in the same figure, we estimate a later peak, a lower normalization and a somewhat larger width. If we naively extrapolate $f(z)$ to lower redshifts (dotted red line), we would find number densities consistent with recent estimates at $z\approx0.5$ \citep{Park26,Lin26}.

Below redshift $z\approx4$, the LRD evolution is in stark contrast to the evolution of X-ray AGN, which rises to a peak at $z\approx1$. This clearly shows that the galaxy-AGN co-evolution framework developed to explain such evolution \citep[e.g.][]{Caplar18} is not applicable to LRDs. To obtain an intuitive interpretation of the evolution of the LRD number density, we compare it to the evolution of the halo mass function (using the code presented in \citealt{Murray13}). In particular, we find that a naive time-derivative of the number densities of halos with a mass $\approx10^{10.5-11.5}$ M$_{\odot}$ matches the shape of the evolution remarkably well (see dashed lines in Fig. $\ref{fig:numberdensity}$). This halo mass range is similar to the halo mass studies usually report based on galaxy-LRD cross-correlation measurements \citep{Arita25,Carranza25,Lin26_clustering}. Below $z\lesssim1$ the derivative becomes negative (i.e. the number densities of halos with mass $10^{10.5-11.5}$ M$_{\odot}$ decrease with decreasing redshift below $z\approx1$ due to merging on more massive halos). Higher mass halos show similar behavior, but with a later and steeper peak. We normalize dn$_{\rm halo}$/dt approximately to the peak number density at $z\sim5$ by multiplying it with a $7\times10^4$ yr timescale. While the normalization is arbitrarily chosen and could similarly be retrieved with a longer timescale but a lower duty cycle, timescales that are too long would practically flatten the decline at lower redshifts too much. This simple comparison suggests that the physical conditions that give rise to the currently observed LRDs have a relatively short lifetime and occur in newly formed structures with halo masses of $\sim10^{11}$ M$_{\odot}$, which are increasingly rare at later times. Nevertheless, this naive comparison warrants a much more detailed investigation of the evolving number density of LRDs in a structure formation framework that we defer to future work.

Additionally, the evolution of the shape of the LF can provide key insights into the physical properties of LRDs. Recently, \cite{YMa25b} argued that the steep drop of the bright end of the LRD LF at $z\approx5$ suggests that LRDs are a population of relatively low mass AGNs with uniformly high Eddington ratios (see also \citealt{Tucci17} and \citealt{Zhang23} for similarly steep bolometric LFs for the subset of AGN powered by low mass SMBHs in their models). At $z\approx2$, the knee of the galaxy stellar mass function has evolved significantly compared to $z\approx5$ \citep{Leja20,Chaikin26}, while the number density of galaxies with stellar masses $\approx3\times10^8$ M$_{\odot}$ (the typical host galaxy masses of LRDs; \citealt{maiolino2024a,Matthee25clustering,WSun26}) is relatively constant. This evolution of the stellar mass function can explain the evolution of the X-ray AGN number density: if X-ray AGNs are primarily hosted by relatively massive galaxies with relatively high ($\approx10^8$ M$_{\odot}$) black hole masses that span a large range in Eddington ratios, we expect a relatively flat AGN LF whose normalization increases rapidly within this redshift interval (as observed).

The fact that there are fewer LRDs at $z\approx2$ than at $z\approx5$, while there could in principle be similar number of galaxies hosting them, suggests that stellar mass is not the key driver of the presence of LRDs. As can be seen in Fig. $\ref{fig:M5100_LF}$, we find indications that the faint-end of the LF evolves less than the bright end. At $z\approx2$, LRDs have a $\sim4$ times lower number density at faint magnitudes than at $z\approx5$, whereas the difference is $\approx10$ times at M$_{5100}\approx-21$. Although this could be due to possible incompleteness issues in measurements of the faint-end at $z\approx5$, it could imply a decrease in the characteristic luminosity of LRDs at lower redshifts, which could be linked to the available (metal-poor) gas content to fuel LRD activity \citep[e.g.][]{Baggen26,Maiolino26}. A more detailed investigation of such changes requires better constraints of the faint and bright-ends of the $z\approx2$--$5$ LRD LF than currently are available. The upcoming Roman telescope is excellently suited to measure the bright end of the LRD LF at $z\approx1$--2 and constrain the rapid evolution from $z\approx1$--2. JWST spectroscopy will be required to improve constraints on the faint end, either by extending an analysis similar to ours to lensing cluster fields, or by surveying specifically for the intrinsically stronger H$\alpha$ line, which at $z\approx1$--2 can be identified in medium-band data with NIRCam (such as available through the MINERVA survey; \citealt{Muzzin25}) or NIRISS slitless spectroscopy \citep[e.g.][]{Iani26}.

\section{Summary} \label{sec:summary}
In this paper we used deep JWST/NIRCam imaging and wide field slitless spectroscopic data from the EIGER survey \citep{Kashino23} to identify a sample of little red dots at $z\approx2$. Our main results are the following:

\begin{itemize}
    \item Out of a sample of 53 blindly identified broad-line emitters across redshifts $z=0.8 - 5.0$, we here focus on the 19 broad-line emitters at $z=1.55-3.18$ selected based on the rest-frame near-infrared (Pa$\beta$, \ion{O}{i} and \ion{He}{i}) lines. The classification between LRDs and non-LRDs is based on a combination of the spectral energy distribution (with limited 2-3 band coverage depending on the redshift; up to 5 bands for a subset), the morphology and the \ion{He}{i} and Pa$\gamma$ line-profile. Five of these are classified as LRDs at $z=1.93 - 3.18$, one faint source is challenging to classify and is marked a transitionary source, whereas the others are AGNs with classical (obscured) quasar-like characteristics. The non-LRDs generally are more luminous, are more spatially extended and some show X-ray detections. [Sect. $\ref{sec:method}$, Table $\ref{tab:spectral}$, Figs. $\ref{fig:searchmethod}$ and $\ref{fig:M5100_Rcirc}$]

    \item Based on template-fitting, we estimate that the LRDs have Balmer break strengths encompassing the range observed at high-redshift, $f_{\nu, 4050}/f_{\nu, 3650} \approx1 - 4$. The broad-lines originate from a point-source, but separate clumps and a resolved rest-frame UV morphology are found in the majority. [Sect. $\ref{sec:colors}$, Figs. $\ref{fig:sources1}$ and $\ref{fig:sources2}$]

    \item We find that the broad wings in the \ion{He}{i} and Pa$\gamma$ line-profiles (in one case also Pa$\beta$) have typical Gaussian FWHM of $1400-2100$ km s$^{-1}$, with broader lines found in the Paschen lines and blue-shifted \ion{He}{i} absorption detected in the two reddest sources in the sample. Similar to other studies, the \ion{He}{i} absorbing gas has stronger blue-shifts than the shifts usually found in Balmer absorption lines. We also show that the \ion{He}{i}/Pa$\gamma$ line ratio clearly distinguishes LRDs from classical AGNs, with a lower \ion{He}{i}/Pa$\gamma$ ratio associated with a stronger Balmer break strength, which may be tracing stronger \ion{He}{i} self-absorption in higher column density regimes. [Sects. $\ref{sec:linefitting}$ and $\ref{sec:HeIPaGratio}$, Figs. $\ref{fig:lineprofiles}$ and $\ref{fig:HeIPaG_BBreak}$]

    \item The LRDs span a similar optical luminosity range as high-redshift LRDs with M$_{5100}\approx-21$. Above luminosities of M$_{5100}\lesssim-23$ virtually all broad-line sources are classical AGNs within our surveyed volume, suggesting that the LRD fraction among broad-line sources is strongly luminosity-dependent. Assuming recently derived bolometric conversions applicable to LRDs, our sample spans bolometric luminosities from $(5.88 - 80.34)\times10^{43}$ erg s$^{-1}$ that corresponds to masses of $(4.7-64)\times10^{5}$ M$_{\odot}$ at the Eddington luminosity. [Sect. $\ref{sec:Lbolvalues}$]

    \item We measure the luminosity function of the LRDs and non-LRDs and we find number densities $\approx7\times10^{-6}$ cMpc$^{-3}$ at $z=1.9-2.5$. These independently confirm the reported drop in the number density of LRDs compared to $z\approx5$, albeit with a somewhat more gentle pace. At fixed bolometric luminosity, LRDs only span a small ($\lesssim3$ \%) subset of the AGN population at $z\approx2$. [Sect. $\ref{sec:LF}$, Table $\ref{tab:ndsen}$, Figs. $\ref{fig:M5100_LF}$, $\ref{fig:Lbol_LF}$ and $\ref{fig:numberdensity}$]
\end{itemize}

Our discovery of a new sample of emission-line selected LRDs at $z\approx2$ is an important step towards better understanding the nature of these sources. In the future, follow-up spectroscopy of these sources will enable more precise characterization of their basic properties as Balmer break strengths and Balmer line luminosities. The opportunity to cover the rest-frame near-infrared with sensitive NIRSpec spectroscopy enables detailed studies that are much more challenging at $z\sim5$, such as measurements of gravity-sensitive features like CaT absorption \citep{Lin25_Lowz,Greene26CaT} as well as joint studies of Paschen and Balmer decrements \citep[e.g.][]{Reddy26} that would enable us to distinguish radiative transfer effects from dust attenuation \citep[e.g.][]{Chang26}. Our results also motivate the extension of this methodology to the faint and bright regimes using data over lensing clusters and wide areas, respectively. The current data suggest a very rapid evolution below $z<2$ that can be constrained better with slitless spectroscopy with the Roman Space Telescope. Finally, our results motivate the need for modeling efforts to place the evolution of the LRD LF into a cosmological framework.

\begin{acknowledgements}
We thank an anonymous referee for constructive comments that helped improving the quality of our manuscript.
We thank Anna de Graaff, Federica Loiacono and Marta Volonteri for insightful discussions.
We acknowledge funding by the European Union (ERC, AGENTS,  101076224).
This work is based in part on observations made with the NASA/ESA/CSA James Webb Space Telescope. The data were obtained from the Mikulski Archive for Space Telescopes at the Space Telescope Science Institute, which is operated by the Association of Universities for Research in Astronomy, Inc., under NASA contract NAS 5-03127 for JWST. These observations are associated with programs \#1243.
Some of the data products presented herein were retrieved from the Dawn JWST Archive (DJA). DJA is an initiative of the Cosmic Dawn Center (DAWN), which is funded by the Danish National Research Foundation under grant DNRF140.
This research has used data obtained from the Chandra Source Catalog provided by the Chandra X-ray Center (CXC).
Software used in this work includes:
Python, \texttt{matplotlib} \citep{matplotlib}, \texttt{numpy} \citep{numpy}, \texttt{scipy} \citep{scipy}, \texttt{Astropy} \citep{astropy}, \texttt{lmfit} \citep{Erwin2015}, \texttt{TOPCAT} \citep{topcat}. The authors used Claude (Anthropic) to assist with code development. All AI-assisted content was reviewed, verified, and edited by the authors, who take full responsibility for the accuracy and integrity of this work.
\end{acknowledgements}

\bibliographystyle{aa}
\bibliography{my_bibliography_cleaned}

\begin{appendix}

\renewcommand{\topfraction}{0.95}
\renewcommand{\bottomfraction}{0.95}
\renewcommand{\textfraction}{0.05}
\renewcommand{\floatpagefraction}{0.85}
\renewcommand{\dbltopfraction}{0.95}
\renewcommand{\dblfloatpagefraction}{0.85}
\setcounter{topnumber}{4}
\setcounter{bottomnumber}{2}
\setcounter{totalnumber}{6}
\setcounter{dbltopnumber}{4}

\section{Completeness curves}
\begin{figure}[!htbp]
    \centering
    \includegraphics[width=\hsize]{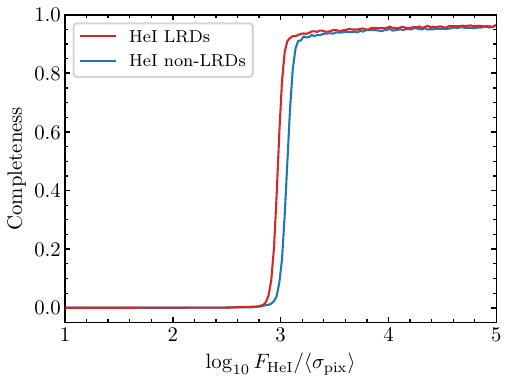}
    \caption{Completeness function of \ion{He}{i} emitters as a function of the ratio of the \ion{He}{i} line flux to the average pixel error at the position of the line. The red curve shows the result for the LRD-like line profile, while the blue curve shows the result for a broader line profile that is characteristic of the non-LRD broad-line emitters. }
    \label{fig:completeness}
\end{figure}

Figure $\ref{fig:completeness}$ shows the completeness function of broad \ion{He}{i}+Pa$\gamma$ emitters that we derived as described in Sect. $\ref{sec:LF}$.

\section{Photometry} \label{app:A}
The available HST/ACS and JWST/NIRCam photometry of all broad-line sources identified in this work at $z=1.5-3.2$ is listed in Table $\ref{tab:photometry}$. The data reduction and methods are explained in the EIGER survey papers \citep{Kashino23,Kashino26}.

\begin{table*}[!t]
\centering
\caption{Photometry of the full sample.}
\label{tab:photometry}
\setlength{\tabcolsep}{4pt}
\begin{tabular}{lcccccccc}
\hline
ID & LRD & $z_\mathrm{spec}$ & F606W & F775W & F814W & F115W & F200W & F356W \\
\hline
J0100-18107 &  & 1.552 &  &  &  & $22.26 \pm 0.05$ & $22.35 \pm 0.05$ & $21.59 \pm 0.05$ \\
J1148-21539 & \checkmark & 1.931 & $26.91 \pm 0.11$ & $26.99 \pm 0.15$ &  & $25.90 \pm 0.06$ & $23.22 \pm 0.05$ & $22.85 \pm 0.05$ \\
J1120-11834 &  & 2.046 & $22.89 \pm 0.05$ & $22.74 \pm 0.05$ & $23.27 \pm 0.05$ & $22.58 \pm 0.05$ & $22.08 \pm 0.05$ & $21.65 \pm 0.05$ \\
J0148-7135 &  & 2.130 &  &  &  & $21.32 \pm 0.05$ & $20.94 \pm 0.05$ & $20.95 \pm 0.05$ \\
J0148-15902 &  & 2.178 &  &  &  & $23.16 \pm 0.05$ & $22.46 \pm 0.05$ & $22.25 \pm 0.05$ \\
J1148-21459 &  & 2.192 & $25.30 \pm 0.09$ & $24.83 \pm 0.08$ &  & $23.25 \pm 0.06$ & $22.00 \pm 0.05$ & $21.68 \pm 0.05$ \\
J1148-22359 &  & 2.201 & $24.87 \pm 0.06$ & $24.42 \pm 0.06$ &  & $22.77 \pm 0.05$ & $21.58 \pm 0.05$ & $21.50 \pm 0.05$ \\
J0100-1751 &  & 2.219 &  &  &  & $23.90 \pm 0.06$ & $22.48 \pm 0.05$ & $21.90 \pm 0.05$ \\
J159-6107 & \checkmark & 2.243 &  &  &  & $26.10 \pm 0.07$ & $24.22 \pm 0.05$ & $23.57 \pm 0.05$ \\
J159-6127 &  & 2.247 &  &  &  & $23.00 \pm 0.05$ & $21.39 \pm 0.05$ & $20.70 \pm 0.05$ \\
J1030-2735 & \checkmark & 2.328 &  &  &  & $25.26 \pm 0.06$ & $22.43 \pm 0.05$ & $22.30 \pm 0.05$ \\
J1148-3444 &  & 2.334 & $20.59 \pm 0.05$ & $20.42 \pm 0.05$ &  & $20.03 \pm 0.05$ & $19.72 \pm 0.05$ & $19.63 \pm 0.05$ \\
J0148-12798 &  & 2.357 &  &  &  & $22.59 \pm 0.05$ & $21.88 \pm 0.05$ & $21.47 \pm 0.05$ \\
J1030-12378 &  & 2.378 &  &  &  & $23.41 \pm 0.06$ & $22.30 \pm 0.05$ & $21.90 \pm 0.05$ \\
J0148-18339 &  & 2.400 &  &  &  & $24.45 \pm 0.06$ & $23.59 \pm 0.05$ & $23.10 \pm 0.05$ \\
J1030-9732 &  & 2.409 &  & $25.32 \pm 0.07$ &  & $24.49 \pm 0.05$ & $24.69 \pm 0.05$ & $23.97 \pm 0.05$ \\
J0148-10704 &  & 2.435 &  &  &  & $21.95 \pm 0.05$ & $19.63 \pm 0.05$ & $19.32 \pm 0.05$ \\
J1030-2545 & \checkmark & 2.498 &  & $26.81 \pm 0.14$ &  & $26.61 \pm 0.08$ & $25.07 \pm 0.05$ & $23.94 \pm 0.05$ \\
J0148-9325 & \checkmark & 3.177 &  &  &  & $24.82 \pm 0.06$ & $23.32 \pm 0.05$ & $23.02 \pm 0.05$ \\ \hline
\end{tabular}
\tablefoot{F606W, F775W and F814W are based on HST/ACS imaging data, whereas F115W, F200W and F356W are from JWST/NIRCam. Magnitudes are in the AB system. An error floor of 5 \% was added in quadrature to account for systematic uncertainties in zero-point calibration as well as aperture corrections. This dominates the error budget in most measurements.}
\end{table*}

\section{Additional SEDs and line-profiles}\label{app:B}
\begin{figure*}[!tp]
    \centering
       \includegraphics[width=\linewidth]{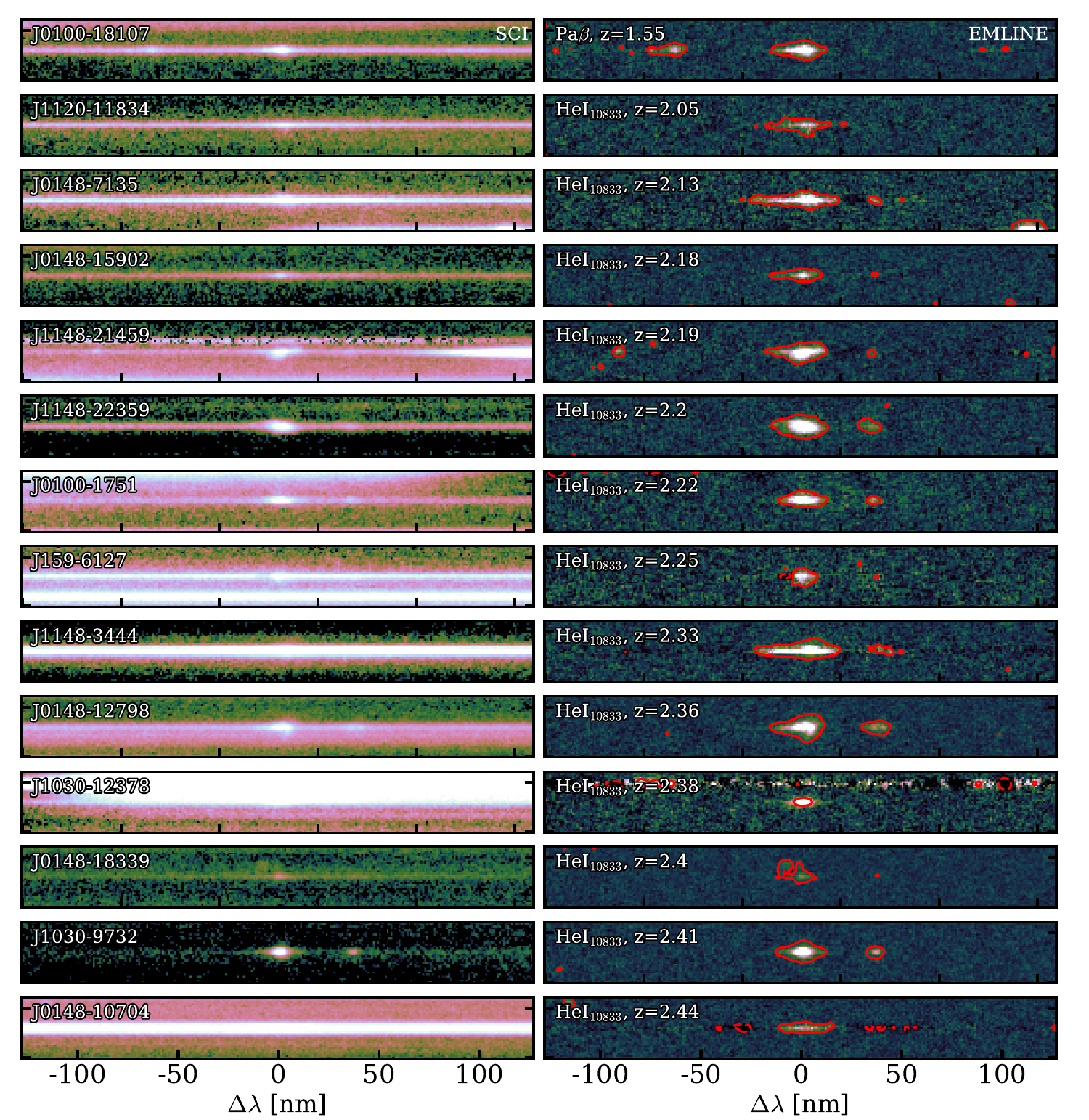}
    \caption{As Fig. $\ref{fig:searchmethod}$, now showing the non-LRD broad line sources. }
    \label{fig:searchmethod_nonLRD}
\end{figure*}

\begin{figure}[!htbp]
    \centering
    \includegraphics[width=\hsize]{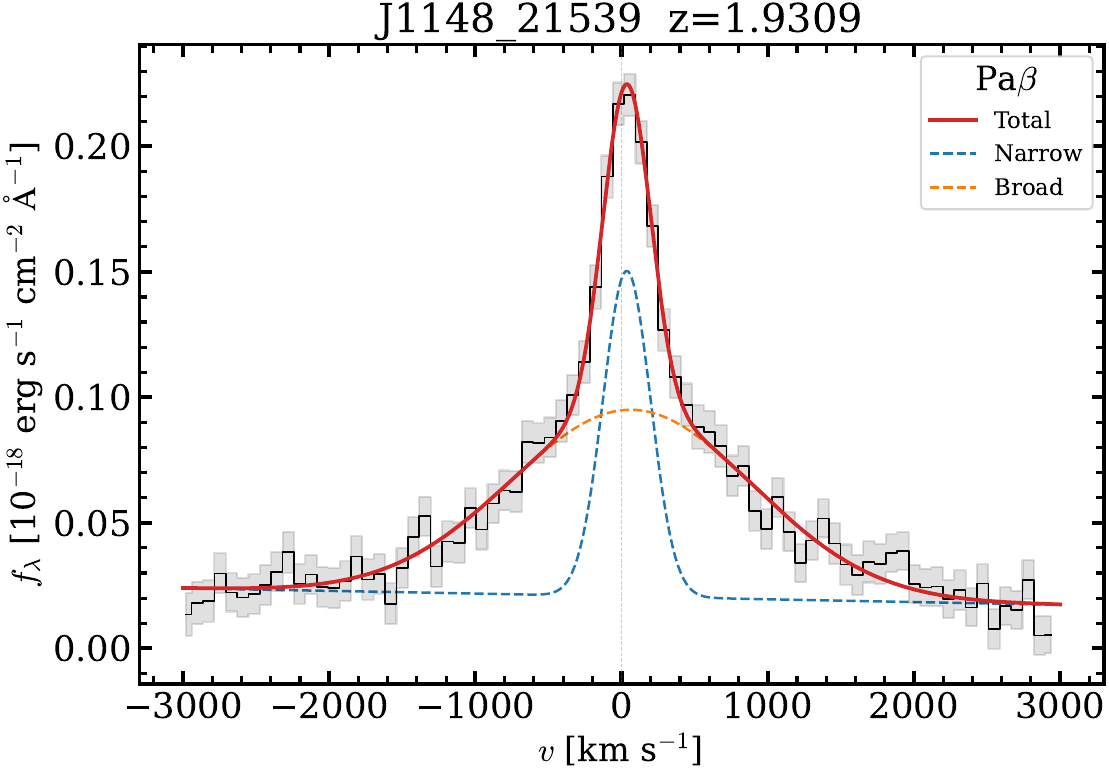}
    \caption{The Pa$\beta$ line-profile of J1148$\_$21539 (as Fig. $\ref{fig:lineprofiles}$). The profile is well described by a single narrow + broad profile.} \label{fig:lineprofiles_PaB}
\end{figure}

\begin{figure*}[!tp]
\centering
    \begin{tabular}{ccc}
    \hspace{-0.8cm} \includegraphics[width=5.6cm]{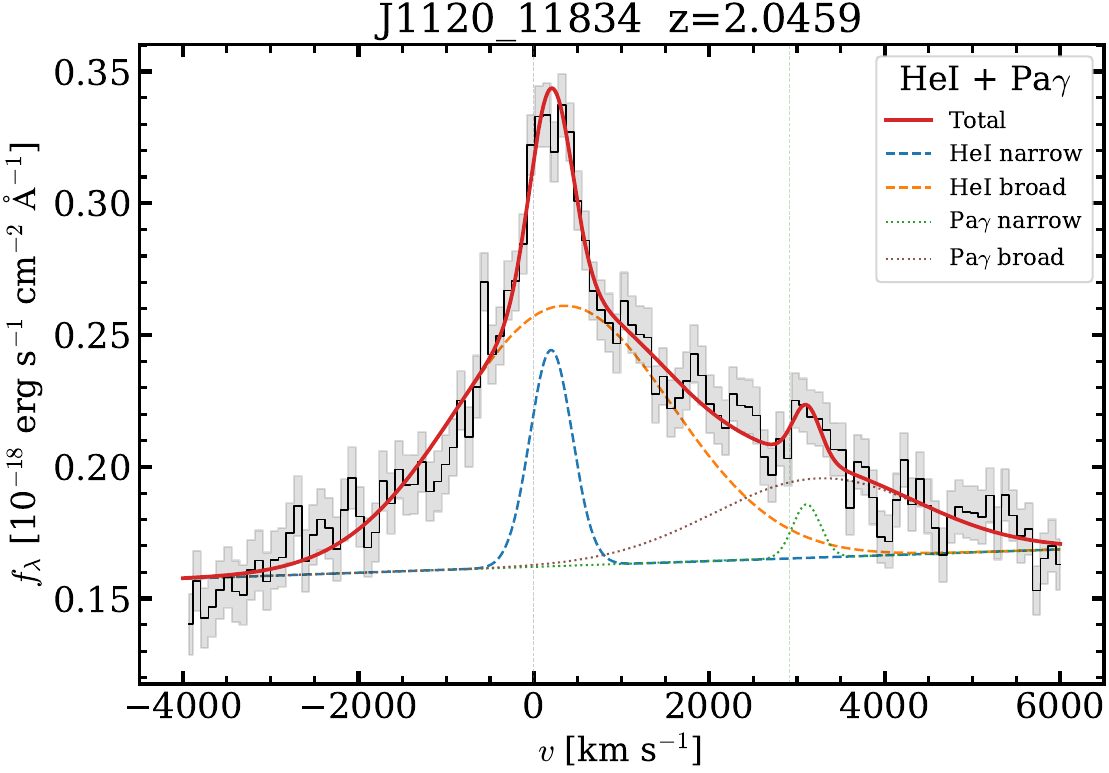} &
     \includegraphics[width=5.6cm]{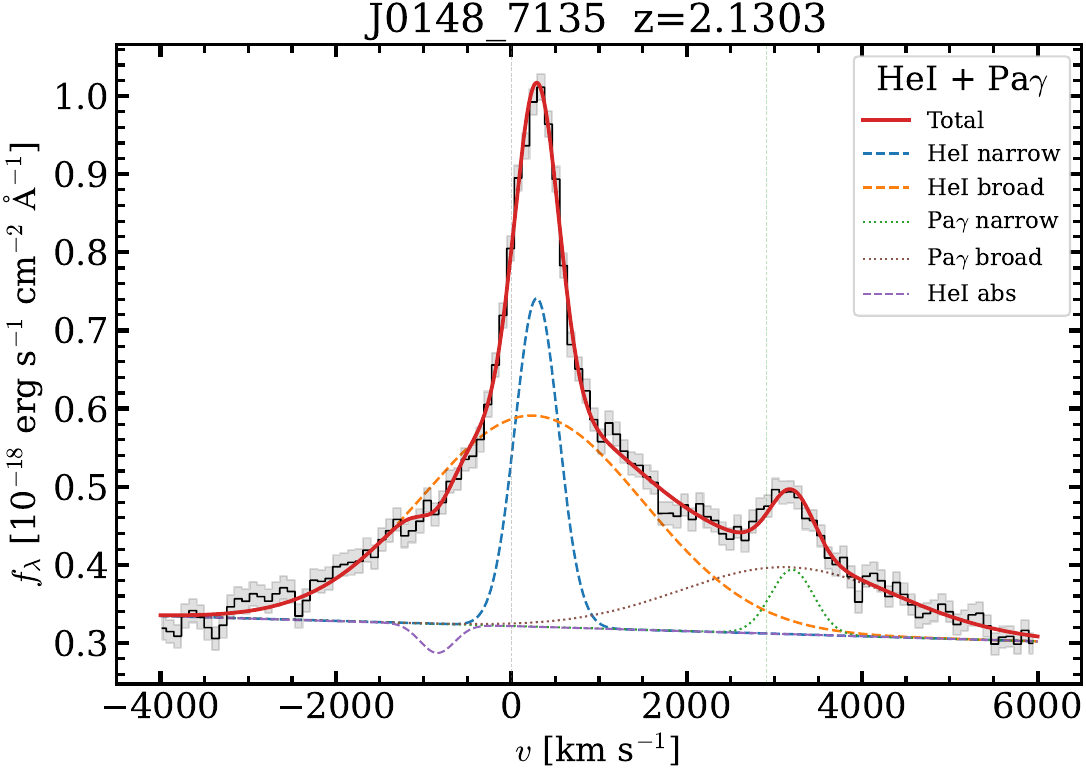} &   \includegraphics[width=5.6cm]
    {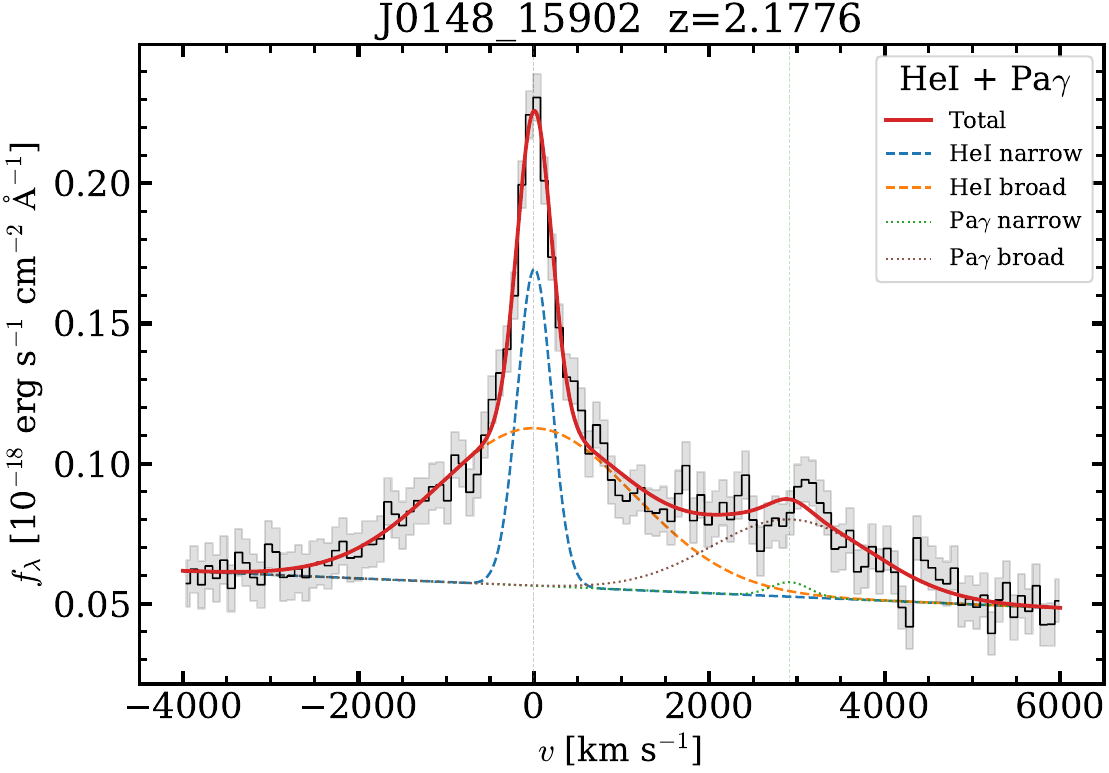} \\
     \hspace{-0.8cm}\includegraphics[width=5.6cm]{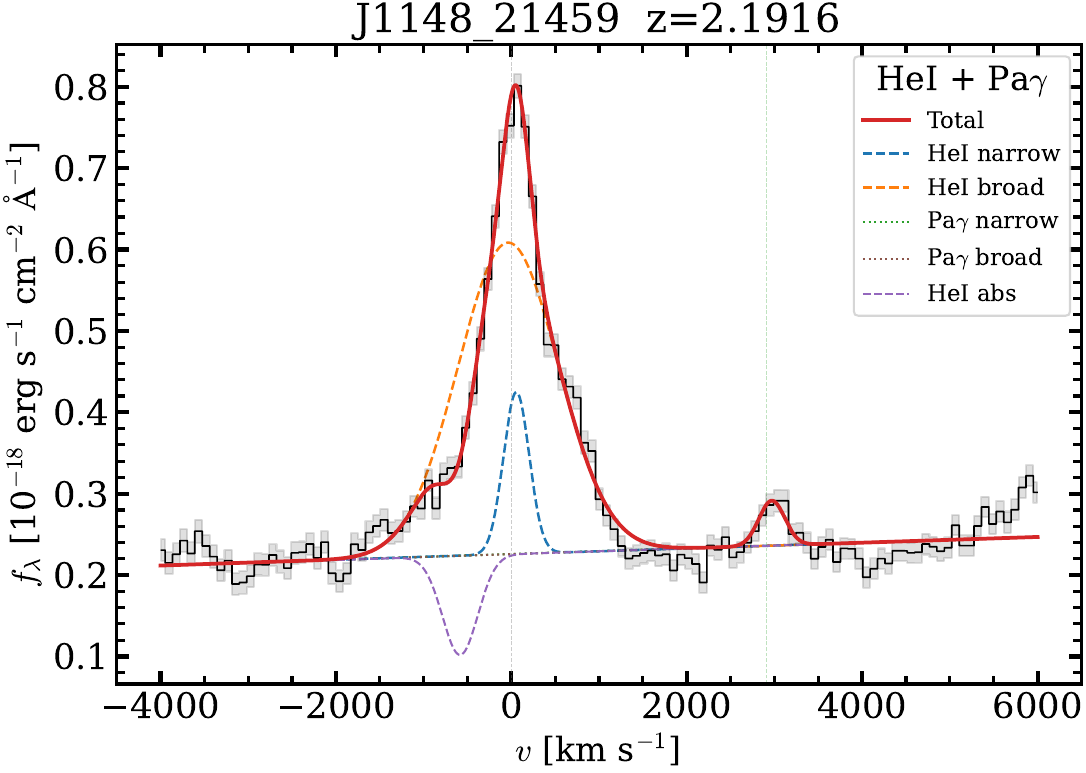} &  \includegraphics[width=5.6cm]{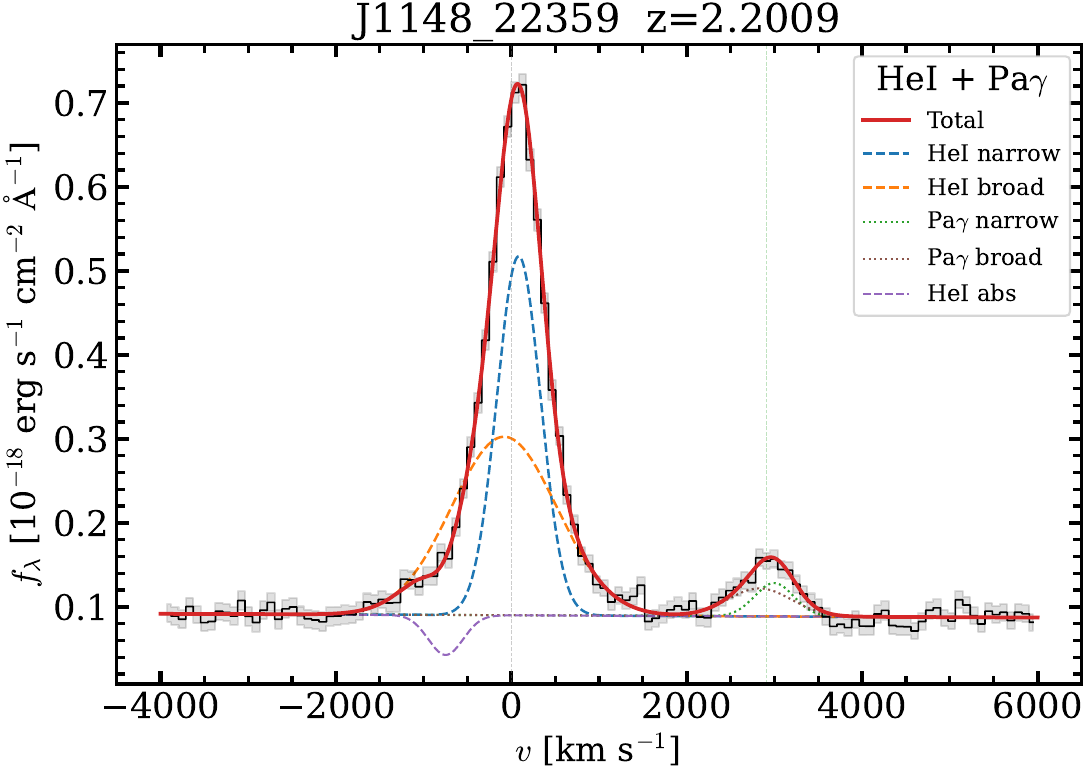} &
     \includegraphics[width=5.6cm]{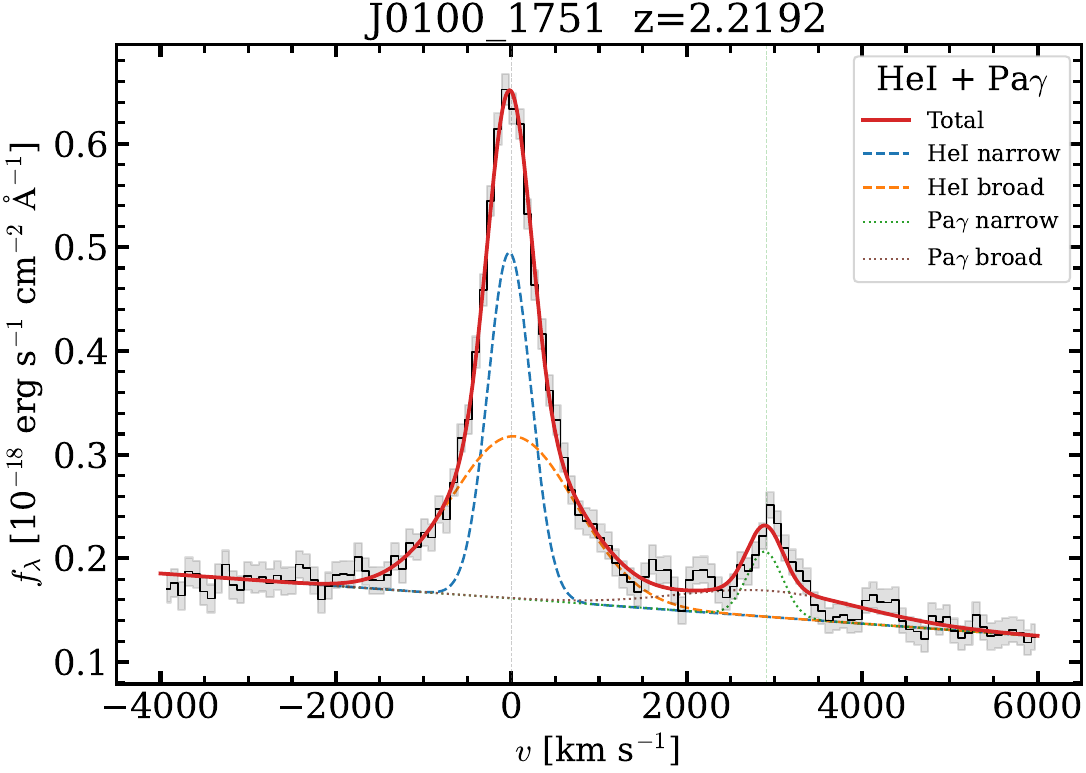} \\
    \hspace{-0.8cm} \includegraphics[width=5.6cm]
     {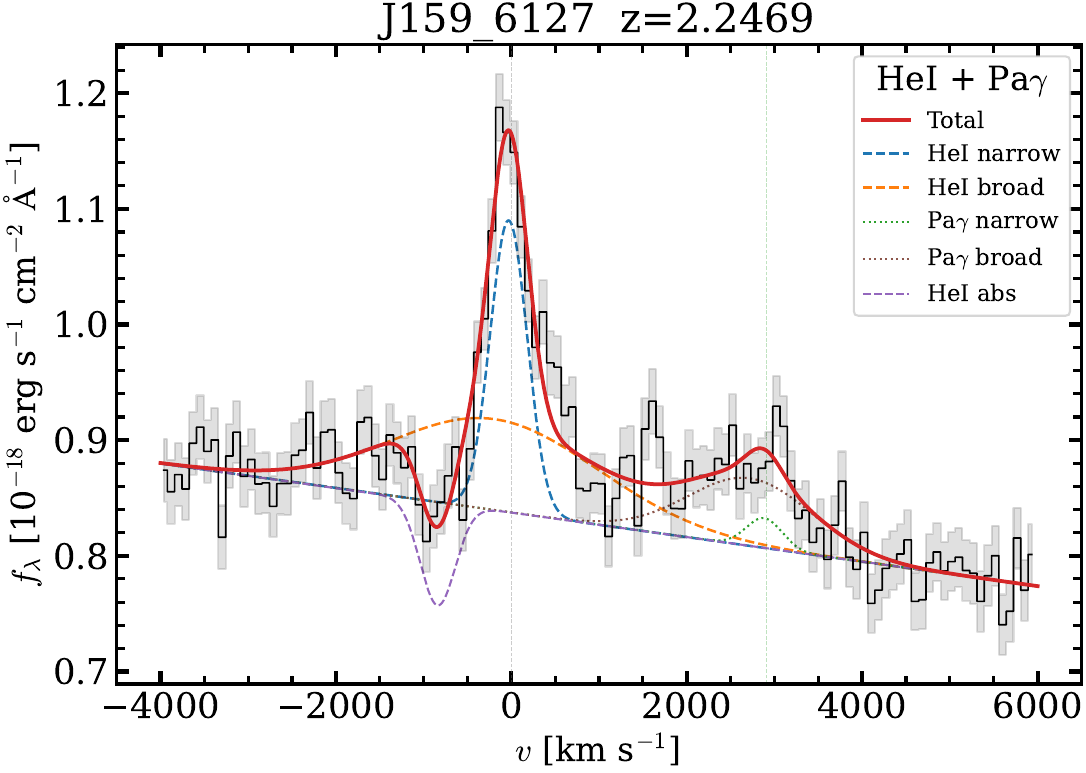} &
    \includegraphics[width=5.6cm]{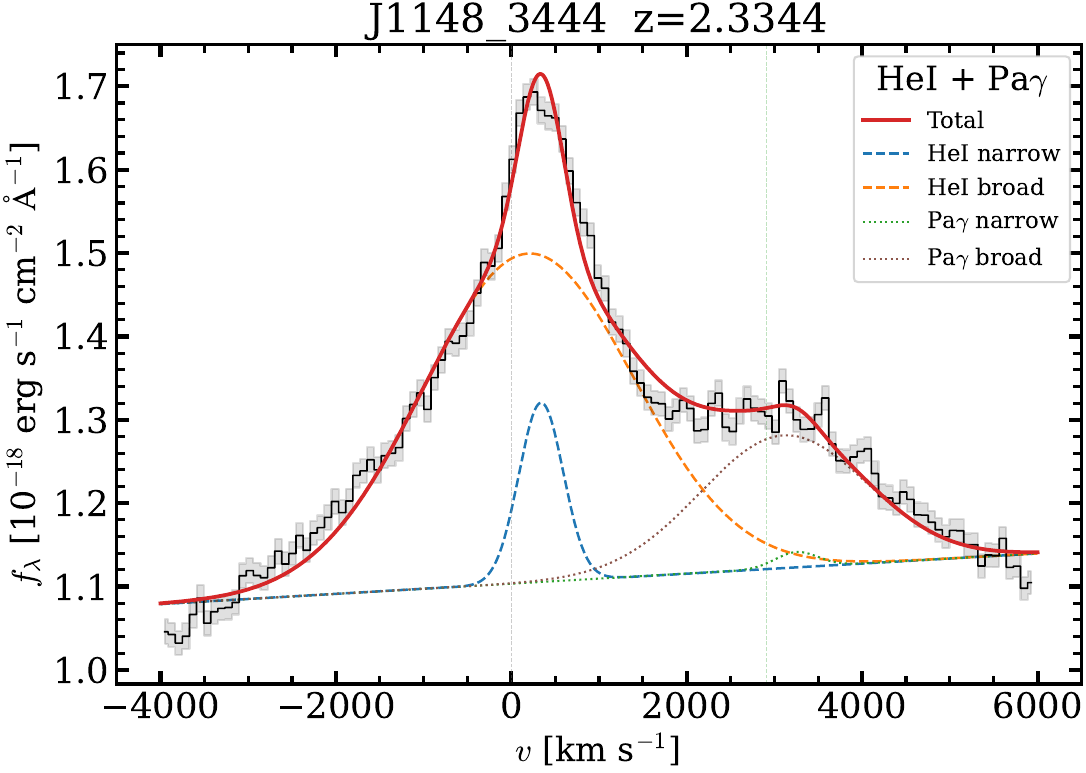} &
    \includegraphics[width=5.6cm]{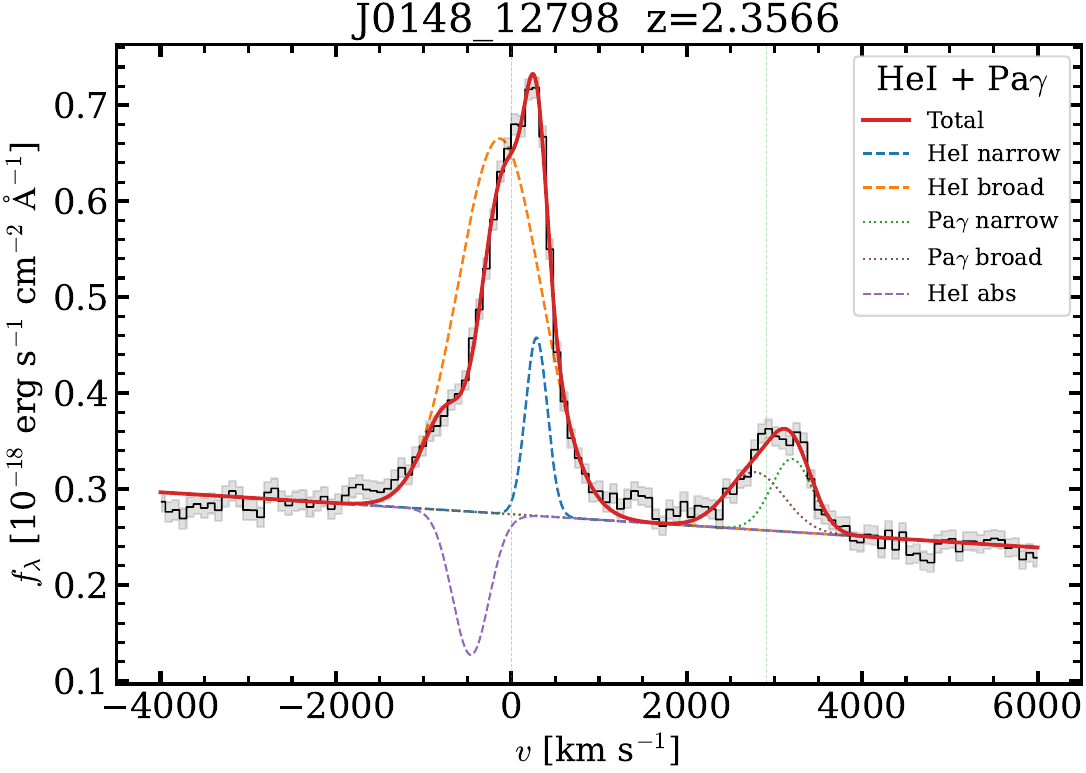} \\
    \hspace{-0.8cm}\includegraphics[width=5.6cm]{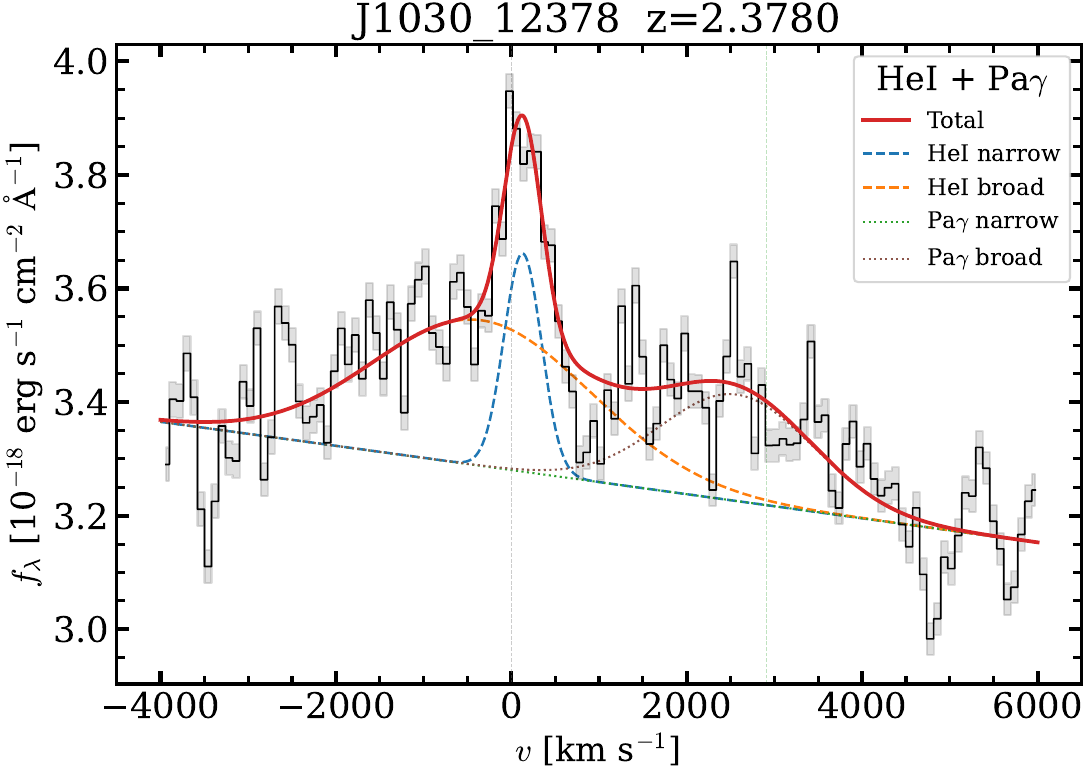} &
    \includegraphics[width=5.6cm]{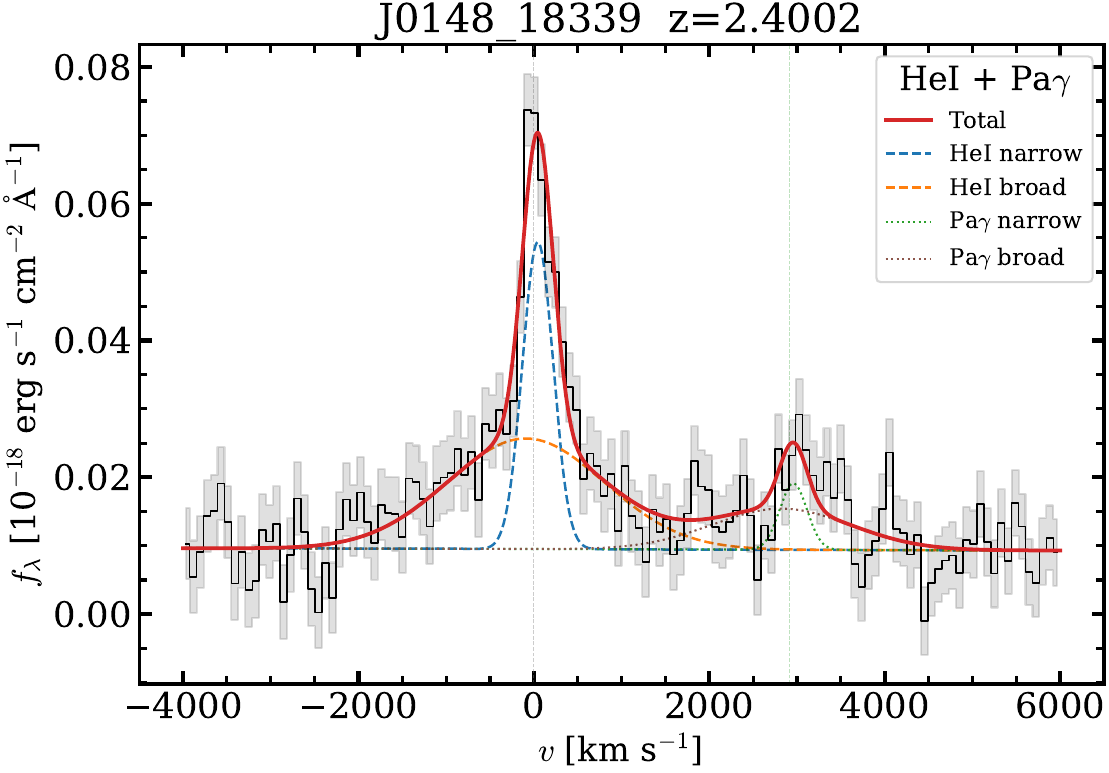} &
    \includegraphics[width=5.6cm]{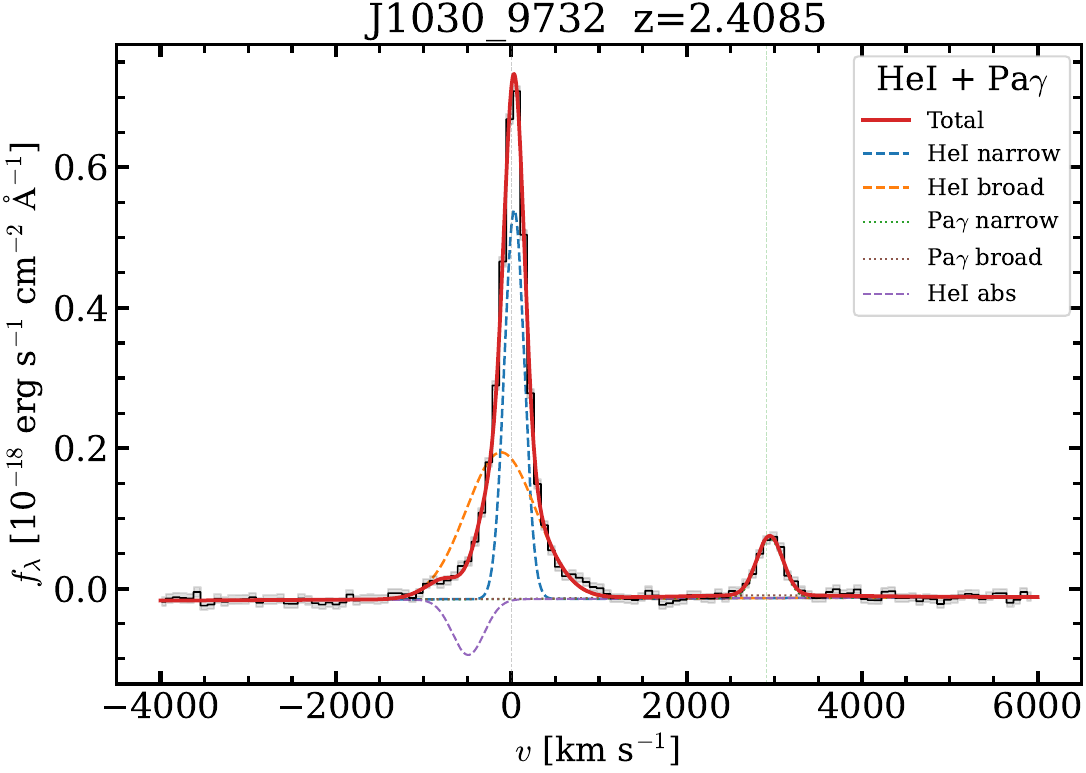} \\
   \hspace{-0.8cm} \includegraphics[width=5.6cm]{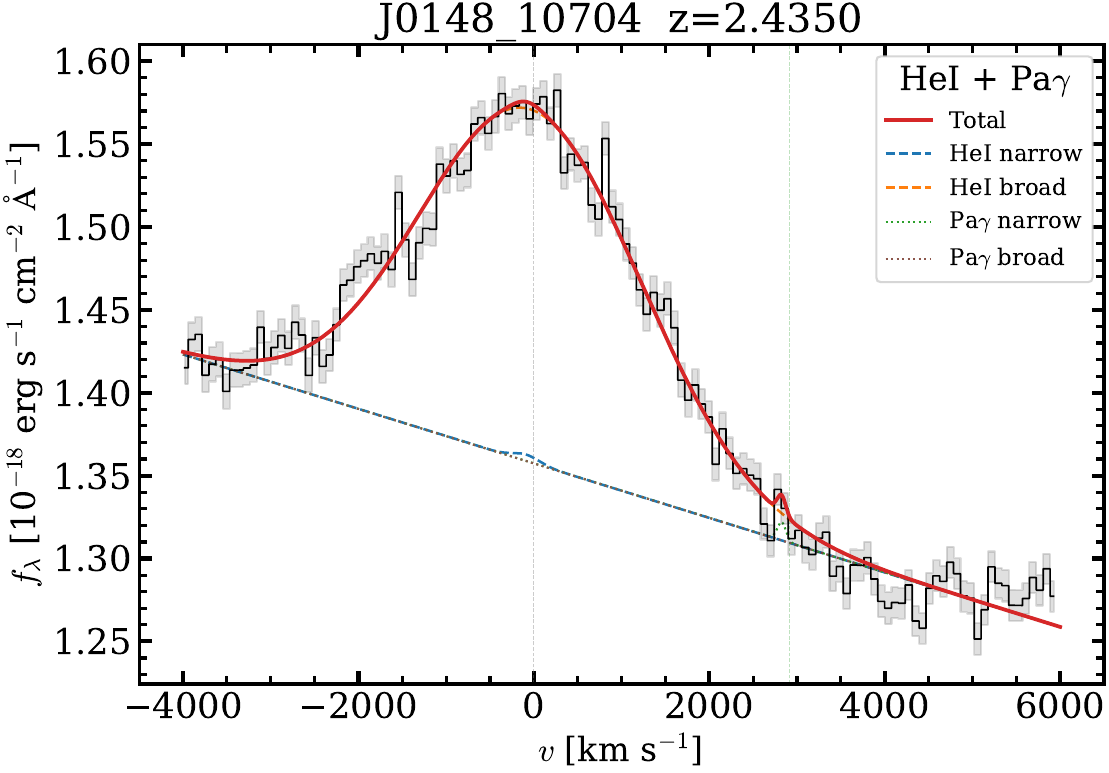} &
    \includegraphics[width=5.6cm]{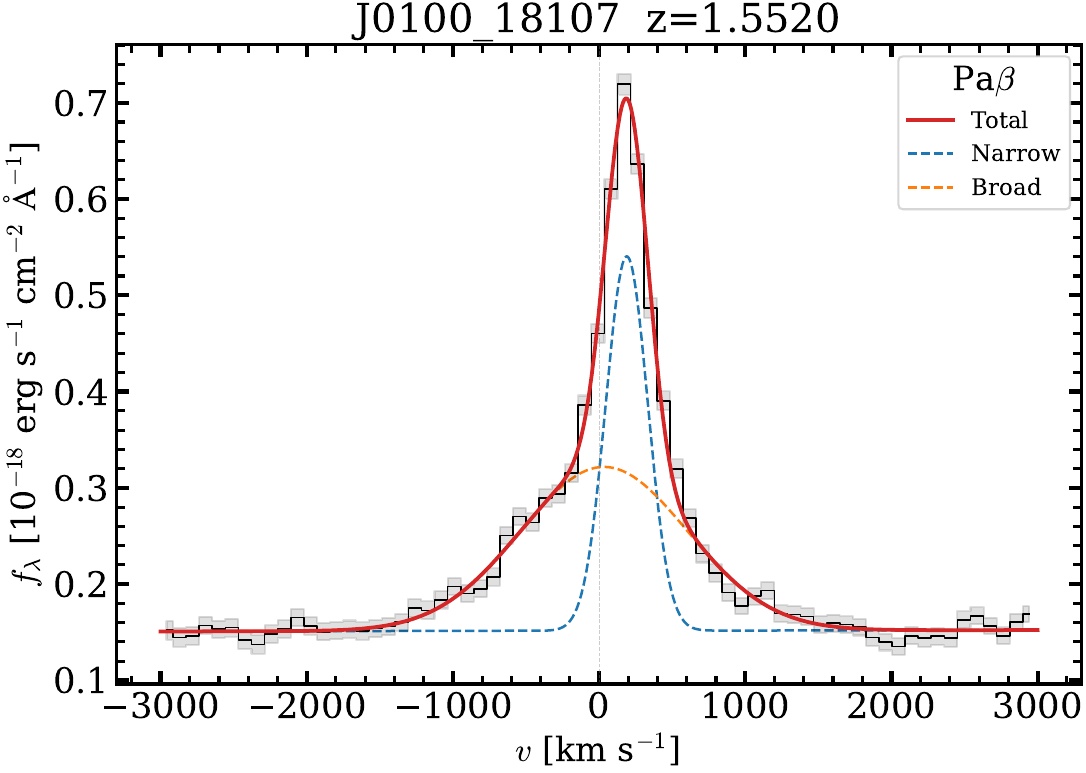}
    \\
    \end{tabular}
    \caption{The \ion{He}{i}+Pa$\gamma$ profiles of the 14 non-LRDs in the sample. Red lines show the total fitted profiles, that are composed of narrow and broad Pa$\gamma$ (dotted green and purple, respectively), narrow and broad \ion{He}{i} emission (blue and orange dashed lines) for objects with \ion{He}{i}+Pa$\gamma$ emission. The last panel also shows the fit Pa$\beta$ for J0100-18107, which is well described by a single narrow+broad profile. We note that the line-profiles were fitted on the SCI spectra.}
    \label{fig:profiles_nonLRDs}
\end{figure*}

For completeness to the figures in the main text, we here provide complementary information, particularly on the non-LRD broad-line emitters. In Fig. $\ref{fig:searchmethod_nonLRD}$ we show the NIRCam grism SCI and EMLINE spectra of the identified lines, similar to Fig. $\ref{fig:searchmethod}$. This figure shows that the continuum emission of most non-LRDs is detected in the grism data, highlighting their brighter continuum magnitudes and lower line EWs compared to LRDs. In Figs. $\ref{fig:sources_nonLRDs}$ and $\ref{fig:sources_nonLRDs2}$ we show the SEDs, the false color NIRCam images and the line profiles of the non-LRDs, as Fig. $\ref{fig:sources1}$ in the main text. In Fig. $\ref{fig:lineprofiles_PaB}$, we show the Pa$\beta$ line of the LRD J1148-21539 and its best-fit narrow+broad Gaussian model (see Sect. $\ref{sec:linefitting}$). Finally, in Fig. $\ref{fig:profiles_nonLRDs}$ we show the best-fit emission-line profiles to our sample of non-LRDs, based on the SCI spectra. Note that these spectra may be contaminated by continuum emission from other sources in the field, impacting the continuum level and slope, but not the line-shapes.

\begin{figure*}[!tp]
    \centering
    \begin{tabular}{ccc}
\hspace{-0.8cm} \includegraphics[width=6.3cm]{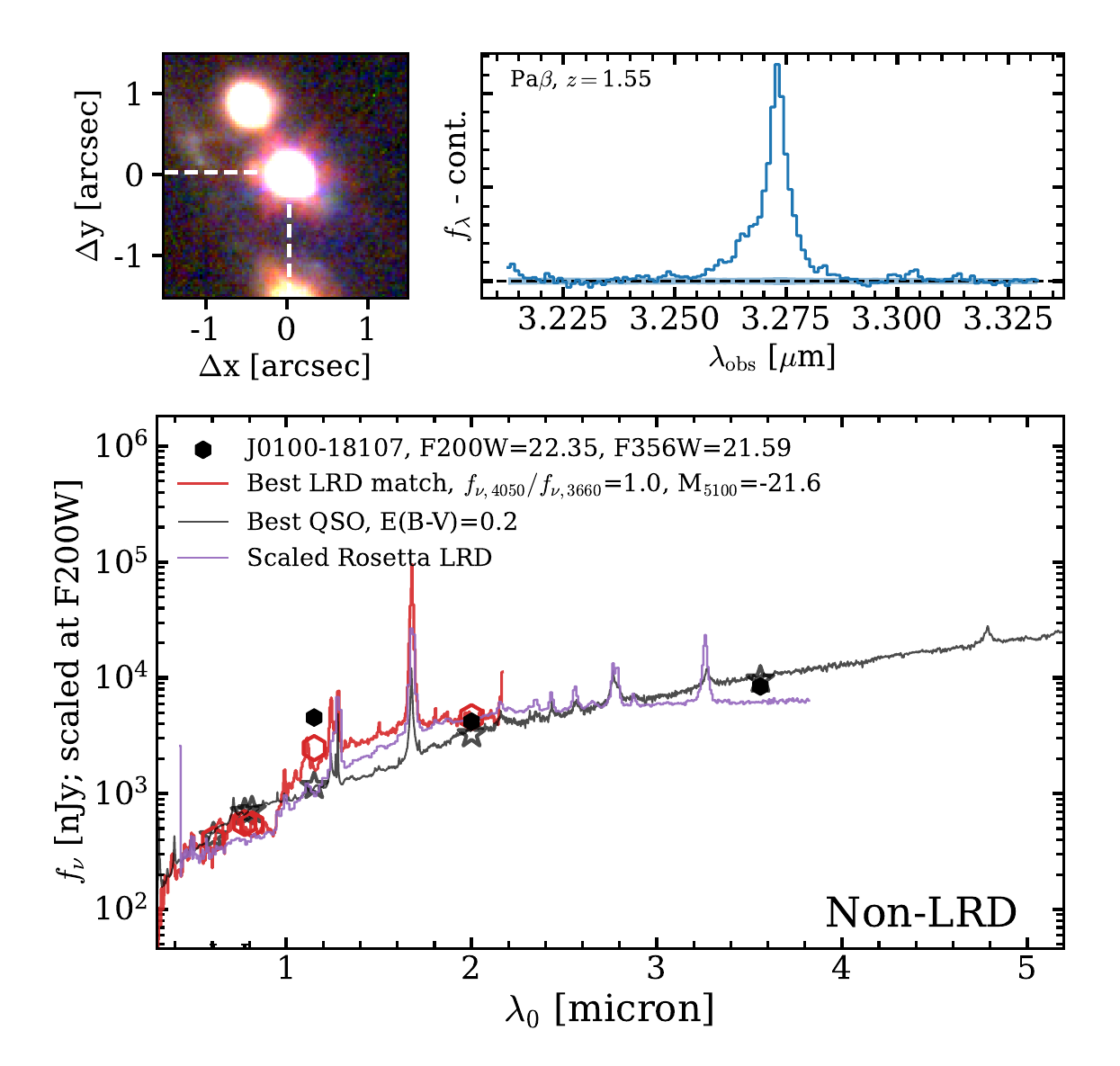} &
 \hspace{-0.8cm} \includegraphics[width=6.3cm]{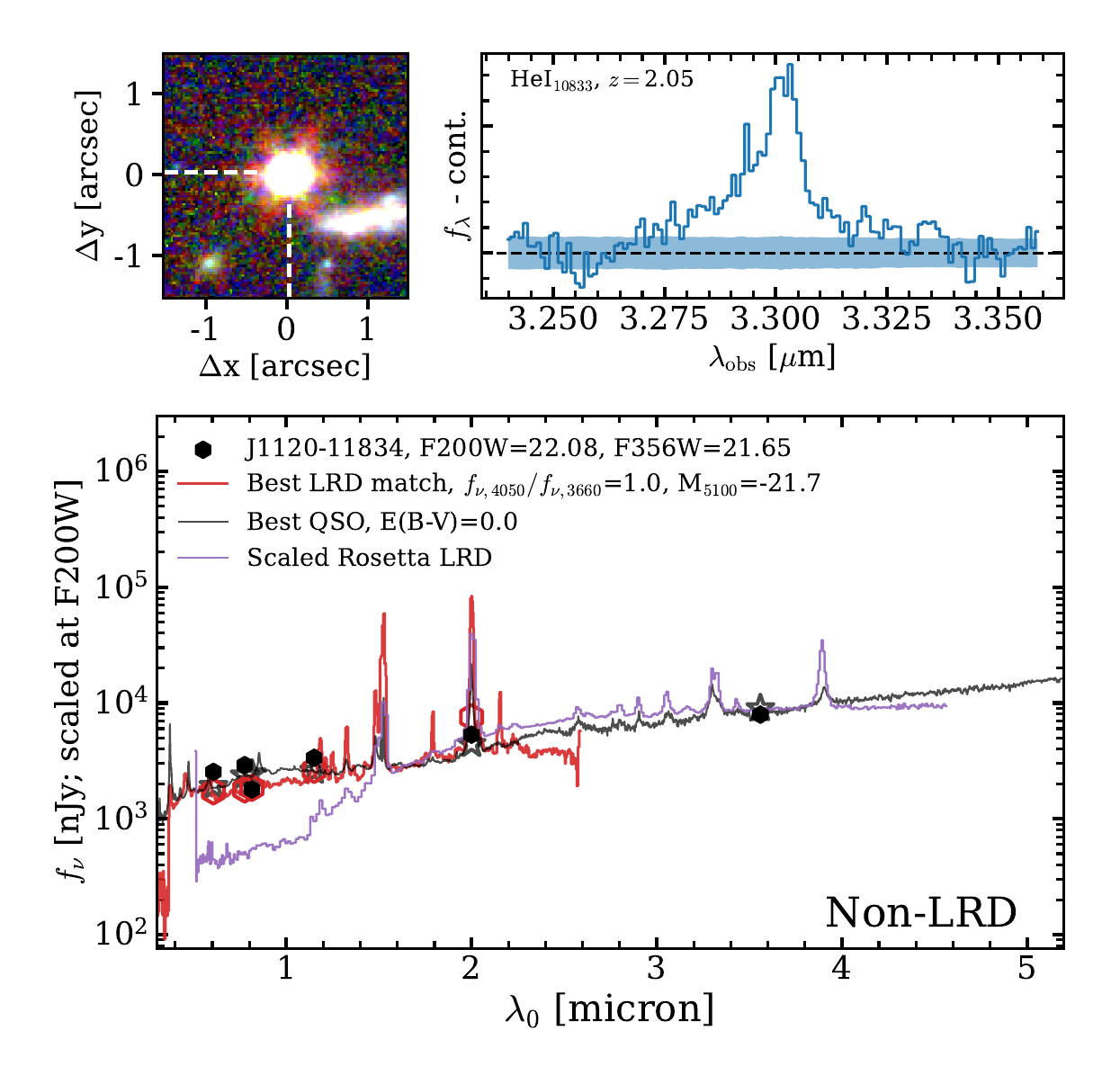} &
  \hspace{-0.8cm} \includegraphics[width=6.3cm]{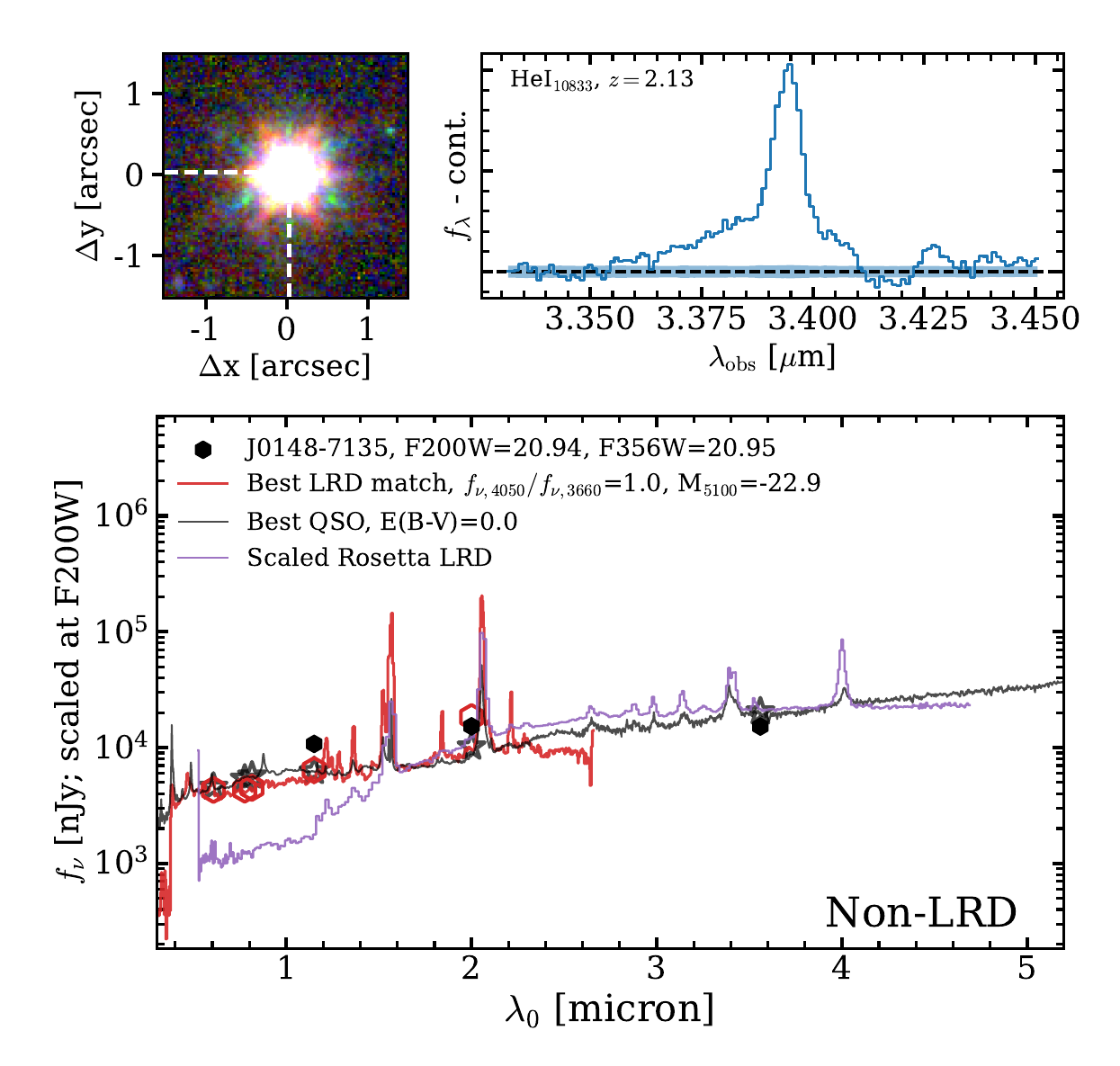} \\

\hspace{-0.8cm} \includegraphics[width=6.3cm]{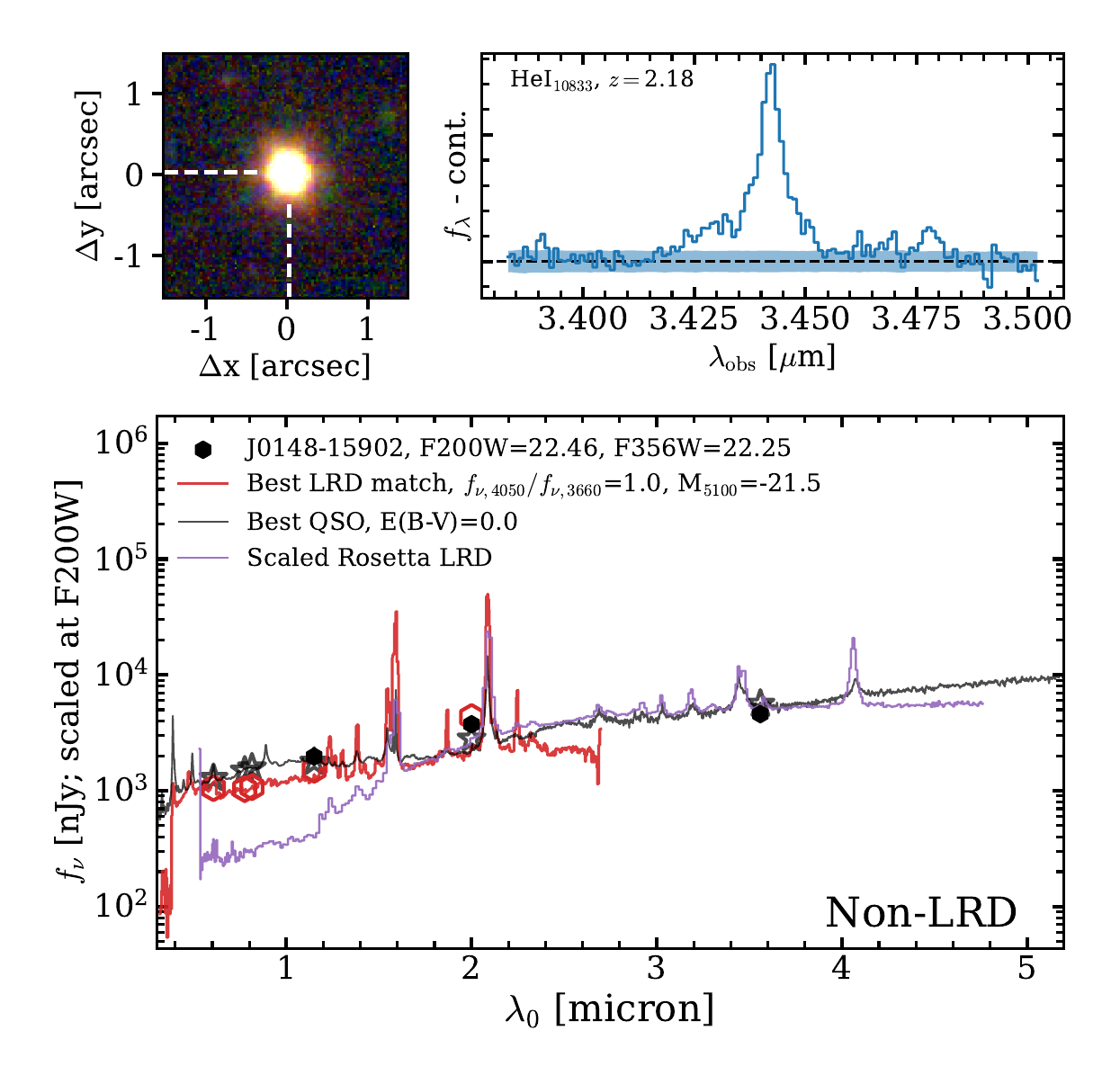} &
 \hspace{-0.8cm} \includegraphics[width=6.3cm]{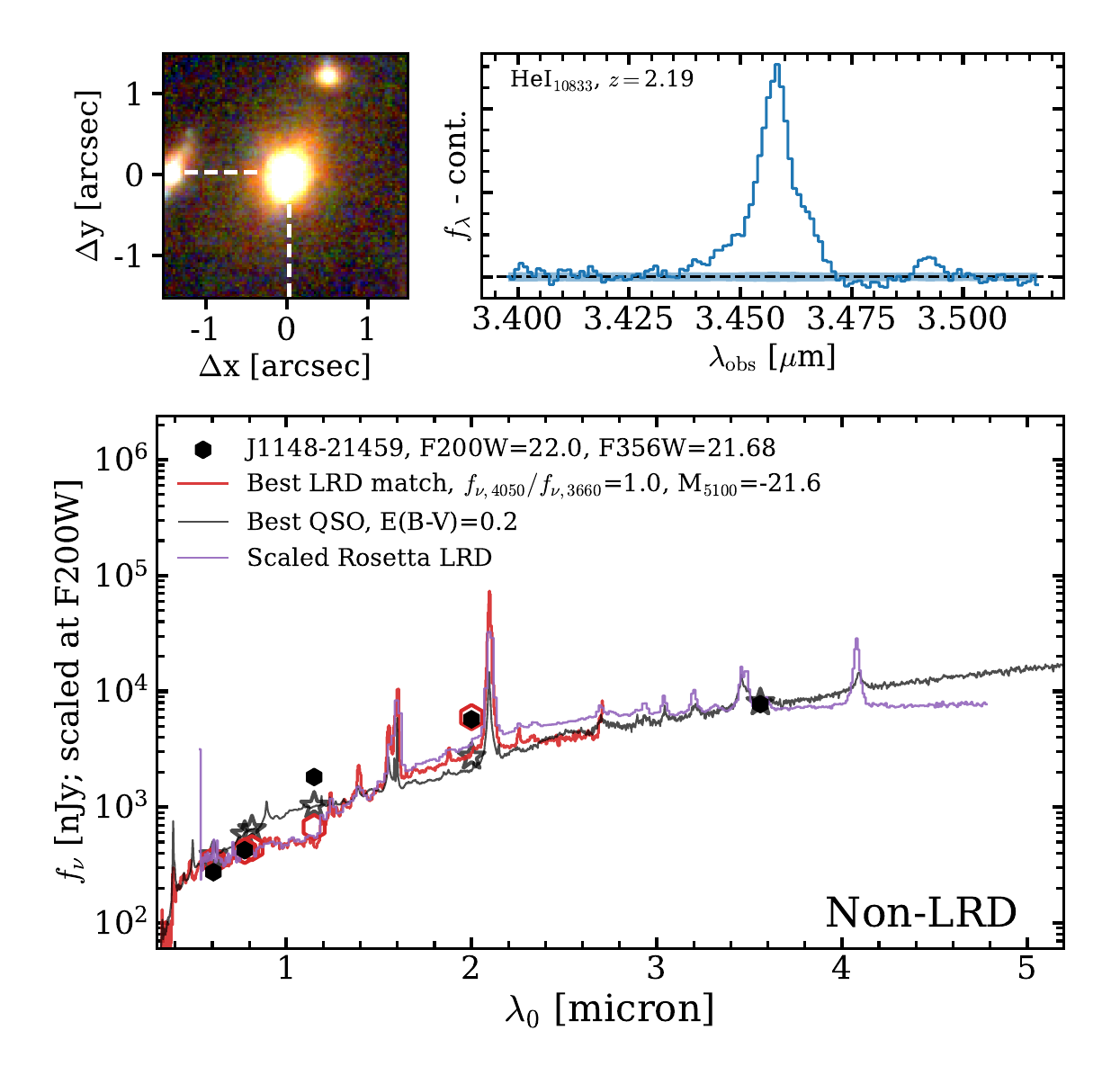} &
  \hspace{-0.8cm} \includegraphics[width=6.3cm]{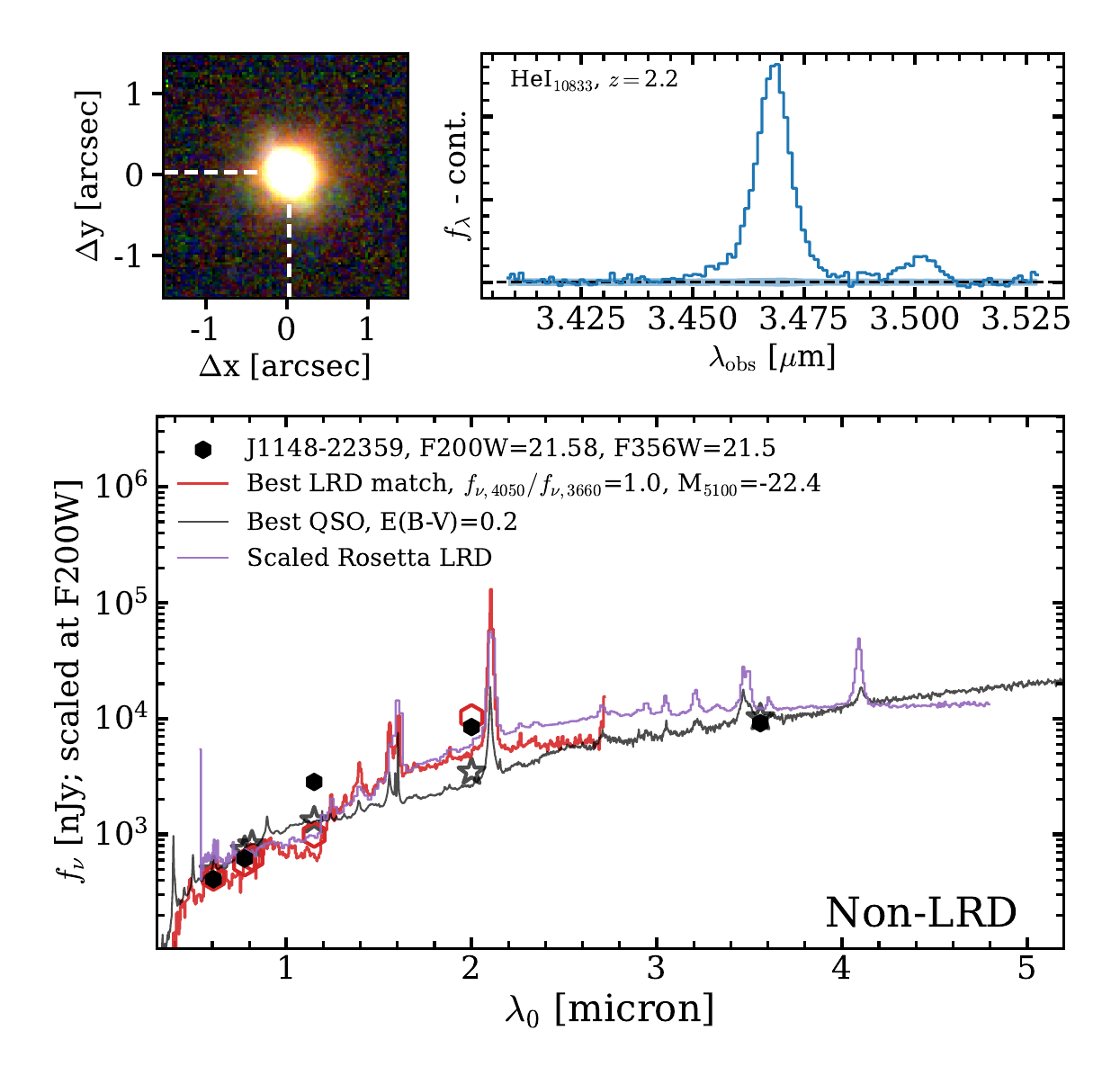} \\

\hspace{-0.8cm} \includegraphics[width=6.3cm]{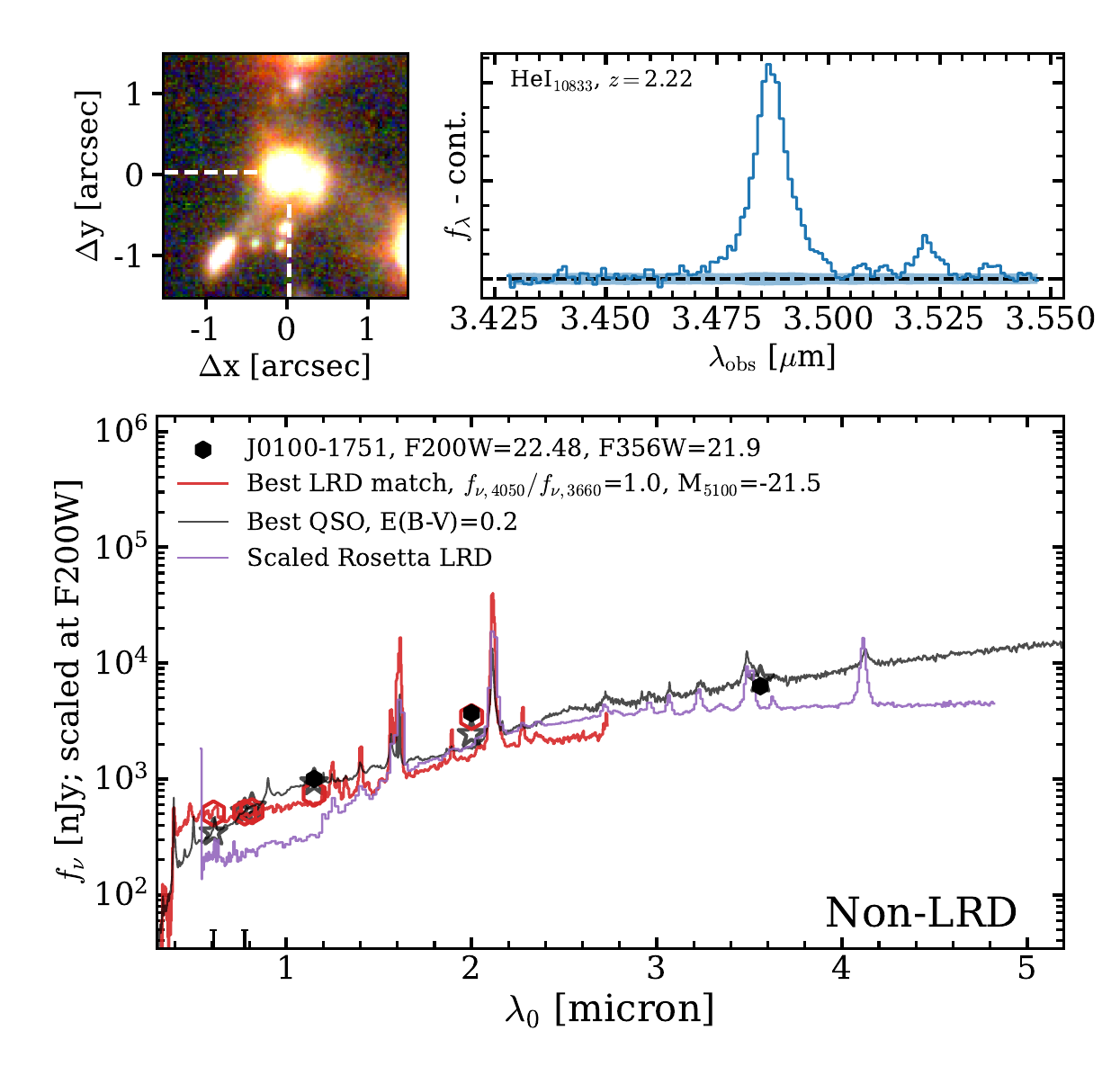} &
 \hspace{-0.8cm} \includegraphics[width=6.3cm]{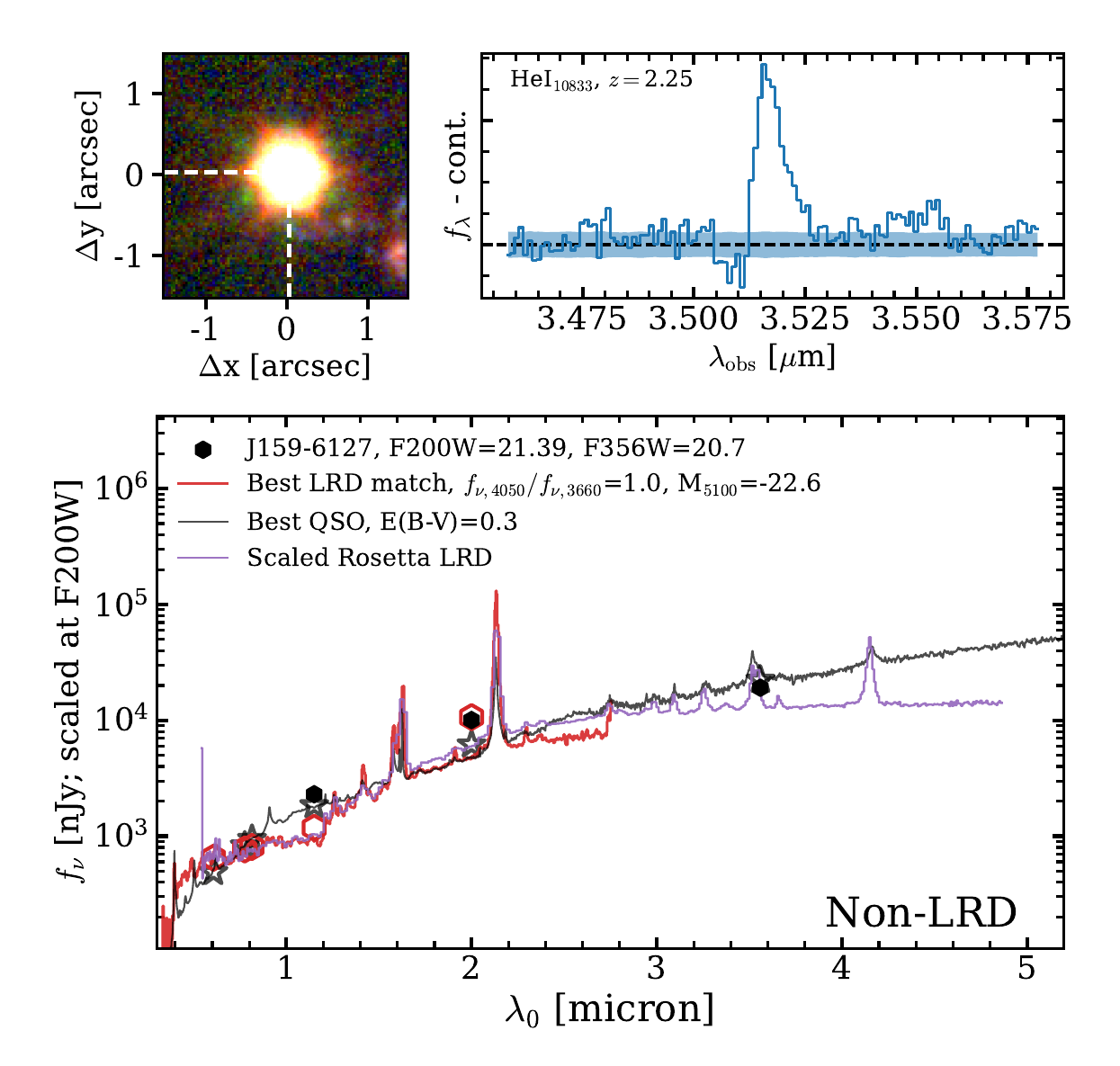} &
  \hspace{-0.8cm} \includegraphics[width=6.3cm]{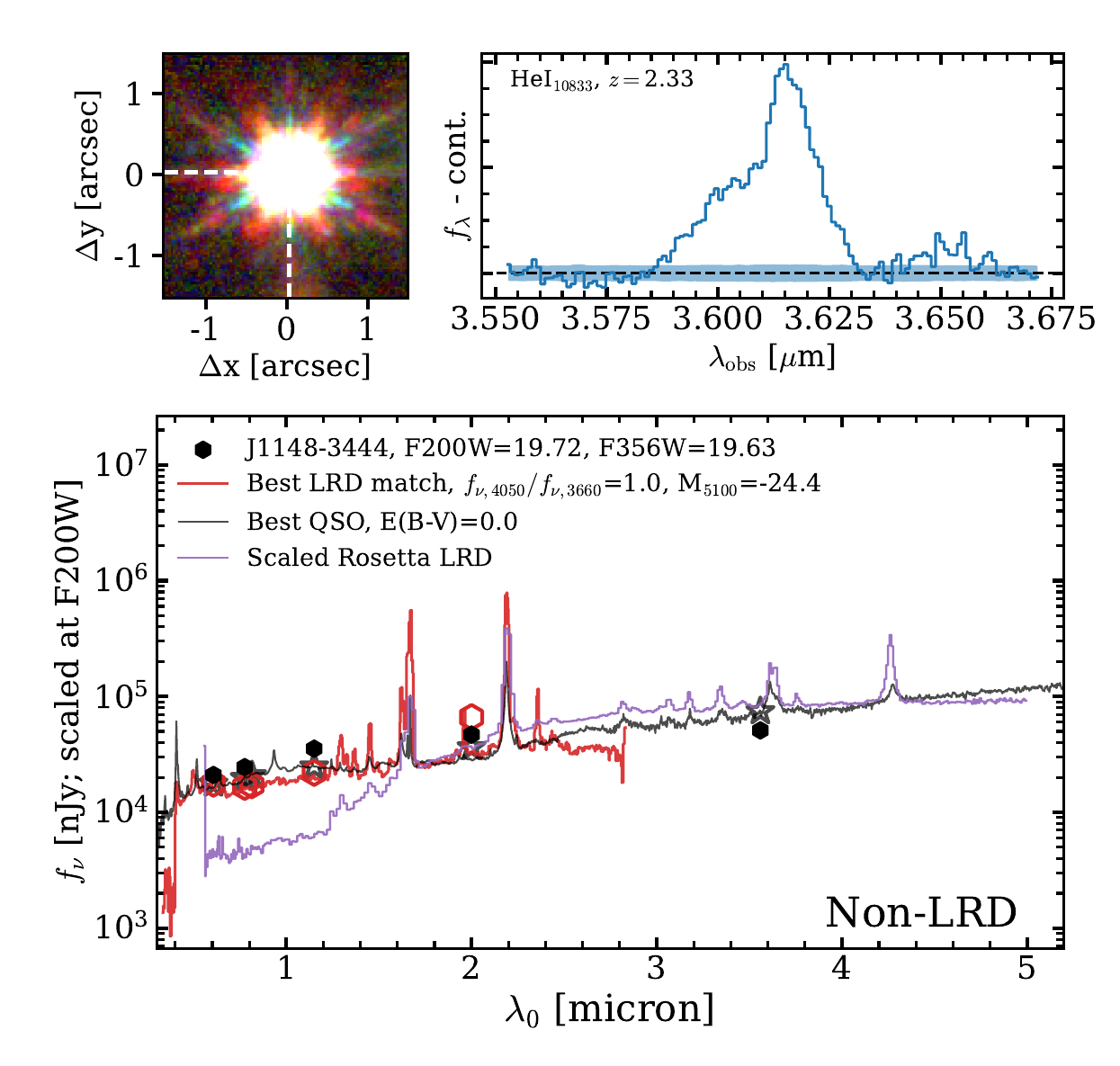} \\
    \end{tabular}
    \caption{As Fig. $\ref{fig:sources1}$, but now showing sources classed as non-LRDs.}
    \label{fig:sources_nonLRDs}
\end{figure*}

\begin{figure*}[!tp]
    \centering
    \begin{tabular}{ccc}
\hspace{-0.8cm} \includegraphics[width=6.3cm]{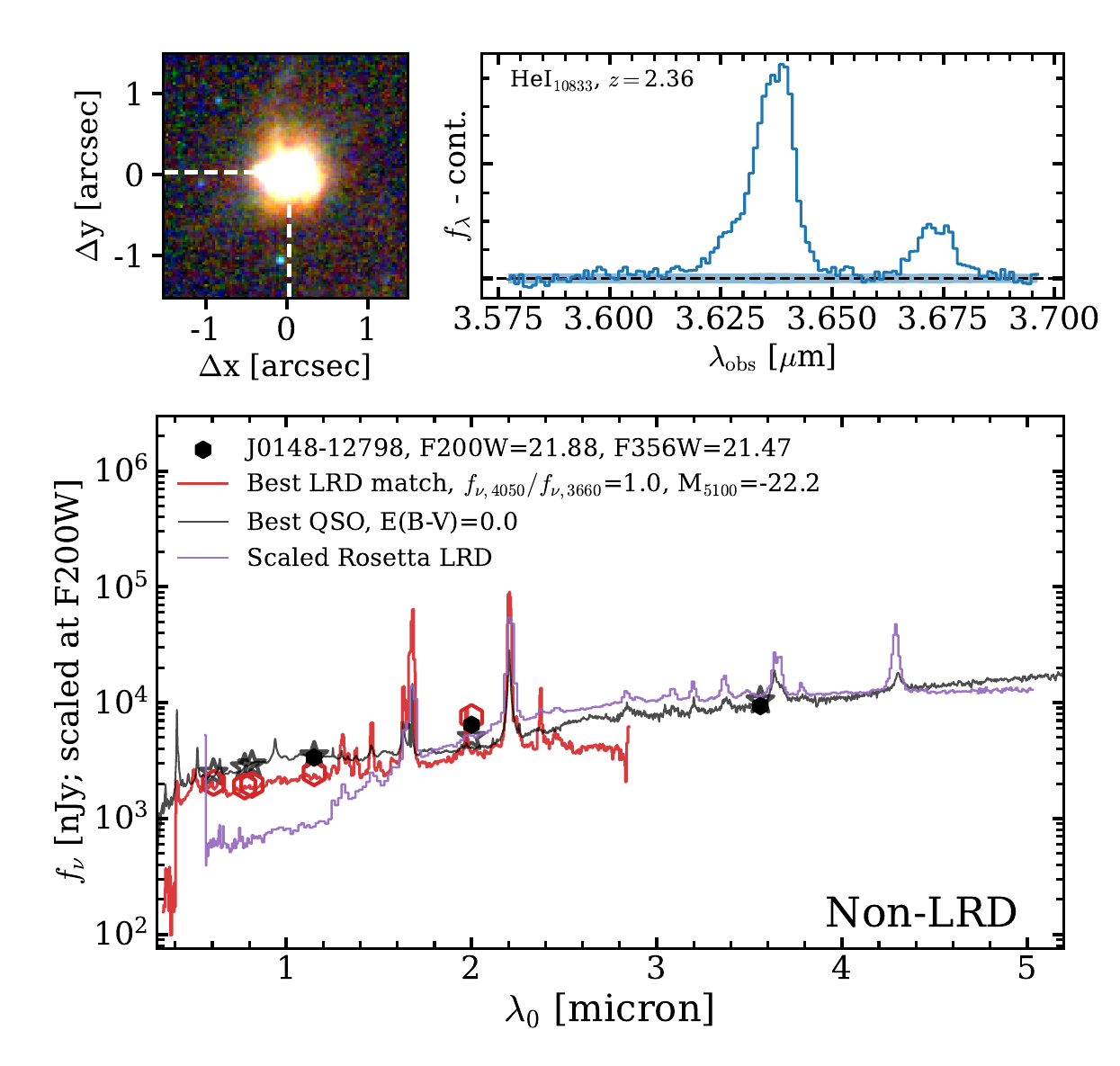} &
 \hspace{-0.8cm} \includegraphics[width=6.3cm]{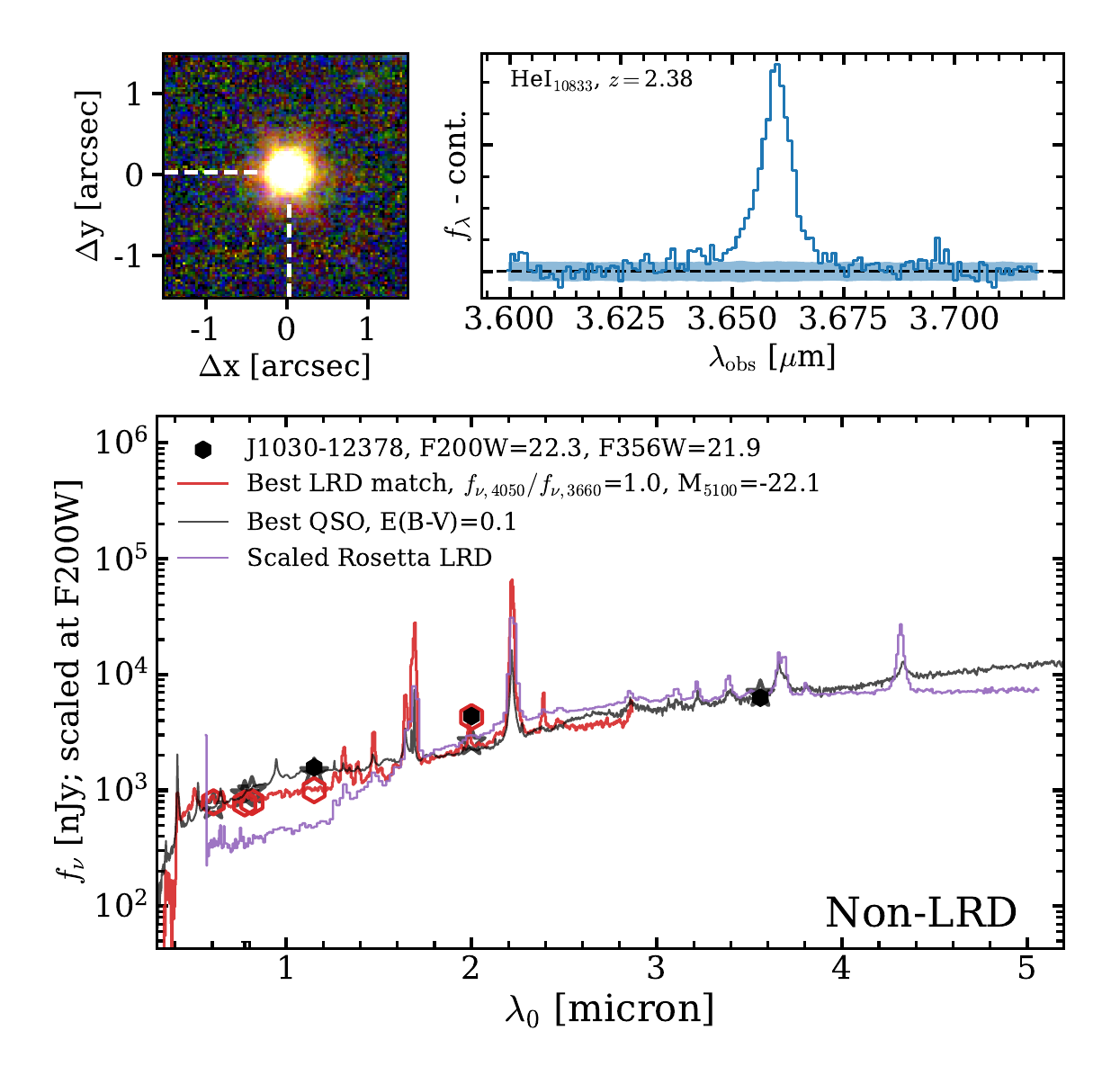} &
  \hspace{-0.8cm} \includegraphics[width=6.3cm]{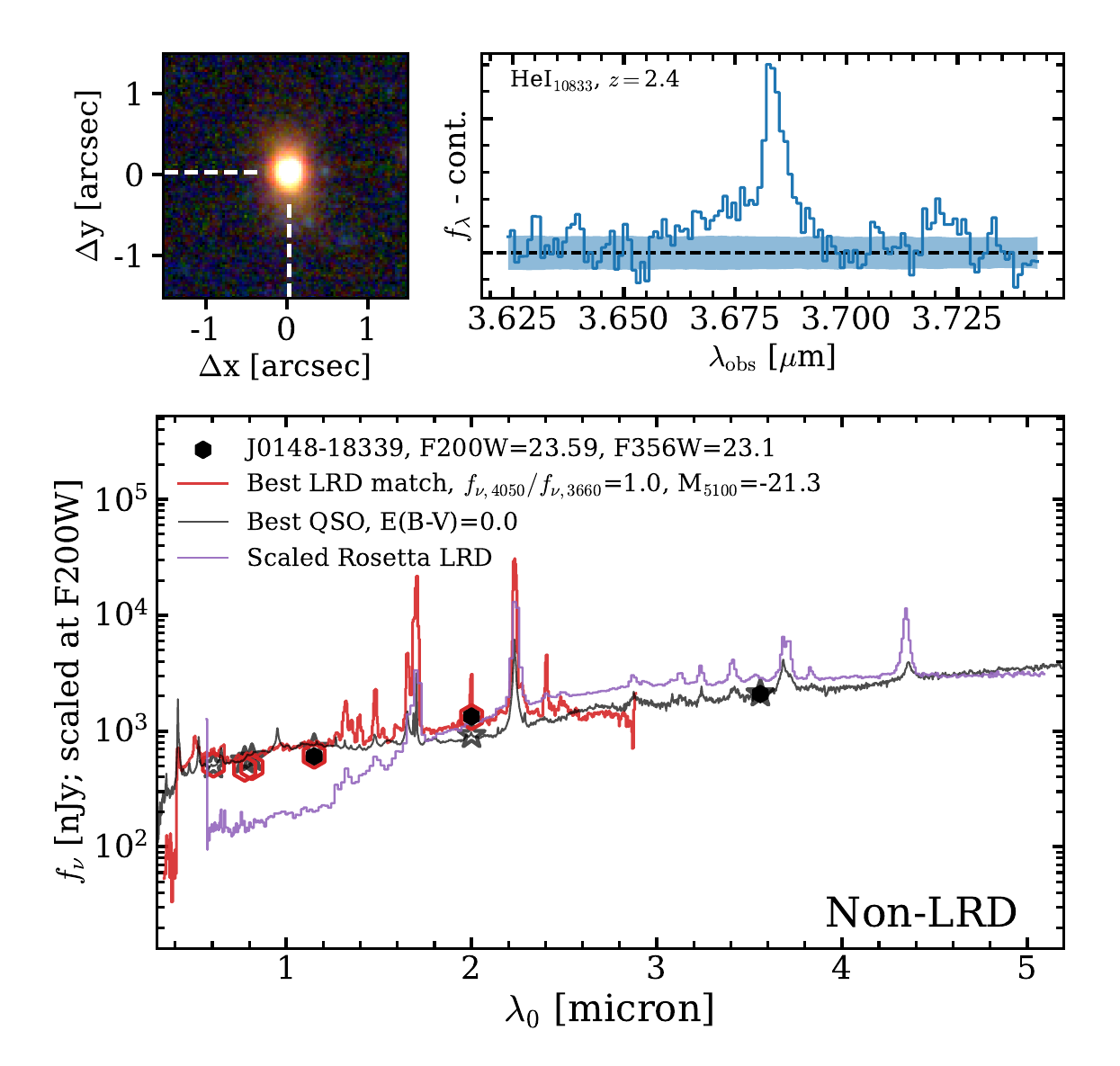} \\
\hspace{-0.8cm} \includegraphics[width=6.3cm]{figures/sources/INSPEC_J1030_9732.pdf} &
  \hspace{-0.8cm} \includegraphics[width=6.3cm]{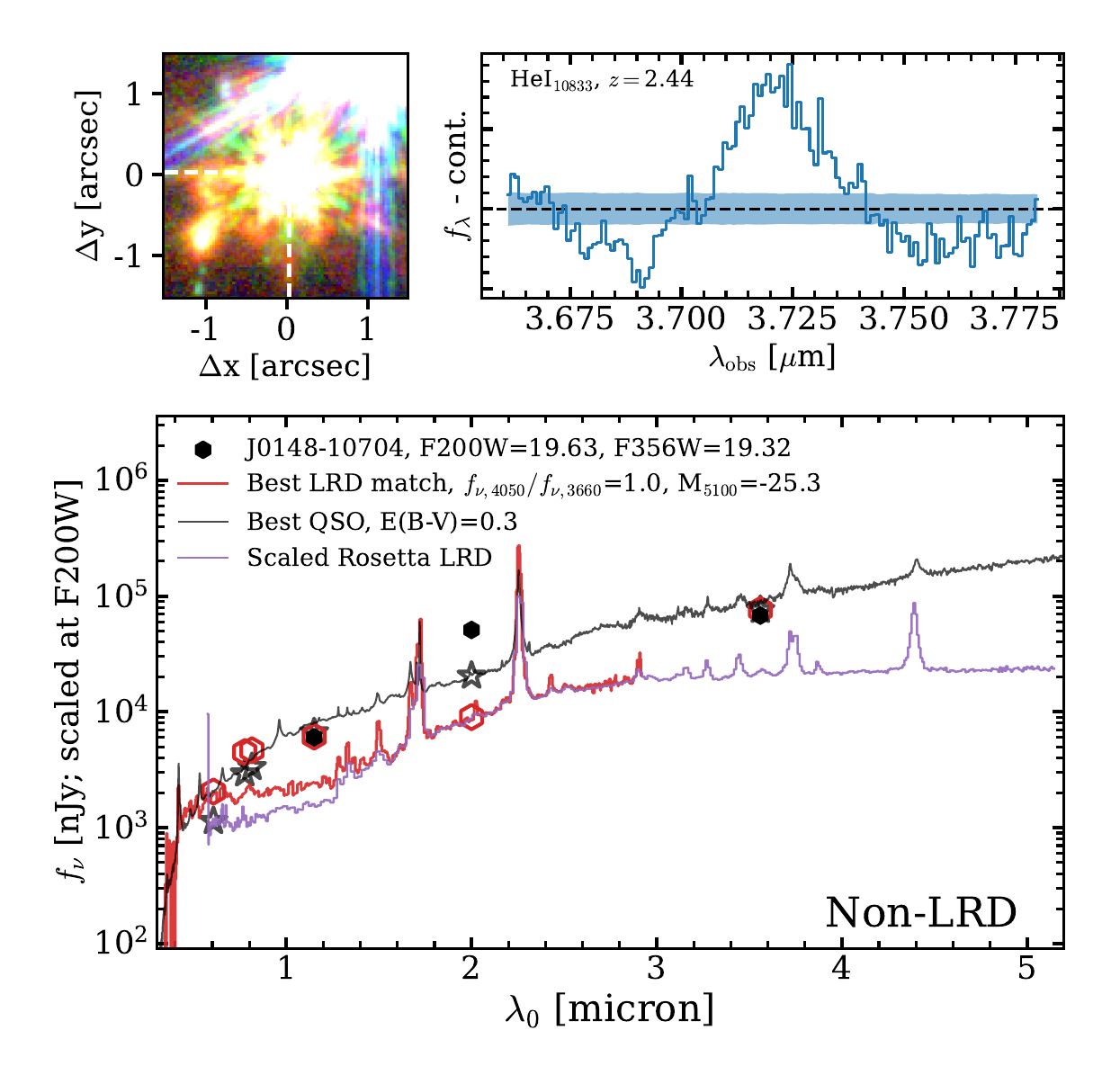} \\
    \end{tabular}
    \caption{Continued from Fig. $\ref{fig:sources_nonLRDs}$.}
    \label{fig:sources_nonLRDs2}
\end{figure*}

\end{appendix}

\end{document}